\documentclass[10pt,a4paper]{article}
\usepackage{times}
\usepackage{mathrsfs}
\usepackage{latexsym,amsopn,amsmath,amssymb}
\usepackage{amsthm}
\usepackage{amsfonts,amsbsy,amscd,stmaryrd}
\usepackage{paralist}
\usepackage{geometry}
\newcommand{\mbc}{\mathbb{C}}

\newcommand{\mbr}{\mathbb{R}}

\newcommand{\mbz}{\mathbb{Z}}
\newcommand{\ca}{\mathcal{A}}
\newcommand{\cb}{\mathcal{B}}
\newcommand{\cc}{\mathcal{C}}

\newcommand{\ce}{\mathcal{E}}
\newcommand{\cf}{\mathcal{F}}
\newcommand{\cg}{\mathcal{G}}
\newcommand{\ch}{\mathcal{H}}
\newcommand{\ci}{\mathcal{I}}

\newcommand{\ck}{\mathcal{K}}
\newcommand{\cl}{\mathcal{L}}
\newcommand{\cm}{\mathcal{M}}
\newcommand{\cn}{\mathcal{N}}

\newcommand{\cp}{\mathcal{P}}

\newcommand{\cs}{\mathcal{S}}
\newcommand{\ct}{\mathcal{T}}
\newcommand{\cu}{\mathcal{U}}
\newcommand{\cv}{\mathcal{V}}
\newcommand{\cw}{\mathcal{W}}
\newcommand{\cx}{\mathcal{X}}
\def\rond{\mathscr}
\newcommand{\ra}{\rond{A}}
\newcommand{\rb}{\rond{B}}
\newcommand{\rc}{\rond{C}}
\newcommand{\rd}{\rond{D}}
\newcommand{\re}{\rond{E}}

\newcommand{\rh}{\rond{H}}
\newcommand{\ri}{\rond{I}}

\newcommand{\rk}{\rond{K}}
\newcommand{\rl}{\rond{L}}
\newcommand{\rM}{\rond{M}}
\newcommand{\mr}{\rond{M}}
\newcommand{\rn}{\rond{N}}

\newcommand{\rp}{\rond{P}}
\newcommand{\rr}{\rond{R}}
\newcommand{\rs}{\rond{S}}
\newcommand{\rt}{\rond{T}}

\newcommand{\rx}{\rond{X}}
\def\rmb{\mathrm{b}}
\def\rmc{\mathrm{c}}
\def\rmd{\text{d}}
\def\rme{\mathrm{e}}
\def\rmo{\mathrm{o}}
\def\rmu{\mathrm{u}}
\def\rmw{\mathrm{w}}

\newcommand{\Cbu}{\cc_{\rmb}^{\rmu}}

\newcommand{\Co}{\cc_{\rmo}}
\newcommand{\Cc}{\cc_{\rmc}}
\renewcommand{\proof}{\noindent{\bf Proof: }}

\def\cchi{\raisebox{.45 ex}{$\chi$}} 

\def\bra#1{\langle{#1}|}
\def\ket#1{|{#1}\rangle}
\def\braket#1#2{\langle{#1}|{#2}\rangle}

\def\jap#1{\langle {#1} \rangle}
\def\spe{\mathrm{Sp_{ess}}}
\def\sp{\mathrm{Sp}}
\def\se{\sigma_{\mathrm{ess}}}

\def\rarrow{\rightarrow}

\def\what{\widehat}
\def\what#1{\widehat{ #1\,}}
\def\wtilde{\widetilde}

\def\nin{\notin}

\def\supp{\mbox{\rm supp\! }}

\def\nin{\notin}
\def\pprod{\textstyle \prod}
\def\ooplus{\textstyle{\bigoplus}}

\def\ccup{\textstyle{\bigcup}}
\def\ccap{\textstyle{\bigcap}}

\def\qed{\hfill \vrule width 8pt height 9pt depth-1pt \medskip}

\def\build#1_#2^#3{\mathrel{\mathop{\kern 0pt#1}\limits_{#2}^{#3}}}
\def\thesection{\arabic{section}}
\renewcommand{\theequation}{\arabic{section}.\arabic{equation}}

\newcounter{PAR}[section]
\def\thePAR{\thesection.\arabic{PAR}}
\def\PAR{\par\addvspace{2 ex}
\noindent 
\refstepcounter{PAR}{\bf \thePAR}
\hspace{1 mm}
}

\newtheorem{theorem}{Theorem}[section]
\newtheorem{lemma}[theorem]{Lemma}
\newtheorem{proposition}[theorem]{Proposition}
\newtheorem{corollary}[theorem]{Corollary}

{\theoremstyle{remark}
\newtheorem{definition}[theorem]{\bf Definition}
\newtheorem{remark}[theorem]{\bf Remark}
\newtheorem{remarks}[theorem]{\bf Remarks}
\newtheorem{example}[theorem]{\bf Example}

}
\long\def\symbolfootnote[#1]#2{\begingroup%
\def\thefootnote{\fnsymbol{footnote}}\footnote[#1]{#2}\endgroup} 
\begin{document}
\title{Hilbert $C^*$-modules and spectral analysis \\ 
of many-body systems} 
\author{Mondher DAMAK\thanks{\ University of Sfax, 3029 Sfax, Tunisia.
    E-mail: \texttt{mondher.damak@fss.rnu.tn}} 
\hspace{2mm} and 
    Vladimir GEORGESCU\thanks{\ CNRS
    and University of Cergy-Pontoise, 95000 Cergy-Pontoise, France.
    E-mail: \texttt{vlad@math.cnrs.fr}}}
\maketitle 
\vspace{-10mm}
\begin{abstract}
\noindent
We study the spectral properties of a class of many channel
Hamiltonians which contains those of systems of particles
interacting through $k$-body and field type forces which do not
preserve the number of particles.  Our results concern the essential
spectrum, the Mourre estimate, and the absence of singular
continuous spectrum.  The appropriate formalism involves graded
$C^*$-algebras and Hilbert $C^*$-modules as basic tools.
\end{abstract}
\tableofcontents

\section{Introduction and main results}
\label{s:intro}
\protect\setcounter{equation}{0}

In this section, after some general comments on the algebraic approach
that we shall use, we describe our main results in a slightly
simplified form. For notations and terminology, see Subsections
\ref{ss:not}, \ref{ss:group} and \ref{ss:grca}

\PAR\label{ss:intro} {\bf An algebraic approach}

By \emph{many-body system} we mean a system of particles interacting
between themselves through $k$-body forces with arbitrary $k\geq1$
but also subject to interactions which allow the system to make
transitions between states with different numbers of particles. The
second type of interactions consists of creation-annihilation
processes as in quantum field theory so we call them field type
interactions.

We use the terminology \emph{$N$-body system} in a rather loose
sense. Strictly speaking this should be a system of $N$ particles
which may interact through $k$-body forces with $1\leq k \leq N$.
However we also speak of $N$-body system when we consider the
following natural abstract version: the configuration space of the
system is a locally compact abelian group $X$, so the momentum space
is the dual group $X^*$, and the ``elementary Hamiltonians''
(cf. below) are of the form $h(P)+\sum_Y v_Y(Q)$.  Here $h$ is a
real function on $X^*$, the $Y$ are closed subgroups of $X$, and
$v_Y\in\Co(X/Y)$. One can give a meaning to the number $N$ even in
this abstract setting, but this is irrelevant here.

Similarly, we shall give a more general meaning to the notion of
many-body system: these are systems obtained by coupling a certain
number (possibly infinite) of $N$-body systems.  Our framework is
abstract and allows one to treat quite general examples which, even if
they do not have an immediate physical meaning, are interesting
because they furnish Hamiltonians with a rich many channel
structure. Note that here and below we do not use the word ``channel''
in the scattering theory sense, speaking about ``phase structure''
could be more appropriate.

The Hamiltonians we want to analyze are rather complicated objects and
standard Hilbert space techniques seem to us inefficient in this
situation.  Instead, we shall adopt a strategy proposed in
\cite{GI1,GI2} based on the observation that often the $C^*$-algebra
generated\symbolfootnote[2]{\ A self-adjoint operator $H$ on a Hilbert
  space $\ch$ is affiliated to a $C^*$-algebra $\rc$ of operators on
  $\ch$ if $(H+i)^{-1}\in\rc$. If $\re$ is a set of self-adjoint
  operators, the smallest $C^*$-algebra such that all $H\in\re$ are
  affiliated to it is the $C^*$-algebra generated by $\re$.  }  by the
Hamiltonians we want to study (we call them \emph{admissible}) has a
quite simple and remarkable structure which allows one to describe its
quotient with respect to the ideal of compact operators in more or
less explicit terms. And this suffices to get the qualitative spectral
properties which are of interest to us.  We shall refer to this
$C^*$-algebra as the \emph{Hamiltonian algebra} (or $C^*$-algebra of
Hamiltonians) of the system.

To clarify this we consider the case of $N$-body systems \cite{DG1}.
Let $X$ be a finite dimensional real vector space (the configuration
space).  Let $\ct$ be a set of subspaces of $X$.  In the
non-relativistic case an Euclidean structure is given on $X$ and the
simplest Hamiltonians are of the form
\begin{equation}\label{eq:hex}
H=\Delta+\sum_{Y\in\ct}v_Y(\pi_Y(x))
\end{equation}
where $\Delta$ is the Laplace operator, $v_Y$ is a continuous function
with compact support on the quotient space $X/Y$, and $\pi_Y:X\to X/Y$
is the canonical surjection (only a finite number of $v_Y$ is not
zero). Such Hamiltonians should clearly be admissible. On the other
hand, if a Hamiltonian $h(P)+V$ is considered as admissible then
$h(P+k)+V$ should be admissible too because the zero momentum $k=0$
should not play a special role. In other terms, translations in
momentum space should leave invariant the set of admissible
Hamiltonians.  We shall now describe the smallest $C^*$-algebra
$\rc_X(\cs)$ such that the operators \eqref{eq:hex} are affiliated to
it and which is stable under translations in momentum space.  Let
$\cs$ be the set of finite intersections of subspaces from $\ct$ and
\begin{equation*}
\cc_X(\cs)=\textstyle\sum^\rmc_{Y\in\cs}\Co(X/Y)\equiv
\text{ norm closure of } \textstyle\sum_{Y\in\cs}\Co(X/Y).
\end{equation*}
Note that one may think of $\cc_X(\cs)$ as a $C^*$-algebra of
multiplication operators on $L^2(X)$. Let $C^*(X)$ be the group
$C^*$-algebra of $X$ (see \S \ref{ss:group}). Then Corollary
\ref{co:cop2} gives:
\begin{equation*}
\rc_X(\cs)=\cc_X(\cs)\cdot C^*(X)\equiv
\text{closed linear subspace generated by the } S T \text{ with }
S\in\cc_X(\cs), T\in C^*(X).
\end{equation*}
It turns out that this algebra is canonically isomorphic with the
crossed product $\cc_X(\cs)\rtimes X$.  This example illustrates our
point: the Hamiltonian algebra of an $N$-body system is a remarkable
mathematical object. Moreover, $\cc_X(\cs)$ contains the ideal of
compact operators and its quotient with respect to it can be computed
by using general techniques from the theory of crossed products
\cite{GI1}.  On the other hand, $\cc_X(\cs)$ is equipped with an
$\cs$-graded $C^*$-algebra structure \cite{BG1,Ma,Ma2} and this gives
a method of computing the quotient which is more convenient in the
framework of the present paper.

The main difficulty in this algebraic approach is to isolate the
correct $C^*$-algebra.  Of course, we could accept an a priori given
$\rc$ as $C^*$-algebra of energy observables but we stress that a
correct choice is of fundamental importance: if the algebra $\rc$ we
start with is too large, then its quotient with respect to the
compacts will probably be too complicated to be useful. On the other
hand, if it is too small then physically relevant Hamiltonians will
not be affiliated to it. We refer to \cite{GI1,GI2,GI4,Geo} for
examples of Hamiltonian algebras of physical interest.

The basic object of this paper is the $C^*$-algebra $\rc$ defined in
Theorem \ref{th:C}. This is the Hamiltonian algebra of interest here,
in fact for us a many-body Hamiltonian is just a self-adjoint
operator affiliated to $\rc$. We shall see that this is a very large
class. On the other hand, it turns out that $\rc$ is generated by a
rather small class of ``elementary'' Hamiltonians involving only
quantum field like interactions, analogs in our context of the
Pauli-Fierz Hamiltonians. 

As in the $N$-body case \cite{ABG} the natural framework for the study
of many-body Hamiltonians is that of $C^*$-algebras graded by
semilattices. In fact, we are able to make a systematic spectral
analysis of the self-adjoint operators affiliated to $\rc$ because
$\rc$ is graded with respect to a certain semilattice $\cs$.  We shall
see that the channel structure and the formulas for the essential
spectrum and the threshold set which appears in the Mourre estimate
are completely determined by $\cs$, cf. Remark \ref{re:NM}.

Hilbert $C^*$-modules play an important technical role in the
construction of $\rc$, for example the component $\rc_{XY}$ of $\rc$
is a Hilbert $\rc_Y$-module where $\rc_Y$ is an $N$-body type algebra
(i.e. a crossed product as above). But they also play a more
fundamental role in a kind of second quantization formalism, see
\S\ref{ss:comments}.

We mention that the algebra $\rc$ is not adapted to symmetry
considerations, in particular in applications to physical systems
consisting of particles one has to assume them distinguishable. The
Hamiltonian algebra for systems of identical particles interacting
through field type forces (both bosonic and fermionic case) is
constructed in \cite{Geo}.

\PAR\label{ss:sintro} {\bf The Hamiltonian $C^*$-algebra $\rc$}

Let $\cs$ be a set of locally compact
abelian (lca) groups such that for  $X,Y\in\cs$:
\begin{compactenum}
\item[(i)] if $X\supset Y$ then the topology and the group structure
  of $Y$ coincide with those induced by $X$,
\item[(ii)] $X\cap Y \in \cs$,
\item[(iii)] there is $Z\in\cs$ such that $X\cup Y\subset Z$ and
  $X+Y$ is closed in $Z$,
\item[(iv)] $X\supsetneq Y \Rightarrow X/Y$ is not compact.
\end{compactenum}
If the first three conditions are satisfied we say that $\cs$ is an
\emph{inductive semilattice of compatible groups}.  Condition (iii) is
not completely stated, a compatibility assumption should be added (see
Definition \ref{df:iscg}).  However, this supplementary assumption is
automatically satisfied if all the groups are $\sigma$-compact
(countable union of compact sets).

The groups $X\in\cs$ should be thought as configuration spaces of
physical systems and the purpose of our formalism is to provide a
mathematical framework for the description of the coupled system.
If the systems are of the standard $N$-body type one may think that
the $X$ are finite dimensional real vector spaces. This, however,
will not bring any significative simplification of the proofs.

The following are the main examples one
should have in mind.

\begin{compactenum}
\item
Let $\cx$ be a $\sigma$-compact lca group and let $\cs$ be a set of
closed subgroups of $\cx$ with $\cx\in\cs$ and such that if
$X,Y\in\cs$ then $X\cap Y\in\cs$, $X+Y$ is closed, and $X/Y$ is not
compact if $X\supsetneq Y$.
\item
One may take $\cs$ \emph{equal to the set of all finite dimensional
vector subspaces of a vector space} over an infinite locally compact
field: this is the main example in the context of the many-body
problem.
\item \label{p:eu} The natural framework for the
  \emph{nonrelativistic many-body problem} is: $\cx$ is a real
  prehilbert space and $\cs$ a set of finite dimensional subspaces
  of $\cx$ such that if $X,Y\in\cs$ then $X\cap Y\in\cs$ and $X+Y$
  is included in a subspace of $\cs$ (there is a canonical choice,
  namely the set of \emph{all} finite dimensional subspaces of
  $\cx$). Then each $X\in\cs$ is an Euclidean space hence much more
  structure is available.
\item
One may consider an extension of the usual $N$-body problem by
taking as $\cx$ in example 1 above a finite dimensional real vector
space.  In the standard framework \cite{DeG1} the semilattice $\cs$
consists of linear subspaces of $\cx$ or here we allow them to be
closed additive subgroups. We mention that the closed additive
subgroups of $\cx$ are of the form $X=E+L$ where $E$ is a vector
subspace of $\cx$ and $L$ is a lattice in a vector subspace $F$ of
$\cx$ such that $E\cap F=\{0\}$.  More precisely, $L=\sum_k\mbz f_k$
where $\{f_k\}$ is a basis in $F$. Thus $F/L$ is a torus and if $G$
is a third vector subspace such that $\cx=E\oplus F\oplus G$ then
the space $\cx/X\simeq (F/L)\oplus G$ is a cylinder with $F/L$ as
basis.
\end{compactenum}

We assume that each $X\in\cs$ is equipped with a Haar measure, so
the Hilbert space $\ch(X)\equiv L^2(X)$ is well defined: this is the
state space of the system with $X$ as configuration space.  We
define the Hilbert space of the total system as the Hilbertian
direct sum
\begin{equation}\label{eq:hs}
\ch\equiv\ch_\cs=\oplus_X \ch(X).
\end{equation}
If $O=\{0\}$ is the zero group we take $\ch(O)=\mbc$.  There is no
particle number observable like in the Fock space formalism but
there is a remarkable $\cs$-valued observable \cite[\S 8.1.2]{ABG}
defined by associating to $X\in\cs$ the orthogonal projection
$\Pi_X$ of $\ch$ onto the subspace $\ch(X)$.

We shall identify $\Pi_X^*$ with the canonical embedding of $\ch(X)$
into $\ch$. We abbreviate\symbolfootnote[2]{\ $L(\ce,\cf)$ and
  $K(\ce,\cf)$ are the spaces of bounded and compact operators
  respectively between two Banach spaces $\ce,\cf$.}
\[
\rl_{XY}=L(\ch(Y),\ch(X)), \quad \rk_{XY}=K(\ch(Y),\ch(X)),
\quad \text{and} \quad \rl_X=\rl_{XX}, \quad \rk_X=\rk_{XX}.
\]
One may think of an operator $T$ on $\ch$ as a matrix with components
$T_{XY}=\Pi_X T\Pi_Y^*\in\rl_{XY}$ and write $T=(T_{XY})_{X,Y\in\cs}$.
We will be interested in subspaces of $L(\ch)$ constructed as direct
sums in the following sense. Assume that for each couple $X,Y$ we are
given a closed subspace $\rr_{XY}\subset\rl_{XY}$. Then we define
\begin{equation}\label{eq:sumc}
\rr\equiv(\rr_{XY})_{X,Y\in\cs}= {\textstyle\sum^\rmc_{X,Y\in\cs}}
\Pi_X^*\rr_{XY}\Pi_Y
\end{equation}
where $\sum^\rmc$ means closure of the sum. We say that the
$\rr_{XY}$ are the components of $\rr$.

For an arbitrary pair $X,Y\in\cs$ we define a closed subspace
$\rt_{XY}\subset\rl_{XY}$ as follows. Chose $Z\in\cs$ such that
$X\cup Y\subset Z$ and let $\varphi$ be a continuous function with
compact support on $Z$. It is easy to check that
$(T_{XY}(\varphi)u)(x)=\int_Y\varphi(x-y)u(y) \rmd y$ defines a
continuous operator $\ch(Y)\to\ch(X)$. Let $\rt_{XY}$ be the norm
closure of the set of these operators. This space is independent of
the choice of $Z$ and $\rt_{XX}=\cc^*(X)$ is the group $C^*$-algebra
of $X$.  Let $\rt\equiv\rt_\cs=(\rt_{XY})_{X,Y\in\cs}$ be defined as
in \eqref{eq:sumc}.  This is clearly a closed self-adjoint subspace
of $L(\ch)$ but is not an algebra in general.

If $X,Y\in\cs$ and $Y\subset X$ let $\pi_Y: X\to X/Y$ be the natural
surjection and let $\cc_X(Y)\cong\Co(X/Y)$ be the $C^*$-algebra of
bounded uniformly continuous functions on $X$ of the form
$\varphi\circ\pi_Y$ with $\varphi\in\Co(X/Y)$. If $X,Y\in\cs$ and
$Y\not\subset X$ let $\cc_X(Y)=\{0\}$. Then let
$\cc_X=\sum^\rmc_{Y}\cc_X(Y)$, this is also a $C^*$-algebra of
bounded uniformly continuous functions on $X$.  We embed
$\cc_X\subset\rl_X$ by identifying a function with the operator on
$\ch(X)$ of multiplication by that function. Then let
\begin{equation}\label{eq:imp0}
\cc\equiv\cc_\cs=\oplus_X\cc_X,
\end{equation}
this is a $C^*$-algebra of operators on $\ch$.  Moreover, for each
$Z\in\cs$ let 
\begin{equation}\label{eq:impz}
\cc(Z)\equiv\cc_\cs(Z)=\oplus_X\cc_X(Z)=\oplus_{X\supset Z}\cc_X(Z),
\end{equation}
this is a $C^*$-subalgebra of $\cc$ and we clearly have
$\cc=\sum^\rmc_Z\cc(Z)$.

\begin{theorem}\label{th:C}
The space\symbolfootnote[2]{\ If $\ce,\cf,\cg$ are Banach spaces and
  $(e,f)\mapsto ef$ is a bilinear map $\ce\times\cf\to\cg$ and if
  $E\subset\ce,F\subset\cf$ are linear subspaces then $EF$ is the
  linear subspace of $\cg$ generated by the elements $ef$ with $e\in
  E,f\in F$ and $E\cdot F$ is its closure.  } $\rc=\rt\cdot\rt$ is a
$C^*$-algebra of operators on $\ch$ and we have
\begin{equation}\label{eq:imp1}
\rc=\rt\cdot\cc=\cc\cdot\rt
\end{equation}
For each $Z\in\cs$ let
\begin{equation}\label{eq:imp2}
\rc(Z)=\rt\cdot\cc(Z)=\cc(Z)\cdot\rt.
\end{equation} 
This is a $C^*$-subalgebra of $\rc$ and $\{\rc(Z)\}_{Z\in\cs}$ is a
linearly independent family of $C^*$-subalgebras of $\rc$ such that
$\sum^\rmc_Z \rc(Z)=\rc$ and $\rc(Z')\rc(Z'')\subset\rc(Z'\cap Z'')$
for all $Z',Z''\in\cs$.
\end{theorem}

This is the main technical result of our paper.  Indeed, by using
rather simple techniques involving graded $C^*$-algebras and the
Mourre method one may deduce from Theorem \ref{th:C} important
spectral properties of many-body Hamiltonians.  The last assertion of
the theorem is an explicit description of the fact that $\rc$ \emph{is
  equipped with an $\cs$-graded $C^*$-algebra structure.}  We set
$\rc=\rc_\cs$ when needed.

The choice of $\rc$ may seem arbitrary but in fact is quite natural in
our context: not only all the many-body Hamiltonians of interest for
us are self-adjoint operators affiliated to $\rc$, but also $\rc$ is
the smallest $C^*$-algebra with this property, cf. Theorem
\ref{th:motiv} for a precise statement.

\begin{remark}\label{re:exp}
Note that $\rc_{XY}=\sum^\rmc_Z\rc_{XY}(Z)$. In matrix notation we
have
\[
\rc=(\rc_{XY})_{X,Y\in\cs} \quad \text{where} \quad
\rc_{XY}=\cc_X\cdot\rt_{XY}=\rt_{XY}\cdot\cc_Y
\]
and $\rc(Z)=(\rc_{XY}(Z))_{X,Y\in\cs}$ where
\[
\rc_{XY}(Z)=\cc_X(Z)\cdot\rt_{XY}=\rt_{XY}\cdot\cc_Y(Z) \hspace{2mm}
\text{if } Z\subset X\cap Y \hspace{2mm} \text{and }
\rc_{XY}(Z)=\{0\} \hspace{2mm} \text{if }  Z\not\subset X\cap Y.
\]
We mention that if $Z$ is complemented in $X$ and $Y$ then
$\rc_{XY}(Z)\simeq \cc^*(Z)\otimes \rk_{X/Z,Y/Z}$.
\end{remark}

\begin{remark}\label{re:rief}
If $X\supset Y$ then the space $\rt_{XY}$ is a ``concrete''
realization of the Hilbert $C^*$-module introduced by Rieffel in
\cite{Ri} which implements the Morita equivalence between the group
$C^*$-algebra $\cc^*(Y)$ and the crossed product $\Co(X/Y)\rtimes
X$. More precisely, $\rt_{XY}$ is equipped with a natural Hilbert
$\cc^*(Y)$-module structure such that its imprimitivity algebra is
canonically isomorphic with $\Co(X/Y)\rtimes X$. In Section
\ref{s:pair} we shall see that for arbitrary $X,Y\in\cs$ the space
$\rt_{XY}$ has a canonical structure of Hilbert $(\Co(X/(X\cap
Y))\rtimes X,\Co(Y/(X\cap Y))\rtimes Y)$ imprimitivity
bimodule. This fact is technically important for the proof of our
main results but plays no role in this introduction.
\end{remark}

\begin{remark}\label{re:dirac}
A simple extension of our formalism allows one to treat particles
with arbitrary spin. Indeed, if $E$ is a complex Hilbert then the
last part of Theorem \ref{th:C} remains true if $\rc$ is replaced by
$\rc^E=\rc\otimes K(E)$ and the $\rc(Z)$ by $\rc(Z)\otimes K(E)$. If
$E$ is the spin space then it is finite dimensional and one obtains
$\rc^E$ exactly as above by replacing the $\ch(X)$ by $\ch(X)\otimes
E=L^2(X;E)$. Then in our later results one may consider instead of
scalar kinetic energy functions $h$ self-adjoint operator valued
functions $h:X^*\to L(E)$. For example, we may take as one particle
kinetic energy operators the Pauli or Dirac Hamiltonians.
\end{remark}

The preceding definition of $\rc$ is quite efficient for theoretical
purposes but much less for practical questions: for example, it is not
obvious how to decide if a self-adjoint operator is affiliated to
it. Our next result is an ``intrinsic'' characterization of
$\rc_{XY}(Z)$ which is relatively easy to check. Since $\rc$ is
constructed in terms of the $\rc_{XY}(Z)$, we get simple affiliation
criteria.

For $x\in X$ and $k\in X^*$ (dual group) we define unitary operators
in $\ch(X)$ by $(U_xu)(x')=u(x'+x)$ and $(V_k u)(x)=k(x)u(x)$. These
correspond to the momentum and position observables $P\equiv P_X$
and $Q\equiv Q_X$ of the system.  If $X,Y\in\cs$ then one can
associate to an element $z\in X\cap Y$ a translation operator in
$\ch(X)$ and a second one in $\ch(Y)$. We shall however denote both
of them by $U_z$ since which of them is really involved in some
relation will always be obvious from the context. If $X$ and $Y$ are
subgroups of a lca group $G$ (equipped with the topologies induced
by $G$) then we have canonical surjections $G^*\to X^*$ and $G^*\to
Y^*$ defined by restriction of characters. So a character $k\in G^*$
defines an operator of multiplication by $k|_X$ on $\ch(X)$ and an
operator of multiplication by $k|_Y$ on $\ch(Y)$. Both will be
denoted $V_k$. In our context the lca group $X+Y$ is well defined
(but generally does not belong to $\cs$) and we may take $G=X+Y$,
cf. Remark \ref{re:cpt}.  Below we denote $Z^\perp$ the polar set of
$Z\subset X$ in $X^*$.

\begin{theorem}\label{th:imp3}
If $Z\subset X\cap Y$ then $\rc_{XY}(Z)$ is the set of
$T\in\rl_{XY}$ satisfying $U_z^*T U_z=T$ if $z\in Z$ and such that
\begin{compactenum}
\item[{\rm(i)}] 
$\|(U_x-1)T\|\to 0$ if $x\to 0$ in $X$ and 
$\|T(U_y-1)\|\to 0$ if $y\to 0$ in $Y$, 
\item[{\rm(ii)}]
$\|V^*_k T V_k-T\|\to 0$ if $k\to 0$ in $(X+Y)^*$ and 
$\|(V_k-1)T\|\to 0$ if $k\to 0$ in $Z^\perp$.
\end{compactenum}
\end{theorem}

Theorem \ref{th:imp3} becomes simpler and can be improved in the
context of Example \ref{p:eu} page \pageref{p:eu}. So let us assume
that $\cs$ consists of finite dimensional subspaces of a real
prehilbert space. Then each $X$ is equipped with an Euclidean
structure and this allows to identify $X^*=X$ such that $V_k$ becomes
the operator of multiplication by the function
$x\mapsto\rme^{i\braket{x}{k}}$ where the scalar product
$\braket{x}{k}$ is well defined for any $x,k$ in the ambient
prehilbert space. For $X\supset Y$ we identify $X/Y=X\ominus Y$, the
orthogonal of $Y$ in $X$.

\begin{corollary}\label{co:imp3}
Under the conditions of Example \ref{p:eu} page \pageref{p:eu} the
space $\rc_{XY}(Z)$ is the set of $T\in\rl_{XY}$ satisfying the next
two conditions:
\begin{compactenum} 
\item[{\rm(i)}]
$U_z^*T U_z=T$ for $z\in Z$ and
$\|V^*_z T V_z-T\|\to 0$ if $z\to 0$ in $Z$,
\item[{\rm(ii)}] 
$\|T(U_y-1)\|\to 0$ if $y\to 0$ in $Y$ and $\|T(V_k-1)\|\to 0$ if
$k\to 0$ in $Y/Z$.
\end{compactenum}
Condition 2 may be replaced with:
\begin{compactenum}
\item[\rm{(iii)}] 
$\|(U_x-1)T\|\to 0$ if $x\to 0$ in $X$ and $\|(V_k-1)T\|\to 0$ if
$k\to 0$ in $X/Z$.
\end{compactenum}
\end{corollary}

\PAR\label{ss:motiv} {\bf Elementary Hamiltonians }

Our purpose in this subsection is to show that $\rc$ is a
$C^*$-algebra of Hamiltonians in a rather precise sense, according to
the terminology used in \cite{GI1,GI2}: we show that $\rc$ is the
$C^*$-algebra generated by a simple class of Hamiltonians which have a
natural quantum field theoretic interpretation.  Since our desire is
only to motivate our construction, in this subsection we shall make
two simplifying assumptions: $\cs$ is finite and if $X,Y\in\cs$ with
$X\supset Y$, then $Y$ is complemented in $X$.

For each couple $X,Y\in\cs$ such that $X\supset Y$ we chose a closed
subgroup $X/Y$ of $X$ such that $X=(X/Y)\oplus Y$. Moreover, we
equip $X/Y$ with the quotient Haar measure which gives us a
factorization $\ch(X)=\ch(X/Y)\otimes\ch(Y)$. Then we define
$\Phi_{XY}\subset\rl_{XY}$ as the closed linear subspace consisting
of ``creation operators'' associated to states from $\ch(X/Y)$,
i.e. operators $a^*(\theta):\ch(Y)\to\ch(X)$ with
$\theta\in\ch(X/Y)$ which act as $u\mapsto\theta\otimes u$.  We set
$\Phi_{YX}=\Phi_{XY}^*\subset\rl_{YX}$, this is the space of
``annihilation operators'' $a(\theta)=a^*(\theta)^*$ defined by
$\ch(X/Y)$.  This defines $\Phi_{XY}$ when $X,Y$ are comparable,
i.e. $X\supset Y$ or $X\subset Y$, which we abbreviate by $X\sim
Y$. If $X\not\sim Y$ then we take $\Phi_{XY}=0$. Note that
$\Phi_{XX}=\mbc 1_X$, where $1_X$ is the identity operator on
$\ch(X)$, because $\ch(O)=\mbc$.

The space $\Phi_{XY}$ for $X\supset Y$ clearly depends on the choice
of the complement $X/Y$. On the other hand, according to Definition
\ref{df:ryz} and Proposition \ref{pr:def3}, we have
\begin{equation}\label{eq:Phi}
\cc^*(X)\cdot\Phi_{XY}=\Phi_{XY}\cdot \cc^*(Y)=\rt_{XY} \quad
\text{if } X\sim Y.
\end{equation}
This seems to us a rather remarkable feature because not only
$\rt_{XY}$ is independent of $X/Y$ but is also well defined even if
$Y$ is not complemented in $X$.

Now we define $\Phi=(\Phi_{XY})_{X,Y\in\cs}\subset L(\ch)$. This is
a closed self-adjoint linear space of bounded operators on $\ch$.  A
symmetric element $\phi\in\Phi$ will be called \emph{field
operator}, this is the analog of a field operator in the present
context. Giving such a $\phi$ is equivalent to giving a family 
$\theta=(\theta_{XY})_{X\supset Y}$ of elements
$\theta_{XY}\in\ch(X/Y)$, the components of the operator
$\phi\equiv\phi(\theta)$ being given by:
$\phi_{XY}=a^*(\theta_{XY})$ if $X\supset Y$, then
$\phi_{XY}=a(\theta_{YX})$ if $X\subset Y$, and finally
$\phi_{XY}=0$ if $X\not\sim Y$. Note that $\Phi_{XX}=\mbc 1_X$
because $\ch(O)=\mbc$. If $u=(u_X)_{X\in\cs}$ then we have
\[
\braket{u}{\phi u} = {\textstyle\sum_{X\supset Y}} 2\Re
\braket{\theta_{XY}\otimes u_Y}{u_X}.
\]
A \emph{standard kinetic energy operator} is an operator on $\ch$ of
the form $K=\oplus_X h_X(P)$ where $h_X:X^*\to\mbr$ is continuous and
$\lim_{k\to\infty}|h_X(k)|=\infty$. The operators of the form
$K+\phi$, where $K$ is a standard kinetic energy operator and
$\phi\in\Phi$ is a field operator, will be called \emph{Pauli-Fierz
  Hamiltonians}.

The proof of the next theorem may be found in the Appendix.

\begin{theorem}\label{th:motiv}
Assume that $\cs$ is finite and that $Y$ is complemented in $X$ if
$X\supset Y$. Then $\rc$ coincides with the $C^*$-algebra generated
by the Pauli-Fierz Hamiltonians.
\end{theorem}

\begin{remark}\label{re:fnb}
It is interesting and important to note that $\rc$ is generated by a
class of Hamiltonians involving only an elementary class of field type
interactions. However, as we shall see in \S\ref{ss:ex}, the class of
Hamiltonians affiliated to $\rc$ is very large and covers $N$-body
systems interacting between themselves (i.e. for varying $N$) with
field type interactions. In particular, the $N$-body type interactions
are generated by pure field interactions and this thanks to the
semilattice structure of $\cs$.
\end{remark}

\PAR\label{ss:grad} {\bf Essential spectrum of operators affiliated
  to $\rc$}

The main assertion of Theorem \ref{th:C} is that $\rc$ is an
$\cs$-graded $C^*$-algebra. The class of $C^*$-algebras graded by
finite semilattices has been introduced and their role in the spectral
theory of $N$-body systems has been pointed out in
\cite{BG1,BG2}. Then the theory has been extended to infinite
semilattices in \cite{DG2}. A much deeper study of this class of
$C^*$-algebras is the subject of the thesis \cite{Ma} of Athina
Mageira (see also \cite{Ma2,Ma3}) whose results allowed us to consider
a semilattice $\cs$ of arbitrary abelian groups (and this is important
in certain applications that we do not mention in this paper). We
mention that her results cover non-abelian groups and the assumption
(iv) (on non-compact quotients) is not necessary in her
construction. This could open the way to interesting extensions of our
formalism.

In \S\ref{ss:grca} we recall some basic facts concerning graded
$C^*$-algebras. Our main tool for the spectral analysis of the
self-adjoint operators affiliated to $\rc$ is Theorem \ref{th:ga}. For
example, it is easy to derive from it the abstract HVZ type
description of the essential spectrum given in Theorem \ref{th:gas}.
Here we give a concrete application in the present framework, more
general results may be found in Sections \ref{s:grad} and
\ref{s:af}.

For each $X\in\cs$ we define a closed subspace of $\ch$ by
\begin{equation}\label{eq:he}
\ch_{\geq X}={\textstyle\bigoplus_{Y\supset X}}\ch(Y).
\end{equation}
This is associated to the semilattice 
$\cs_{\geq X}=\{Y\in\cs \mid Y\supset X\}$ in the same way
as $\ch$ is associated to $\cs$. Let $\rc_{\geq X}$ be the
$C^*$-subalgebra of $\rc$ given by
\begin{equation}\label{eq:ce}
\rc_{\geq X}={\textstyle\sum^\rmc_{Y\supset X}\rc(Y)}\cong
\big({\textstyle\sum^\rmc_{Y\supset X}} \rc_{EF}(Y)\big)_{E\cap F
\supset X}
\end{equation}
and note that $\rc_{\geq X}$ lives on the subspace $\ch_{\geq X}$ of
$\ch$. Moreover, $\rc$ and $\rc_{\geq X}$ are nondegenerate algebras
of operators on the Hilbert spaces $\ch$ and $\ch_{\geq X}$
respectively. It can be shown that there is a unique linear
continuous projection $\rp_{\geq X}:\rc\to\rc_{\geq X}$ such that 
$\rp_{\geq X}(T)=0$ if $T\in\rc(Y)$ with $Y\not\supset X$ and that
this projection is a morphism, cf. Theorem \ref{th:ga}.

Let $H$ be a self-adjoint operator on a Hilbert space $\ch$
affiliated to a $C^*$-algebra of operators $\ra$ on $\ch$.  Then
$\varphi(H)\in\ra$ for all $\varphi\in\Co(\mbr)$. If $\ra$ is the
closed linear span of the elements $\varphi(H)A$ with
$\varphi\in\Co(\mbr)$ and $A\in\ra$, we say that $H$ is
\emph{strictly affiliated to $\ra$}.

Assume that the semilattice $\cs$ has a smallest element
$\min\cs$. Then $X\in\cs$ is an atom if the only element of $\cs$
strictly included in $X$ is $\min\cs$.  Let $\cp(\cs)$ be the set of
atoms of $\cs$. We say that $\cs$ is \emph{atomic} if each of its
elements not equal to $\min\cs$ contains an atom. It is clear that
if the zero group $O$ belongs to $\cs$ then $O$ is the smallest
element of $\cs$ and $\rc(O)=K(\ch)$.

\begin{theorem}\label{th:imp4}
If $H$ is a self-adjoint operator on $\ch$ strictly affiliated to
$\rc$ then for each $X\in\cs$ there is a unique self-adjoint
operator $H_{\geq X}\equiv\rp_{\geq X}(H)$ on $\ch_{\geq X}$ such
that $\rp_{\geq X} (\varphi(H))=\varphi(H_{\geq X})$ for all
$\varphi\in\Co(\mbr)$. The operator $H_{\geq X}$ is strictly
affiliated to $\rc_{\geq X}$.  If $O\in\cs$ and $\cs$ is atomic then
the essential spectrum of $H$ is given by
\begin{equation}\label{eq:esi}
\spe(H)=\overline{\ccup}_{X\in\cp(\cs)}\sp(H_{\geq X}).
\end{equation} 
\end{theorem}

\PAR\label{ss:ex} {\bf Hamiltonians affiliated to $\rc$}

We shall give now examples of self-adjoint operators strictly
affiliated to $\rc$.  The argument is relatively straightforward
thanks to Theorem \ref{th:imp3} but the fact that $\cs$ is allowed to
be infinite brings some additional difficulties.  We are interested in
Hamiltonians of the form $H=K+I$ where $K$ is the kinetic energy
operator of the system and $I$ is the interaction term. Formally $H$
is a matrix of operators $(H_{XY})_{X,Y\in\cs}$, the operator $H_{XY}$
is defined on a subspace of $\ch(Y)$ and has values in $\ch(X)$, and
we have $H_{XY}^*=H_{YX}$ (again formally).  Then
$H_{XY}=K_{XY}+I_{XY}$ and our assumptions will be that $K$ is
diagonal, so $K_{XY}=0$ if $X\neq Y$ and $K_{XX}\equiv K_X$.  The
interactions will be of the form $I_{XY}=\sum_{Z\subset X\cap
  Y}I_{XY}(Z)$, this expresses the $N$-body structures of the various
systems (with various $N$, of course).  Then $H_{XX}=K_X+I_{XX}$ will
be a generalized $N$-body type Hamiltonian ($I_{XX}$ may depend on the
momentum). The non-diagonal operators $H_{XY}=I_{XY}$ define the
interaction between the systems $X$ and $Y$ (these operators too may
depend on the momentum of the systems $X,Y$).  We give now a rigorous
construction of such Hamiltonians.

{\bf(a)} For each $X$ we choose a kinetic energy operator $K_X=h_X(P)$
for the system having $X$ as configuration space. The function
$h_X:X^*\to\mbr$ must be continuous and such that $|h_X(x)|\to\infty$
if $k\to\infty$.  We emphasize the fact that there are no relations
between the kinetic energies $K_X$ of the systems corresponding to
different $X$.  If $\cs$ is infinite, we require
$\lim_X\inf_k|h_X(k)|=\infty$, more explicitly: 
\begin{compactenum} \item[]
$\text{for each real } E \text{ there is a finite set } \ct\subset\cs 
\text{ such that } \inf_k|h_X(k)|>E \text{ if } X\nin\ct.$
\end{compactenum}
This assumption is of the same nature as the non-zero mass condition
in quantum field theory models.

{\bf(b)} We take $K=\oplus_X K_X$ as total kinetic energy of the
system. We denote $\cg=D(|K|^{1/2})$ its form domain equipped with the
norm $\|u\|_\cg=\|\jap{K}^{1/2}u\|$ and observe that
$\cg=\oplus_X\cg(X)$ Hilbert direct sum, where $\cg(X)=D(|K_X|^{1/2})$
is the form domain of $K_X$.

{\bf(c)} The simplest type of interaction terms are given by symmetric
elements $I$ of the multiplier algebra of $\rc$. Then it is easy to
see that $H=K+I$ is strictly affiliated to $\rc$ and that $\rp_{\geq
  X}(H)=K_{\geq X}+\rp_{\geq X} (I)$ where $K_{\geq X}=\oplus_{Y\geq
  X} K_Y$ and $\rp_{\geq X}$ is extended to the multiplier algebras in
a natural way \cite[p. 18]{La}.

{\bf(d)} In order to cover singular interactions (relatively bounded
in form sense with respect to $K$ but not in operator sense) we assume
from now on that the functions $h_X$ are equivalent to regular
weights. This is a quite weak assumption, see page \pageref{p:regw}.
For example, if the $X$ are vector spaces with norms $|\cdot|$ then it
suffices that $a|k|^{\alpha}\leq |h_X(k)|\leq b|k|^{\alpha}$ for some
numbers $a,b,\alpha>0$ (depending on $X$)and all large $k$.  As a
consequence of this fact the $U_x,V_k$ induce continuous operators in
the spaces $\cg(X)$ and their adjoints. These are the operators
involved in the next conditions.

{\bf(e)} For each $X,Y,Z\in\cs$ such that $X\cap Y\supset Z$ let
$I_{XY}(Z):\cg(Y)\to\cg^*(X)$ be a continuous map such that, with
limits in norm in $L(\cg(Y),\cg^*(X))$:
\begin{compactenum}
\item[(i)]
$U_z I_{XY}(Z)=I_{XY}(Z) U_z$ if $z\in Z$ and
$V^*_k I_{XY}(Z) V_k\to I_{XY}(Z)$ if $k\to 0$ in $(X+Y)^*$,
\item[(ii)]
$I_{XY}(Z)(U_y-1)\to 0$ if $y\to 0$ in $Y$ and
$I_{XY}(Z)(V_k-1)\to 0$ if $k\to 0$ in $(Y/Z)^*$.
\end{compactenum}
The conditions of Proposition \ref{pr:tex} are significantly more
general but require more formalism.  We require
$I_{XY}(Z)^*=I_{YX}(Z)$ and set $I_{XY}(Z)=0$ if $Z\not\subset X\cap
Y$.

{\bf(f)} Let $\cg_\rmo$ be the algebraic direct sum of the spaces
  $\cg(X)$ and $\cg_\rmo^*$ the direct product of the adjoint spaces
  $\cg^*(X)$.  Note that $\cg_\rmo$ is a dense subspace of
  $\cg$. The matrix $I(Z)=(I_{XY}(Z))_{X,Y\in\cs}$ can be realized
  as a linear operator $\cg_\rmo\to\cg_\rmo^*$. We shall require
  that this be the restriction of a continuous map
  $I(Z):\cg\to\cg^*$.  Equivalently, the sesquilinear form
  associated to $I(Z)$ should be continuous for the $\cg$
  topology. We also require that $I(Z)$ be norm limit in
  $L(\cg,\cg^*)$ of its finite sub-matrices $\Pi_\ct
  I(Z)\Pi_\ct=(I_{XY}(Z))_{X,Y\in\ct}$.

{\bf(g)} Finally, we assume that there are real positive numbers
  $\mu_Z$ and $a$ with $\sum_Z\mu_Z<1$ and such that either $\pm
  I(Z) \leq\mu_Z|K+ia|$ for all $Z$ or $K$ is bounded from below and
  $ I(Z) \geq -\mu_Z|K+ia|$ for all $Z$. Furthermore, the series
  $\sum_Z I(Z)\equiv I$ should be norm summable in $L(\cg,\cg^*)$.

 Then \emph{the Hamiltonian defined as a form sum $H=K+I$ is a
  self-adjoint operator strictly affiliated to $\rc$, we have
  $H_{\geq X}=K_{\geq X}+{\textstyle\sum_{Z\geq X}} I(Z)$, and the
  essential spectrum of $H$ is given by \eqref{eq:esi}}.

We consider the case when $\cs$ is a set of finite dimensional
subspaces of a real prehilbert space $\cx$ such that if $X,Y\in\cs$
then $X\cap Y\in\cs$ and $X+Y$ is included in a subspace of $\cs$. The
Euclidean structure induced on each $X$ allows us to identify $X^*=X$
and for any two $X,Y\in\cs$ to realize the quotient space $X/Y\cong
X/(X\cap Y)$ as a subspace of $\cx$ by taking
\[
X/Y=X/(X\cap Y)= X\ominus (X\cap Y).
\]
Then for $Z\subset X\cap Y$ we have $X=Z\oplus (X/Z)$ and $Y=Y\oplus
(Y/Z)$  and we identify
\begin{equation}\label{eq:xyzint}
\ch(X)=\ch(Z)\otimes\ch(X/Z)
\quad\text{and}\quad 
\ch(Y)=\ch(Z)\otimes\ch(Y/Z) 
\end{equation}
which gives us canonical tensor decompositions:
\begin{equation}\label{eq:xyzeint}
\rc_{XY}(Z)=\cc^*(Z)\otimes \rk_{X/Z,Y/Z} \quad\text{and}\quad
\rc_{XY}=\rc_{X\cap Y}\otimes \rk_{X/Y,Y/X}.
\end{equation}
When convenient we shall identify
$\ch(Z)\otimes\ch(X/Z)= L^2(Z;\ch(X/Z))$. Let $\cf_Z$ denote the
Fourier transformation in the $Z$ variable. By using
\eqref{eq:xyzeint} and $C^*(Z)=\cf_Z^{-1}\Co(Z)\cf_Z$ we get
\[
\rc_{XY}(Z)=\cf_Z^{-1}\Co(Z;\rk_{X/Z,Y/Z})\cf_Z.
\] 

\begin{example}\label{ex:iint}
We may use this representation to better understand the structure of
the allowed interactions $I_{XY}(Z)$. What follows is a particular
case of Proposition \ref{pr:etex} (cf. the last part of Section
\ref{s:morre}).  We denote $\ch^s(X)$ the usual Sobolev spaces for
$s\in\mbr$. Assume that the form domains of $K_X$ and $K_Y$ are the
spaces $\ch^s(X)$ and $\ch^t(Y)$.  Define $I_{XY}(Z)$ by the
relation
\begin{equation}\label{eq:Ixyz}
\cf_Z I_{XY}(Z) \cf_Z^{-1} \equiv \int_Z^\oplus I_{XY}^Z(k) \rmd k
\end{equation}
where $I_{XY}^Z:Z\to L(\ch^t(Y/Z),\ch^{-s}(X/Z))$ is a continuous
operator valued function satisfying
\begin{equation}\label{eq:Iest}
\sup\nolimits_k
\|(1+|k|+|P_{X/Z}|)^{-s}I_{XY}^Z(k)(1+|k|+|P_{Y/Z}|)^{-t}\| <\infty.
\end{equation} 
The operators $I_{XY}^Z(k)$ must also decay in a weak sense at
infinity, more precisely one of the equivalent conditions must be
satisfied for each $k\in Z$ and some $\varepsilon>0$:
\begin{compactenum}

\item[(i)] $I_{XY}^Z(k):\ch^t(Y/Z) \to \ch^{-s-\varepsilon }(X/Z)$
  is compact,

\item[(ii)] $(V_x-1)I_{XY}^Z(k)\to 0$ in norm in
  $L(\ch^t(Y/Z),\ch^{-s-\varepsilon}(X/Z))$ if $x\to 0$ in $X/Z$.

\end{compactenum}
For $\varepsilon=0$ the condition (ii) is significantly more general
than (i), for example it allows the operator $I_{XY}^Z$ to be of order
$s+t$.  The $I_{XY}(Z)$ with $I_{XY}^Z(k)$ independent of $k$ are
especially simple to define:

\begin{compactenum}
\item[]
Let $I_{XY}^Z:\ch^t(Y/Z)\to\ch^{-s}(X/Z)$ be continuous and such
that, for some $\varepsilon>0$, when considered as a map $\ch^t(Y/Z)
\to \ch^{-s-\varepsilon}(X/Z)$, it becomes compact. Then we take
$I_{XY}(Z)=1_Z\otimes I_{XY}^Z$ relatively to the tensor
factorizations \eqref{eq:xyzint}. 
\end{compactenum}
\end{example}

\PAR\label{ss:mouint} {\bf Non-relativistic many-body Hamiltonians and
Mourre estimate}

Now we shall present our results on the Mourre estimate.  We shall
consider only the non-relativistic many-body problem because in this
case the results are quite explicit. There are serious difficulties
when the kinetic energy is not a quadratic form even in the much
simpler case of $N$-body Hamiltonians, but see \cite{De1,Ger1,DG2} for
some partial results which could be extended to our setting. Note that
the quantum field case is much easier from this point of view because
of the special nature of the interactions: this is especially clear
from the treatments in \cite{Ger2,Geo}, but see also \cite{DeG2}.

For simplicity we shall restrict ourselves to the case when $\cs$ is
a finite semilattice. In fact, the case when $\cs$ is infinite has
already been treated in \cite{DG2} and the extension of the
techniques used there to the case when $\cx$ is infinite dimensional
is rather straightforward.  But the condition
$\lim_X\inf_k|h_X(k)|=\infty$ is quite artificial in the
non-relativistic case since it forces us to replace the Laplacian
$\Delta_X$ by $\Delta_X+E_X$ where $E_X$ is a number which tends to
infinity with $X$.

We denote by $\cs/X$ the set of subspaces $E/X=E\cap X^\perp$, this
is clearly an inductive semilattice of finite dimensional subspaces
of $\cx$ which contains $O=\{0\}$. Hence the $C^*$-algebra
$\rc_{\cs/X}$ and the Hilbert space $\ch_{\cs/X}$ are well defined
by our general rules.  If $X\subset Z\subset E\cap F$ then
\eqref{eq:xyzeint} implies
\[
\rc_{EF}(Z)=\cc^*(Z)\otimes \rk_{E/Z,F/Z}
=\cc^*(X)\otimes \cc^*(Z/X)\otimes \rk_{E/Z,F/Z}.
\]
Moreover, we have $\ch(Y)=\ch(X)\otimes\ch(Y/X)$ for all $Y\supset
X$ hence
\[
\ch_{\geq X}=\ch(X)\otimes\big(\oplus_{Y\supset X}\ch(Y/X)\big).
\]
Thus we have
\begin{equation}\label{eq:f}
\rc_{\geq X}=\cc^*(X)\otimes\rc_{\cs/X} \quad \text{and} \quad
\ch_{\geq X}=\ch(X)\otimes\ch_{\cs/X}.
\end{equation}
Let $\Delta_X$ be the (positive) Laplacian associated to the
Euclidean space $X$ with the convention $\Delta_O=0$. We have
$\Delta_X=h_X(P)$ with $h_X(k)=\|k\|^2$.  We also set
$\Delta_\cs=\oplus_X \Delta_X$ and define $\Delta_{\geq X}$
similarly.  Then for $Y\supset X$ we have 
$\Delta_Y=\Delta_X\otimes 1 +1\otimes\Delta_{Y/X}$ hence we get 
$\Delta_{\geq X}=\Delta_X\otimes 1 +1\otimes\Delta_{\cs/X}$.
The domain and form domain of the operator $\Delta_\cs$ are
given by $\ch_\cs^2$ and $\ch_\cs^1$ where the Sobolev spaces
$\ch_\cs^s\equiv\ch^s$ are defined for any real $s$ by
$\ch^s=\oplus_X \ch^s(X)$.

We define the dilation group on $\ch(X)$ by $(W_\tau
u)(x)=\rme^{n\tau/4}u(\rme^{\tau/2} x)$ where $n$ is the dimension
of $X$. We denote by the same symbol the unitary operator
$\bigoplus_X W_\tau$ on the direct sum $\ch=\bigoplus_X\ch(X)$.  Let
$D$ be the infinitesimal generator of $\{W_\tau\}$, so $D$ is a
self-adjoint operator on $\ch$ such that $W_\tau=\rme^{i\tau D}$. As
usual we do not indicate explicitly the dependence on $X$ or $\cs$
of $W_\tau$ or $D$ unless this is really needed.  The operator $D$
has factorization properties similar to that of the Laplacian, in
particular $D_{\geq X}=D_X\otimes 1 +1\otimes D_{\cs/X}$.

We shall formalize the notion of non-relativistic many-body
Hamiltonian by extending to the present setting Definition 9.1 from
\cite{ABG}. We restrict ourselves to strictly affiliated operators
although the more general case of operators which are only
affiliated covers some interesting physical situations (hard-core
interactions).

Note that since $\cs$ is finite it has a minimal element $\min\cs$
and a maximal element $\max\cs$ (which are in fact the least and the
largest elements) and is atomic.

\begin{definition}\label{df:NR}
\emph{A non-relativistic many-body Hamiltonian of type $\cs$} is a
bounded from below self-adjoint operator $H=H_\cs$ on $\ch=\ch_\cs$
which is strictly affiliated to $\rc=\rc_\cs$ and has the following
property: for each $X\in\cs$ there is a bounded from below
self-adjoint operator $H_{\cs/X}$ on $\ch_{\geq X}$
such that 
\begin{equation}\label{eq:NR}
\rp_{\geq X}(H)\equiv H_{\geq X}=\Delta_X\otimes1+ 1\otimes H_{\cs/X}
\end{equation}
relatively to the tensor factorization from \eqref{eq:f}. Moreover,
when $X=\max\cs$ is the maximal element of $\cs$, hence
$\ch_{\cs/\max\cs}=\ch(O)=\mbc$, we require $H_{\cs/\max\cs}=0$.
\end{definition}

From Theorem \ref{th:imp4} it follows that each $H_{\cs/X}$ is a
non-relativistic many-body Hamiltonian of type $\cs/X$.

\begin{example}\label{ex:NR}
We give here the main example of non-relativistic many-body
Hamiltonians.  As before we take $H=K+I$ but this time the kinetic
energy is $K=\Delta_\cs=\sum_X\Delta_X$. With the notations of point
(b) from \S\ref{ss:ex} we now have $\cg=\ch^1=\oplus_X \ch^1(X)$
and the adjoint space is $\cg^*=\ch^{-1}=\oplus_X \ch^{-1}(X)$.
The interaction term is a continuous operator
$I:\ch^1\to\ch^{-1}$ of the form
\[
I=(I_{XY})_{X,Y\in\cs}={\textstyle\sum_{Z\in\cs}} I(Z)=
{\textstyle\sum_{Z\in\cs}} (I_{XY}(Z))_{X,Y\in\cs}
\]
with $I_{XY}:\ch^1(Y)\to\ch^{-1}(X)$ of the form
$I_{XY}=\sum_{Z\in\cs}I_{XY}(Z)$. If $Z\subset X\cap Y$ we take
$I_{XY}(Z)=1_Z\otimes I_{XY}^Z$ relatively to the tensor
factorization \eqref{eq:xyzint}, where
$I_{XY}^Z:\ch^1(Y/Z)\to\ch^{-1}(X/Z)$ is continuous and such that
when considered as a map $\ch^1(Y/Z) \to \ch^{-1-\varepsilon}(X/Z)$
with $\varepsilon>0$ it is compact.  We set $I_{XY}(Z)=0$ if
$Z\not\subset X\cap Y$ and we require $I_{XY}(Z)^*=I_{YX}(Z)$ for
all $X,Y,Z$. Finally, we assume that there are positive numbers
$\mu_Z,a$ with $\sum\mu_Z<1$ such that $I(Z)\geq -\mu_Z\Delta_\cs
-a$ for all $Z$. Then $H=K+I$ defined in the quadratic form sense is
a non-relativistic many-body Hamiltonian of type $\cs$ and we have
$H_{\geq X}=\Delta_{\geq X}+\sum_{Z\supset X} I(Z)$.
\end{example}

Let us denote $\tau_X=\min H_{\cs/X}$ the bottom of the spectrum of
$H_{\cs/X}$.  From \eqref{eq:NR} we get
\begin{equation}\label{eq:speint}
\sp(H_{\geq X})=[0,\infty) + \sp(H_{\cs/X})=
[\tau_X,\infty) \quad
\text{if } X\neq O.
\end{equation}
Then Theorem \ref{th:imp4} implies (observe that the assertion of the
proposition is obvious if $O\nin\cs$):

\begin{proposition}\label{pr:hvznrint}
If $H$ is a non-relativistic many-body Hamiltonian of type $\cs$
then its essential spectrum is $\spe(H)=[\tau,\infty)$ with
$\tau=\min_{X\in\cp(\cs)} \tau_X$ where $\tau_X=\min H_{\cs/X}$.
\end{proposition}

We refer to Subsection \ref{ss:mest} for terminology related to the
Mourre estimate. We take $D$ as conjugate operator and only mention
that we denote by $\what\rho_H(\lambda)$ the best constant (which
could be infinite) in the Mourre estimate at point $\lambda$. The
\emph{threshold set} $\tau(H)$ of $H$ with respect to $D$ is the set
where $\what\rho_H(\lambda)\leq0$. Note that $\tau(H)$ is always
closed, the nontrivial fact proved below is that it is countable. 

If $A$ is a closed real set then $N_A:\mbr\to[-\infty,\infty[$ is
defined by $N_A(\lambda)=\sup\{ x\in A \mid x\leq\lambda\}$ with
the convention $\sup\emptyset=-\infty$.
Denote $\mathrm{ev}(T)$ the set of eigenvalues of an operator $T$.

\begin{theorem}\label{th:thrintr}
Assume $O\in\cs$ and let $H=H_\cs$ be a non-relativistic many-body
Hamiltonian of type $\cs$ and of class $C^1_\rmu(D)$. Then
$\what\rho_H(\lambda)=\lambda-N_{\tau(H)}(\lambda)$ for all real
$\lambda$ and
\begin{equation}\label{eq:thrintr}
\tau(H)=\ccup_{X\neq O}\mathrm{ev}(H_{\cs/X}).
\end{equation}
In particular $\tau(H)$ is a closed \emph{countable} real set. The
eigenvalues of $H$ which do not belong to $\tau(H)$ are of finite
multiplicity and may accumulate only to points from $\tau(H)$.
\end{theorem}

\begin{example}\label{ex:c1d}
We give examples of Hamiltonians of class $C^1_\rmu(D)$.  We keep
the notations of Example \ref{ex:NR} but to simplify the statement
we consider only interactions which are relatively bounded in
\emph{operator} sense with respect to the kinetic energy. Recall
that the domain of $K=\Delta_\cs$ is
$\ch^2=\oplus_X\ch^2(X)$. The interaction operator $I$ is
constructed as in Example \ref{ex:NR} but we impose stronger
conditions on the operators $I^Z_{XY}$. More precisely, we assume:
\begin{compactenum}

\item[(i)] If $Z\subset X\cap Y$ then $I^Z_{XY}:\ch^2(Y/Z)\to\ch(X/Z)$
  is a compact operator satisfying $(I^{Z}_{XY})^*\supset I^Z_{YX}$
  and we set $I^Z_{XY}=0$ if $Z\not\subset X\cap Y$.  Then all the
  conditions of Example \ref{ex:NR} are satisfied and $I:\ch^2\to\ch$
  is relatively bounded with respect to $K$ in operator sense with
  relative bound zero.

\item[(ii)] Under the assumption (i) the operator
\begin{equation}\label{eq:di}
[D,I^Z_{XY}]\equiv 
D_{X/Z}I^Z_{XY}-I^Z_{XY}D_{Y/Z}:
\ch^2_{\mathrm{loc}}(Y/Z)\to\ch^{-1}_{\mathrm{loc}}(X/Z)
\end{equation}
is well defined.  We require it to be a compact operator
$\ch^2(Y/Z)\to\ch^{-2}(X/Z)$.
\end{compactenum}
Then \emph{the operator $H$ is self-adjoint on $\ch^2$ and of
class $C^1_\rmu(D)$}. We indicated by a subindex the space where the
operator $D$ acts and, for example, we used
\[
D_X=D_Z\otimes 1 + 1\otimes D_{X/Z} \quad \text{relatively to }
\ch(X)=\ch(Z)\otimes\ch(X/Z).
\]
Note also that
\begin{equation}\label{eq:fed}
2iD_X=x\cdot\nabla_x+n/2= \nabla_x\cdot x-n/2 \quad 
\text{if $n$ is the dimension of } X.
\end{equation}
\end{example}

\begin{remark}\label{re:cid}
If we set $E=(X\cap Y)/Z$ then $Y/Z=E\oplus(Y/X)$ and
$X/Z=E\oplus(X/Y)$ hence
\begin{equation*}\label{eq:ixyzci}
\ch(X/Z)=\ch(E)\otimes\ch(X/Y), \quad 
\ch^2(Y/Z)=\big(\ch^2(E)\otimes\ch(Y/X)\big)\cap
\big(\ch(E)\otimes\ch^2(Y/X)\big).
\end{equation*}
Let $\rk^2_{MN}= K(\ch^2(N),\ch(M))$ for arbitrary Euclidean spaces
$M,N$. Then condition (i) of Example \ref{ex:c1d} can be written
$I_{XY}^Z\in \rk^2_{X/Z,Y/Z}$. On the other hand we have
\begin{equation*}
\rk^2_{X/Z,Y/Z} = \rk^2_E\otimes \rk_{X/Y,Y/X} +
\rk_E\otimes\rk^2_{X/Y,Y/X}.
\end{equation*}
See \S\ref{ss:ha} for details concerning these tensor products.To
simplify notations we set $X \boxplus Y = X/Y\times Y/X$.  Then if we
identify a Hilbert-Schmidt operator with its kernel we get
\[
\rk^2_E\otimes \rk_{X/Y,Y/X}\supset \rk^2_E\otimes L^2(X \boxplus Y)
\supset L^2(X \boxplus Y;\rk^2_E)
\]
Thus $I_{XY}^Z \in L^2(X \boxplus Y;\rk^2_E)$ is an explicit example
of operator $I^Z_{XY}$ satisfying condition (i) of Example
\ref{ex:c1d} (see Section \ref{ss:mex} for improvements and a
complete discussion).  Such an $I_{XY}^Z$ acts as follows. Let
$u\in\ch^2(Y/Z)\equiv L^2(Y/X;\ch^2(E))$. Then
$I_{XY}^Z u\in \ch(X/Z) \equiv L^2(X/Y;\ch(E))$ is given by
\[
(I_{XY}^Z u)(x')={\textstyle\int_{Y/X}} I_{XY}^Z(x',y')u(y') \rmd y'.
\]
\end{remark}

\begin{remark}\label{re:cidd}
It is convenient to decompose the expression of $[D,I^Z_{XY}]$ given
in \eqref{eq:di} as follows:
\begin{align}
[D,I^Z_{XY}] 
&=(D_E+D_{X/Y})I^Z_{XY}-I^Z_{XY}(D_E+D_{Y/X}) \nonumber \\
&=[D_E,I^Z_{XY}] +D_{X/Y}I^Z_{XY} -I^Z_{XY}D_{Y/X}. \label{eq:dec}
\end{align}
The first term above is a commutator and so is of a rather different
nature than the next two. On  the other hand 
$I^Z_{XY}D_{Y/X}=(D_{Y/X}I^Z_{YX})^*$. Thus condition (ii) of Example
\ref{ex:c1d} follows from:
\begin{equation}\label{eq:dii}
[D_E,I^Z_{XY}] \text{ and } D_{X/Y}I^Z_{XY} \text{ are compact
  operators } \ch^2(Y/Z)\to\ch^{-2}(X/Z) \text{ for all } X,Y,Z.
\end{equation}
It is convenient to use the representation of $\ch^2(Y/Z)$ given in
Remark \ref{re:cid} and also
\begin{equation*}\label{eq:klean2i}
\ch^{-2}(X/Z) = \ch^{-2}(E)\otimes\ch(X/Y)
+\ch(E)\otimes\ch^{-2}(X/Y).
\end{equation*}
For example, if $I_{XY}^Z\in L^2(X \boxplus Y;\rk^2_E)$ as in Remark
\ref{re:cid} then the kernel of the operator $[D_E,I^Z_{XY}]$ is the
map $(x',y')\mapsto[D_E,I^Z_{XY}(x',y')]$ so it suffices to ask
\[
[D_E,I^Z_{XY}]\in L^2(X \boxplus Y;K(\ch^2(E),\ch^{-2}(E))
\]
in order to ensure that $[D_E,I^Z_{XY}]$ is a compact operator
$\ch^2(Y/Z)\to\ch^{-2}(X/Z)$. For the term $D_{X/Y}I^Z_{XY}$ it
suffices to require the compactness of the operator
\[
D_{X/Y}I^Z_{XY}\equiv 1_E\otimes D_{X/Y}I^Z_{XY}: \ch^2(Y/Z)\to
\ch(E)\otimes\ch^{-2}(X/Y).
\]
By taking into account \eqref{eq:fed} we see that this is a condition
on the formal kernel $x'\cdot\nabla_{x'}I_{XY}^Z(x',y')$. For example,
it suffices that the operator
$\jap{Q_{X/Y}}I^Z_{XY}:\ch^2(Y/Z)\to\ch(X/Z)$ be compact, which is a
short range assumption. The condition on $I^Z_{XY}D_{Y/X}$ is a
requirement on the formal kernel $y'\cdot\nabla_{y'}I_{XY}^Z(x',y')$.
\end{remark}

Theorem \ref{th:thrintr} has important applications in the spectral
analysis of $H$: absence of singularly continuous spectrum and an
optimal version of the limiting absorption principle. Optimality
refers both to the Besov spaces in which we establish the existence of
the boundary values of the resolvent and to the degree of regularity
of the Hamiltonian with respect to the conjugate operator $D$: it
suffices that $H$ be of Besov class $C^{1,1}(D)$.  We refer to
\S\ref{ss:scs} for these results and present here a less refined
statement.

Let $\ch_{s}=\oplus_X\ch_{s}(X)$ where the $\ch_{s}(X)$ are the
Sobolev spaces associated to the position observable on $X$ (these are
obtained from the usual Sobolev spaces associated to $L^2(X)$ by a
Fourier transformation).  Let $\mbc_+$ be the open upper half plane
and $\mbc^H_+=\mbc_+\cup(\mbr\setminus\tau(H))$. If we replace the
upper half plane by the lower one we similarly get the sets $\mbc_-$
and $\mbc^H_-$.

\begin{theorem}\label{th:c11s}
If $H$ is of class $C^{1,1}(D)$ then its singular continuous spectrum
is empty. The holomorphic maps $\mbc_\pm\ni z\mapsto(H-z)^{-1}\in
L(\ch_{s},\ch_{-s})$ extend to norm continuous functions on
$\mbc^H_\pm$ if $s>1/2$.
\end{theorem}

If $H$ satisfies the conditions of Example \ref{ex:c1d} then
$J\equiv [D,I]\in L(\ch^2,\ch^{-1})$.  Then a very rough sufficient
condition for $H$ to be of class $C^{1,1}(D)$ is that $[D,J]\in
L(\ch^2,\ch^{-2})$.  A much weaker sufficient assumption is the Dini
type condition
\begin{equation}\label{eq:dini}
\int_0^1\|W_\varepsilon^*JW_\varepsilon-J\|_{\ch^2\to\ch^{-2}}
\frac{\rmd\varepsilon}{\varepsilon} <\infty.
\end{equation}
Note that $[D,J]\in L(\ch^2,\ch^{-2})$ is equivalent to
\[
\|W_\tau^*J W_\tau-J\|_{\ch^2\to\ch^{-2}} \leq C |\tau| \quad
\text{for some constant $C$ and all real } \tau
\]
hence \eqref{eq:dini} is indeed a much weaker condition. See
\S\ref{ss:mex} for a discussion of the Dini and $C^{1,1}$ classes in
the present context. 

\begin{remark}\label{re:NM}
We stress that there is no qualitative difference between an $N$-body
Hamiltonian (fixed $N$) and a many-body Hamiltonian involving
interactions which do not preserve $N$ if these notions are defined in
terms of the same semilattice $\cs$.  More precisely the channel
structure and the formulas for the essential spectrum and the
threshold set which appears in the Mourre estimate are identical,
cf. Theorems \ref{th:imp4} and \ref{th:thrintr}. Only the
$\cs$-grading of the Hamiltonian algebra matters.
\end{remark}

\PAR\label{ss:comments} {\bf Comments and examples}

$\rc$ has an interesting class of $\cs$-graded $C^*$-subalgebras (see
the end of Section \ref{s:grass}). If $\ct\subset\cs$ we set
\begin{equation*}
\rc_\ct\equiv{\textstyle\sum_{X,Y\in\ct}^\rmc}\rc_{XY} \quad
\text{and} \quad \ch_\ct\equiv\oplus_{X\in\ct}\ch(X).
\end{equation*}
Then $\rc_\ct$ is a $C^*$-algebra supported by the subspace $\ch_\ct$
of $\ch$, in fact $\rc_\ct=\Pi_\ct\rc\Pi_\ct$ where $\Pi_\ct$ is
the orthogonal projection of $\ch$ onto $\ch_\ct$, and is graded by
the ideal $\ccup_{X\in\ct}\cs(X)$ generated by $\ct$ in $\cs$.

If $\cs$ is a finite semilattice of subspaces of an Euclidean space
and $\ct$ is a totally ordered subset, then the Hamiltonians
considered in \cite{SSZ} are affiliated to $\rc_\ct(\cs)$. Thus the
results from \cite{SSZ} are consequences of the Theorems
\ref{th:thrintr} and \ref{th:c11s}.  

We mention that in the preceding context, due to the fact that $\ct$
is totally ordered, the construction of $\rc_\ct$ and the proof of the
fact that it is an $\cs$-graded $C^*$-algebra do not require the
machinery from Sections \ref{s:hlca}--\ref{s:grass}. In fact, an
alternative abstract framework is much simpler in this case. The main
point is that we can write $\ct$ as a strictly increasing family of
subspaces $X_0\subset\dots\subset X_n$ hence we have tensorial
factorizations $\ch(X_{k})=\ch(X_{k-1})\otimes\ch(X_{k}/X_{k-1})$ for
all $k\geq1$.  If we set $\cg_k=\ch(X_{k}/X_{k-1})$ then we get a
factorization $\ch_n=\otimes_{k=1}^n\cg_k$, where
$\ch_n=\ch(X_n)$. Now let $\cg_1,\dots,\cg_n$ be arbitrary Hilbert
spaces and define
\[
\ch_m=\otimes_{k=1}^m\cg_k \quad \text{and} \quad 
\ch=\oplus_{m=1}^n\ch_m.
\]
Observe that for each couple $i<j$ right tensor multiplication by
elements of $\otimes_{i<k\leq j}\cg_k$ defines a closed linear
subspace $\cu_{ji}\subset L(\ch_i,\ch_j)$ isometrically isomorphic to
$\otimes_{i<k\leq j}\cg_k$. Then we set $\cu_{ij}=\cu_{ji}^*$ and
$\cu_{ii}=\mbc$.  Assume that $\cs$ is an arbitrary semilattice and
$\rc_n$ is an $\cs$-graded $C^*$-algebra on $\ch_n$ and define the
closed self-adjoint space $\rc_m$ of operators on $\ch_m$ by
$\rc_m=\cu_{mn}\cdot\rc_n\cdot\cu_{nm}$. Finally, we define a space of
operators $\rc$ on $\ch$ by the rule  \label{p:htp}
$\rc_{ij}=\rc_i\cdot\cu_{ij}$. The interested reader will easily find
the natural conditions which ensure that $\rc$ is a $C^*$-algebra and
then the compatibility conditions which allow one to equip it with a
rather obvious $\cs$-graded structure (see page \pageref{p:2ex}). In
fact the toy model corresponding to $n=2$ explains everything and has
a nice interpretation in terms of Hilbert $C^*$-modules,
cf. \eqref{eq:link}.

There are extensions of this abstract formalism which are of some
interest and that one can handle. Let $\cs$ be a semilattice such that
for each couple $\sigma',\sigma''\in\cs$ there is $\sigma\in\cs$ which
is larger than both $\sigma'$ and $\sigma''$.  Assume that we are
given a family of Hilbert space
$\{\ch_\sigma\}_{\sigma\in\cs}$. Moreover, assume that for each couple
$\sigma\leq\tau$ we have $\ch_\tau=\ch_\sigma\otimes\ch_\tau^\sigma$
for a given Hilbert space $\ch_\tau^\sigma$. The $\cu_{\tau\sigma}$
are defined as before for $\sigma\leq\tau$ and then one may extend the
definition to any couple $\sigma,\tau$ in a natural way. Finally, if a
family of $\cs$-graded $C^*$-algebras $\rc_\sigma$ is given and a
certain compatibility condition is satisfied, one may construct an
algebra $\rc$ and an $\cs$-grading on it.

A nice but easy example corresponds to the case when $\cs$ is the set
of subsets of a finite set $I$. More generally, it is very easy to
treat the case when $\cs$ is a distributive relatively
ortho-complemented lattice. Such a situation is specific to quantum
field models without symmetry considerations.

We must, however, emphasize the following important point. If
$X,Y\in\cs$ and $Y\subset X$, and if we are in the framework of
Theorem \ref{th:C}, then we do not have a tensor factorization
$\ch(X)=\ch(Y)\otimes\ce$ in any natural way ($Y$ is not complemented
in $X$). Moreover, even if a decomposition $X=Y\oplus Y'$ is possible,
our algebra $\rc$ is independent of the choice of $Y'$.  This seems to
us a quite remarkable property which is lost in the preceding abstract
situations.

We shall make some comments now on the many-body system associated to
a standard $N$-body system by our construction. We shall see that we
get a self-interacting system in which although the number of
particles is not conserved, the total mass is conserved.

We refer to \cite{DeG1} or to \cite[Chapter 10]{ABG} for details on
the following formalism. Let $m_1,\dots,m_N$ be the masses of the $N$
``elementary particles''.  We assume that there are no external fields
and always take as origin of the reference system the center of mass
of the system. Then the configuration space $X$ of the system of $N$
particles is the set of $x=(x_1,\dots,x_N)\in(\mbr^d)^N$ such that
$\sum_km_k x_k=0$, where $\mbr^d$ is the physical space. We equip $X$
with the scalar product $\braket{x}{y}=\sum_{k=1}^N2m_k x_k y_k$. Then
the Laplacian associated to it has the usual physical meaning.

A cluster decomposition is just a partition $\sigma$ of the set
$\{1,\dots,N\}$ and the sets of the partition are called clusters.  We
think about a cluster $a\in\sigma$ as a ``composite particle'' of mass
$m_a=\sum_{k\in a}m_k$.  Let $|\sigma|$ be the number of clusters of
$\sigma$. Then we interpret $\sigma$ as a system of $|\sigma|$
particles with masses $m_a$ hence its configuration space should be
the set of $x=(x_a)_{a\in\sigma}\in(\mbr^d)^{|\sigma|}$ such that
$\sum_a m_a x_a=0$ equipped with the scalar product defined as above.

Let us define $X_\sigma$ as the set of $x\in X$ such that $x_i=x_j$ if
$i,j$ belong to the same cluster and let us equip $X_\sigma$ with the
scalar product induced by $X$. Then there is an obvious isometric
identification of $X_\sigma$ with the configuration space of the
system $\sigma$ as defined before. The advantage now is that all the
spaces $X_\sigma$ are isometrically embedded in the same $X$.  The set
$\mathfrak{S}$ of partitions is ordered as usual in the mathematical
literature (so not as in \cite{ABG}, for example), namely
$\sigma\leq\tau$ means that $\tau$ is finer than $\sigma$. Then
clearly $\sigma\leq\tau$ is equivalent to $X_\sigma\subset
X_\tau$. Moreover, $X_\sigma\cap X_\tau=X_{\sigma\wedge\tau}$.
Thus we see that $\mathfrak{S}$ is isomorphic as semilattice with the
set $\cs=\{X_\sigma\mid \sigma\in\mathfrak{S}\}$ of subspaces of $X$.  

Now we may apply our construction to $\cs$. We get a system whose
state space is $\ch=\oplus_\sigma \ch(X_\sigma)$. If the system is in
a state $u\in\ch(X_\sigma)$ then it consists of $|\sigma|$ particles
of masses $m_a$. Note that $\min\mathfrak{S}$ is the partition
consisting of only one cluster $\{1,\dots,N\}$ with mass
$M=m_1+\dots+m_N$. Since there are no external fields and we decided
to eliminate the motion of the center of mass, this system must be the
vacuum. And its state space is indeed
$\ch(X_{\min\mathfrak{S}})=\mbc$. The algebra $\rc$ in this case
predicts usual inter-cluster interactions associated, for examples, to
potentials defined on $X^\sigma=X/X_\sigma$, but also interactions
which force the system to make a transition from a ``phase'' $\sigma$
to a ``phase'' $\tau$. In other terms, the system of $|\sigma|$
particles with masses $(m_a)_{a\in\sigma}$ is tranformed into a system
of $|\tau|$ particles with masses $(m_b)_{b\in\tau}$. Thus the number
of particles varies from $1$ to $N$ but the total mass existing in the
``universe'' is constant and equal to $M$.

\PAR\label{ss:impr} {\bf On the role of Hilbert $C^*$-modules and
  imprimitivity $C^*$-algebras }

At a technical level, Hilbert $C^*$-modules are involved in a very
natural way in our formalism. For example the space
$\rc_{ij}=\rc_i\cdot\cu_{ij}$ introduced on page \pageref{p:htp} is in
fact the tensor product in the category of such modules of the
$C^*$-algebra $\rc_i$ and of the Hilbert space $\cu_{ij}$ and one
needs this to prove that $\rc$ is graded.

However, the Hilbert $C^*$-modules play an important role at a
fundamental level because they allow us to ``unfold'' a Hamiltonian
algebra $\ra$ such as to construct new Hamiltonian algebras. Indeed,
our results show that \emph{if $\mr$ is a full Hilbert $\ra$-module
  then the imprimitivity $C^*$-algebra $\ck(\mr)$ could also be
  interpreted as Hamiltonian algebra of a system related in some
  natural way to the initial one}.  For example, this is a natural
method of second quantizing $N$-body systems, i.e. introducing
interactions which couple subsystems corresponding to different
cluster decompositions.

We understood the role in our work of the imprimitivity algebra of a
Hilbert $C^*$-module thanks to a discussion with Georges Skandalis: he
recognized (a particular case of) the main $C^*$-algebra $\rc$ we have
constructed as the imprimitivity algebra of a certain Hilbert
$C^*$-module. Theorem \ref{th:mor} is a reformulation of his
observation in the present framework (at the time of the discussion
our definition of $\rc$ was rather different because we were working
in a tensor product formalism, as on page \pageref{p:htp}).

In the physical $N$-body situation discussed in \S\ref{ss:comments} it
is clear that going from $\ra$ to the imprimitivity algebra of $\mr$
may be thought as a ``second quantization'' of the $N$-body system:
this explains our definition \ref{df:sc}. The full Hilbert
$\rc_X$-module $\rn_X$ constructed \`a la Skandalis in Theorem
\ref{th:mor} is such that its imprimitivity algebra is
$\rc_X^\#=\rc_{\cs(X)}$.  So, more generally, given a full Hilbert
$\ra$-module $\mr$ it is natural to call its imprimitivity algebra the
\emph{second quantization of $\ra$ determined by $\mr$}.

We mention that the notion of graded Hilbert $C^*$-module that we use,
cf. \S\ref{ss:gf}, is also due to G. Skandalis. He has also shown us a
nice abstract construction of such modules starting from a given
graded $C^*$-algebra and using tensor product techniques, but this
method is not used in the present paper.

If $\ra$ is graded and $\mr$ is a graded Hilbert $\ra$-module then
$\ck(\mr)$ is equipped with a canonical structure of graded
$C^*$-algebra (Theorem \ref{th:kghm}).  If $\mr$ is an arbitrary full
Hilbert $\ra$-module it is not clear to us if there are general and
natural conditions on $\mr$ which ensure that a grading of $\ra$ can
be transported to $\ck(\mr)$. However, even if the grading is lost,
something can be done thanks to the Rieffel correspondence: the
isomorphism between the lattice of all ideals of $\ra$ and that of
$\ck(\mr)$ defined by $\ri\mapsto\ck(\mr\ri)$.

For example, let $\{\ra_i\}_{i\in I}$ be a family of ideals of $\ra$
which generates $\ra$.  Then $\ck(\mr)$ is equipped with the family of
ideals $\ck(\mr\ra_i)$ such that $\bigcup_i\ck(\mr\ra_i)$ generates
$\ck(\mr)$ and
\begin{equation}\label{eq:mid}
\ccap_i\ck(\mr\ra_i)=\ck(\mr\ccap_i\ra_i).
\end{equation}
Assume that $\ra$ is the $C^*$-algebra of Hamiltonians of a system
whose state space is the Hilbert space $\ch$ and that
$\bigcap_i\ra_i=K(\ch)$. The interest of these assumptions is that it
allows one to compute the essential spectrum of observables affiliated
to $\ra$ in rather complicated situations by using the following
argument.  Let $\cp_i$ be the canonical surjection of $\ra$ onto the
quotient $C^*$-algebra $\ra/\ra_i$.  If $H$ is an observable
affiliated to $\ra$ then $H_i=\cp_i(H)$ is an observable affiliated to
$\ra/\ra_i$ and one has \cite[(2.2)]{GI1}
\begin{equation}\label{eq:essi}
\se(H)=\overline\ccup_i\sigma(H_i).
\end{equation} 
where $\overline\ccup$ means closure of the union.  Now assume that
$\mr$ is realized as a closed linear subspace of $L(\ch,\cg)$ for some
Hilbert space $\cg$ such that $\mr^*\cdot\mr=\ra$ and
$\mr\mr^*\mr\subset\mr$.  Then
$\ck(\mr)\cong\rb\equiv\mr\cdot\mr^*$. If we set $\mr_i=\mr\ra_i$ then
$\mr_i$ is a full Hilbert $\ra_i$-module and we have
$$
\mr_i^*\cdot\mr_i=\ra_i\cdot\mr^*\cdot\mr\cdot\ra_i=
\ra_i\cdot\ra\cdot\ra_i=\ra_i 
$$
and $\mr_i\mr_i^*\mr_i\subset\mr_i$.  So we get
$\ck(\mr_i)\cong\mr_i\cdot\mr_i^*\equiv\rb_i$, hence $\{\rb_i\}$ is
the family of ideals of $\rb$ associated to $\{\ra_i\}$. From
\eqref{eq:mid} we get
\[
\ccap_i\rb_i=\ck(\mr\ccap_i\ra_i)=\ck(\mr K(\ch))=
(\mr K(\ch))\cdot(\mr K(\ch))^*.
\]
It is clear that $\mr K(\ch)$ is the closed linear span in
$L(\ch,\cg)$ of the set of operators of the from $\ket{Mh}\bra{h'}$
with $h,h'\in\ch$.  Thus, \emph{if $\mr\ch$ is dense in $\cg$ then
  $\mr K(\ch)=K(\ch,\cg)$} and from this we clearly get $\ccap_i\rb_i=
K(\cg)$. So we may compute the essential spectrum of an observable
affiliated to the unfolding $\rb$ of $\ra$ with the help its quotients
with respect to the ideals $\rb_i$ by using an analog of
\eqref{eq:essi}.

{\bf Acknowledgments:} We are indebted to Georges Skandalis for very
helpful suggestions and remarks.

\section{Preliminaries on Hilbert $C^*$-modules}
\label{s:pre}
\protect\setcounter{equation}{0}

Hilbert $C^*$-modules are the natural framework for the
constructions of this paper. Some basic knowledge of the theory of
Hilbert $C^*$-modules would be useful for understanding what follows
but is not really necessary. In this section we shall translate the
necessary facts in a purely Hilbert space setting to make them
easily accessible to people working in the spectral theory of
quantum Hamiltonians. Our basic reference for the general theory of
Hilbert $C^*$-modules is \cite{La} but see also \cite{Bl,JT,RW}.

\PAR\label{ss:not} If $E,F$ are Banach spaces then $L(E,F)$ is the
Banach space of linear continuous maps $E\to F$ and $K(E,F)$ the
subspace of compact maps.  We set $L(E)=L(E,E)$ and $K(E)=K(E,E)$.
We denote $1_E$ or just $1$ the identity map on a Banach space $E$.
Sometimes we set $1_{L^2(X)}=1_X$ if $X$ is a lca group.
Two unusual abbreviations are convenient: by \emph{lspan} and
\emph{clspan} we mean ``linear span'' and ``closed linear span''
respectively.  If $\ca_i$ are subspaces of a Banach space then
$\sum^\rmc_i\ca_i$ is the clspan of $\cup_i\ca_i$.

Let $E,F,G,H$ be Banach spaces. If $\ca\subset L(E,F)$ and
$\cb\subset L(F,G)$ are linear subspaces then $\cb\ca$ is the lspan
of the products $BA$ with $A\in\ca,B\in\cb$ and $\cb\cdot\ca$ is
their clspan.  If $\cc\subset L(G,H)$ is a linear subspace then
$\cc\cdot(\cb\cdot\ca)=(\cc\cdot\cb)\cdot\ca\equiv\cc\cdot\cb\cdot\ca$
is the clspan of the products $CBA$.

If $E,F,G$ are Hilbert spaces then $\ca^*$ is the set of operators of
the form $T^*\in L(F,E)$ with $T\in\ca$. Clearly
$(\cb\cdot\ca)^*=\ca^*\cdot\cb^*$ and
$\ca_1\subset\ca_2\Rightarrow\ca_1^*\subset\ca_2^*$. In particular, if
$E=F=G$ and $\ca=\ca^*$ and $\cb=\cb^*$ then
$\ca\cdot\cb\subset\cb\cdot\ca$ is equivalent to
$\ca\cdot\cb=\cb\cdot\ca$.

\PAR\label{ss:nota} By \emph{ideal} in a $C^*$-algebra we mean closed
self-adjoint ideal. A $*$-homomorphism between two $C^*$-algebras will
be called \emph{morphism}. We write $\ra\simeq\rb$ if the
$C^*$-algebras $\ra,\rb$ are isomorphic and $\ra\cong\rb$ if they are
canonically isomorphic (the isomorphism should be clear from the
context).

If $\ra$ is a $C^*$-algebra then a \emph{Banach $\ra$-module} is a
Banach space $\mr$ equipped with a continuous bilinear map
$\ra\times\mr\ni(A,M)\mapsto MA\in\mr$ such that $(MA)B=M(AB)$.  We
denote $\rM\cdot\ra$ the clspan of the elements $MA$ with $A\in\ra$
and $M\in\rM$.  By the Cohen-Hewitt theorem \cite{FD} for each
$N\in\rM\cdot\ra$ there are $A\in\ra$ and $M\in\rM$ such that
$N=MA$, in particular $\mr\cdot\ra=\mr\ra$.  Note that by module we
mean ``right module'' but the Cohen-Hewitt theorem is also valid for
left Banach modules.

A (right) \emph{Hilbert $\ra$-module} is a Banach $\ra$-module $\mr$
equipped with an $\ra$-valued sesquilinear map
$\braket{\cdot}{\cdot}\equiv\braket{\cdot}{\cdot}_\ra$ which is
positive (i.e. $\braket{M}{M}\geq0$) $\ra$-sesquilinear
(i.e. $\braket{M}{NA}=\braket{M}{N}A$) and such that
$\|M\|\equiv\|\braket{M}{M}\|^{1/2}$. Then $\mr=\mr\ra$.  The clspan
of the elements $\braket{M}{M}$ is an ideal of $\ra$ denoted
$\braket{\mr}{\mr}$. One says that $\mr$ is \emph{full} if
$\braket{\mr}{\mr}=\ra$.  If $\ra$ is an ideal of a $C^*$-algebra
$\rc$ then $\mr$ is equipped with an obvious structure of Hilbert
$\rc$-module.

The examples of interest in this paper are the ``concrete'' Hilbert
$C^*$-modules described in \S\ref{ss:scls} as Hilbert $C^*$-submodules
of $L(\ce,\cf)$. A Hilbert $\mbc$-module is a usual Hilbert space. Any
$C^*$-algebra $\ra$ has a canonical structure of Hilbert $\ra$-module:
the $\ra$-module structure of $\ra$ is defined by the action of $\ra$
on itself by right multiplication and the inner product is
$\braket{A}{B}_\ra=A^*B$.

Let $\rM,\rn$ be Hilbert $\ra$-modules. Then $T\in L(\rM,\rn)$ is
called adjointable if there is $T^*\in L(\rn,\rM)$ such that
$\braket{TM}{N}=\braket{M}{T^*N}$ for $M\in\rM$ and $N\in\rn$. The
map $T^*$ is uniquely defined and is called adjoint of $T$. It is
clear that $T$ and $T^*$ are $\ra$-linear, e.g. $T(MA)=T(M)A$ for
all $M\in\rM$ and $A\in\ra$.  The set of adjointable maps is a
closed subspace of $L(\rM,\rn)$ denoted $\cl(\rM,\rn)$.

An important class of adjointable operators is defined as
follows. If $M\in\mr$ and $N\in\rn$ then the map $M'\mapsto
N\braket{M}{M'}$ is an element of $\cl(\rM,\rn)$ denoted
$\ket{N}\bra{M}$ or $NM^*$. Then \emph{$\ck(\mr,\rn)$ is the closed
  linear subspace generated by these elements}.  The space
$\ck(\rM)\equiv\ck(\rM,\rM)$ is a $C^*$-algebra called
\emph{imprimitivity algebra} of the Hilbert $\ra$-module $\rM$.
Clearly $\ck(\ra)=\ra$.

If $\rb$ is a $C^*$-algebra and $\mr$ is a left Banach $\rb$-module
then a left Hilbert $\rb$-module structure on $\mr$ is defined as
above with the help of a $\rb$-valued inner product
${_\rb}\braket{\cdot}{\cdot}$ linear and $\ra$-linear in the first
variable. For example, if $\mr$ is a Hilbert $\ra$-module then
clearly $\mr$ is a left Banach $\ck(\mr)$-module and if we set
${_{\ck(\mr)}}\braket{M}{N}=MN^*$ we get a canonical full left
Hilbert $\ck(\mr)$-module structure on $\mr$.

If $\mr$ is a full right Hilbert $\ra$-module, a full left Hilbert
$\rb$-module, and ${_\rb}\braket{M}{N}P=M\braket{N}{P}_\ra$ for all
$M,N,P\in\mr$, then one says that $\mr$ is a
\emph{$(\rb,\ra)$-imprimitivity bimodule} and that $\ra$ and $\rb$
are \emph{Morita equivalent}.  $\rM$ is a
$(\ck(\rM),\ra)$-imprimitivity bimodule and one can show that there
is a unique isomorphism of $\rb$ onto $\ck(\rM)$ such that
${_\rb}\braket{M}{N}$ is sent into $MN^*$.

\PAR\label{ss:niet}
Assume that $\rn$ is a closed subspace of a Hilbert $\ra$-module $\mr$
and let $\braket{\rn}{\rn}$ be the clspan of the elements
$\braket{N}{N}$ in $\ra$. If $\rn$ is an $\ra$-submodule of $\mr$ then
it inherits an obvious Hilbert $\ra$-module structure from $\mr$.  If
$\rn$ is not an $\ra$-submodule of $\mr$ it may happen that there is a
$C^*$-subalgebra $\rb\subset\ra$ such that $\rn\rb\subset\rn$ and
$\braket{\rn}{\rn}\subset\rb$. Then clearly we get a Hilbert
$\rb$-module structure on $\rn$. On the other hand, it is clear that
such a $\rb$ exists if and only if $\rn\braket{\rn}{\rn}\subset\rn$
and then $\braket{\rn}{\rn}$ is a $C^*$-subalgebra of $\ra$. Under
these conditions we say that \emph{$\rn$ is a Hilbert $C^*$-submodule}
of the Hilbert $\ra$-module $\mr$. Then $\rn$ inherits a Hilbert
$\braket{\rn}{\rn}$-module structure and this defines the
$C^*$-algebra $\ck(\rn)$. Moreover, if $\rb$ is as above then
$\ck(\rn)=\ck_\rb(\rn)$.

If $\rn$ is a closed subspace of a Hilbert $\ra$-module $\rM$ then
let $\ck(\rn|\mr)$ be the closed subspace of $\ck(\mr)$ generated by
the elements $NN^*$ with $N\in\rn$. It is easy to prove that
\emph{if $\rn$ is a Hilbert $C^*$-submodule of $\mr$ then
$\ck(\rn|\mr)$ is a $C^*$-subalgebra of $\ck(\mr)$ and the map
$T\mapsto T|_\rn$ sends $\ck(\rn|\mr)$ onto $\ck(\rn)$ and is an
isomorphism of $C^*$-algebras}. Then we identify $\ck(\rn|\mr)$
with $\ck(\rn)$.

\PAR\label{ss:scls} If $\ce,\cf$ are Hilbert spaces then we equip
$L(\ce,\cf)$ with the Hilbert $L(\ce)$-module structure defined as
follows: the $C^*$-algebra $L(\ce)$ acts to the right by composition
and we take $\braket{M}{N}=M^*N$ as inner product, where $M^*$ is
the usual adjoint of the operator $M$. Note that $L(\ce,\cf)$ is
also equipped with a natural left Hilbert $L(\cf)$-module structure:
this time the inner product is $MN^*$.

Now let $\mr\subset L(\ce,\cf)$ be a closed linear subspace and let
$\mr^*\subset L(\cf,\ce)$ be the set of adjoint operators $M^*$ with
$M\in\mr$. Then \emph{$\mr$ is a Hilbert $C^*$-submodule of
$L(\ce,\cf)$ if and only if $\mr\mr^*\mr\subset\mr$}.

These are the ``concrete'' Hilbert $C^*$-modules we are interested
in.  We summarize below some immediate consequence of the discussion
in \S\ref{ss:niet}.

\begin{proposition}\label{pr:ss}
Let $\ce,\cf$ be Hilbert spaces and let $\mr$ be a Hilbert
$C^*$-submodule of $L(\ce,\cf)$. Then $\ra\equiv\mr^*\cdot\mr$ and
$\rb\equiv\mr\cdot\mr^*$ are $C^*$-algebras of operators on $\ce$ and
$\cf$ respectively and $\mr$ is equipped with a canonical structure of
$(\rb,\ra)$-imprimitivity bimodule. 
\end{proposition}

It is clear that $\mr^*$ will be a Hilbert $C^*$-submodule of
$L(\cf,\ce)$. We mention that $\mr^*$ is canonically identified with
the left Hilbert $\ra$-module $\ck(\mr,\ra)$ dual to $\mr$.

\begin{proposition}\label{pr:clsubmod}
Let $\rn$ be a $C^*$-submodule of $L(\ce,\cf)$ such that
$\rn\subset\mr$ and $\rn^*\cdot\rn=\mr^*\cdot\mr$,
$\rn\cdot\rn^*=\mr\cdot\mr^*$. Then $\rn=\mr$.
\end{proposition}
\proof If $M\in\mr$ and $N\in\rn$ then $MN^*\in\rb=\rn\cdot\rn^*$ and
$\rn\rn^*\rn\subset\rn$ hence $MN^*N\in\rn$. Since $\rn^*\cdot\rn=\ra$
we get $MA\in\rn$ for all $A\in\ra$. Let $A_i$ be an approximate
identity for the $C^*$-algebra $\ra$. Since one can factorize $M=M'A'$
with $M'\in\mr$ and $A'\in\ra$ the sequence $MA_i=M'A'A_i$ converges
to $ M'A'=M$ in norm. Thus $M\in\rn$.
\qed

It is clear that $\ra\cdot\ce=\ce\Rightarrow\mr^*\cdot\cf=\ce$ and
$\rb\cdot\cf=\cf\Rightarrow\mr\cdot\ce=\cf$.  Moreover:
\begin{equation}\label{eq:ndg}
\ra\cdot\ce = \ce \text{ and } \rb\cdot\cf=\cf  \Leftrightarrow
\mr\cdot\ce=\cf \text{ and } \mr^*\cdot\cf=\ce.
\end{equation}
If the relations \eqref{eq:ndg} are satisfied we say that $\mr$ is a
\emph{nondegenerate} Hilbert $C^*$-submodule of $L(\ce,\cf)$. For
such modules we have the following concrete representation of
$\cl(\mr)$, cf.  Proposition 2.3 in \cite{La}. If a symbol like
$S^{(*)}$ appears in a relation this means that the relation holds
for both $S$ and $S^*$.

\begin{proposition}\label{pr:1ss}
If $\rb\cf=\cf$ then
\begin{equation}\label{eq:1ss}
\cl(\mr)\cong\{ S\in L(\cf)\mid S^{(*)}\mr\subset\mr\}=
\{ S\in L(\cf)\mid S^{(*)}\rb\subset\rb\}
\end{equation}
where the canonical isomorphism associates to $S$ the map
$M\mapsto SM$. 
\end{proposition}

The proof of the next proposition is left as an exercise.

\begin{proposition}\label{pr:2ss}
Let $\ce,\cf,\ch$ be Hilbert spaces and let $\mr\subset L(\ch,\ce)$
and $\rn\subset L(\ch,\cf)$ be Hilbert $C^*$-submodules. Let $\ra$ be
a $C^*$-algebra of operators on $\ch$ such that $\mr^*\cdot\mr$ and
$\rn^*\cdot\rn$ are ideals of $\ra$ and let us view $\mr$ and $\rn$ as
Hilbert $\ra$-modules. Then $\ck(\mr,\rn)\cong\rn\cdot\mr^*$ the
isometric isomorphism being determined by the condition
$\ket{N}\bra{M}=NM^*$. 
\end{proposition}

\PAR\label{ss:ha} We recall the definition of the tensor product of
a Hilbert space $\ce$ and a $C^*$-algebra $\ra$ in the category of
Hilbert $C^*$-modules. We equip the algebraic tensor product
$\ce\odot\ra$ with the obvious right $\ra$-module structure and with
the $\ra$-valued sesquilinear map given by
\begin{equation}\label{eq:hoa}
\braket{{\textstyle\sum_{u\in\ce}}u\otimes
  A_u}{{\textstyle\sum_{v\in\ce}}v\otimes B_v} 
={\textstyle\sum_{u,v}}\braket{u}{v}A^*_uB_v
\end{equation}
where $A_u=B_u=0$ outside a finite set. Then the completion of
$\ce\odot\ra$ for the norm $\|M\|\equiv\|\braket{M}{M}\|^{1/2}$ is a
full Hilbert $\ra$-module denoted $\ce\otimes\ra$. Clearly its
imprimitivity algebra is
\begin{equation}\label{eq:cha}
\ck(\ce\otimes\ra)=K(\ce)\otimes\ra.
\end{equation}  
The reader may easily check that if $Y$ is a locally compact space
then $\ce\otimes\Co(Y)\cong\Co(Y;\ce)$. And if $X$ is a locally
compact space equipped with a Radon measure then $L^2(X)\otimes\ra$
is the completion of $\Cc(X;\ra)$ for the norm $\|\int_X F(x)^*F(x)
\rmd x\|^{1/2}$.  Hence $L^2(X)\otimes\Co(Y)$ is the completion of
$\Cc(X\times Y)$ for the norm $\sup_{y\in Y}(\int_Y |F(x,y)|^2 \rmd
x)^{1/2}$. Note that $L^2(X;\ra)\subset L^2(X)\otimes\ra$ strictly
in general. If $\ra\subset L(\cf)$ then the norm on
$L^2(X)\otimes\ra$ we can also be written as follows:
\begin{equation}\label{eq:L2a}
\|{\textstyle\int_X} F(x)^*F(x) \rmd x\|=
{\textstyle\sup_{f\in\cf,\|f\|=1}} {\textstyle\int_X} 
\|F(x)f\|^2 \rmd x.
\end{equation}
Now assume that $\ra$ is realized on a Hilbert space $\cf$. Then we
have a natural embedding
\begin{equation}\label{eq:etensa}
\ce\otimes\ra \subset L(\cf,\ce\otimes\cf)
\end{equation}
which we describe below. For each $u\in\ce$ and $A\in\ra$ let
$\ket{u}\otimes A:\cf\to\ce\otimes\cf$ be the map $f\mapsto
u\otimes(Af)$. Note that if $\ket{u}$ is the map $\mbc\to\ce$ given by
$\lambda\mapsto\lambda u$ then $\ket{u}\otimes A$ is really a tensor
product of operators because $\cf\equiv\mbc\otimes\cf$. Let
$\bra{u}=\ket{u}^*:\ce\to\mbc$ be the adjoint map
$v\mapsto\braket{u}{v}$.  Then $(\ket{u}\otimes A)^*=\bra{u}\otimes
A^*:\ce\otimes\cf\to\cf$ acts on decomposable tensors as follows:
$(\bra{u}\otimes A^*) (v\otimes f)=\braket{u}{v}A^*f$.  From
\eqref{eq:hoa} we easily deduce now that there is a unique continuous
linear map $\ce\otimes\ra\to L(\ce,\ce\otimes\cf)$ such that the image
of $u\otimes A$ be $\ket{u}\otimes A$ and this map is an isometry of
$\ce\otimes\ra$ onto the clspan of the set of operators of the form
$\ket{u}\otimes A$. This defines the canonical identification
\eqref{eq:etensa} of $\ce\otimes\ra$ with a closed linear subspace of
$L(\cf,\ce\otimes\cf)$.

Thus if $\ra\subset L(\cf)$ the Hilbert $\ra$-module $\ce\otimes\ra$
is realized as a Hilbert $C^*$-submodule of $L(\cf,\ce\otimes\cf)$,
the dual module is realized $(\ce\otimes\ra)^*\subset
L(\ce\otimes\cf,\ce)$ as the set of adjoint operators, and the
relations
\begin{equation}\label{eq:etens}
(\ce\otimes\ra)^*\cdot (\ce\otimes\ra)=\ra, \hspace{2mm}
(\ce\otimes\ra)\cdot(\ce\otimes\ra)^*=K(\ce)\otimes\ra
\end{equation} 
are immediate.

We consider now more general tensor products.  If $\ce,\cf,\cg,\ch$
are Hilbert spaces and $\mr\subset L(\ce,\cf)$ and $\rn\subset
L(\cg,\ch)$ are closed linear subspaces then we denote $\mr\otimes\rn$
the closure in $L(\ce\otimes\cg,\cf\otimes\ch)$ of the algebraic
tensor product of $\mr$ and $\rn$. Now suppose that $\mr$ is a
$C^*$-submodule of $L(\ce,\cf)$ and that $\rn$ is a $C^*$-submodule of
$L(\cg,\ch)$ and let $\ra=\mr^*\cdot\mr$ and $\rb=\rn^*\cdot\rn$. Then
$\mr$ is a Hilbert $\ra$-module and $\rn$ is a Hilbert $\rb$-module
hence the exterior tensor product, denoted temporarily
$\mr\otimes_{\text{ext}}\rn$, is well defined in the category of
Hilbert $C^*$-modules \cite{La} and is a Hilbert
$\ra\otimes\rb$-module. On the other hand, it is easy to check that
$(\mr\otimes\rn)^*=\mr^*\otimes\rn^*$ and then that $\mr\otimes\rn$ is
a Hilbert $C^*$-submodule of $L(\ce\otimes\cg,\cf\otimes\ch)$ such
that $(\mr\otimes\rn)^*\cdot(\mr\otimes\rn)=\ra\otimes\rb$. Finally,
it is clear that $L(\ce\otimes\cg,\cf\otimes\ch)$ and
$\mr\otimes_{\text{ext}}\rn$ induce the same $\ra\otimes\rb$-valued
inner product on the algebraic tensor product of $\mr$ and $\rn$.
Thus we we get a canonical isometric isomorphism
$\mr\otimes_{\text{ext}}\rn=\mr\otimes\rn$.

In the preceding framework, it is easy to see that we have a
canonical identification
\begin{equation}\label{eq:comtens}
K(\ce,\cf)\otimes K(\cg,\ch)\cong K(\ce\otimes\cg,\cf\otimes\ch).
\end{equation}
In particular $K(\ce,\cf\otimes\ch)\cong K(\ce,\cf)\otimes\ch$.

It will be convenient for our later needs to introduce a more
intuitive notation for certain tensor products.

\begin{definition}\label{df:ww}
If $X$ is a locally compact space equipped with a Radon measure,
$\ce$ and $\cf$ are Hilbert spaces, and $\mr\subset L(\ce,\cf)$ is a
closed subspace, then $L^2_\rmw(X;\mr)$ is the completion of the
space of functions $F:X\to\mr$ of the form $F(x)=\sum f_k(x)M_i$
with $f_k\in\Cc(X)$ and $ M_k\in\mr$ for the norm
\begin{equation}\label{eq:ww}
\|F\|_{L^2_\rmw}=
\|{\textstyle\int_X} F(x)^*F(x) \rmd x\|^{1/2}=
{\sup_{e\in\ce,\|e\|=1}} \big({\textstyle\int_X} 
\|F(x)e\|^2 \rmd x\big)^{1/2}.
\end{equation}
\end{definition}
The elements of $L^2_\rmw(X;\mr)$ are (equivalence classes of)
strongly measurable $L(\ce,\cf)$ valued functions on $X$ and we have
$L^2(X;\mr)\subset L^2_\rmw(X;\mr)$ strictly. For the needs of our
examples $L^2(X;\mr)$ is largely sufficient but $L^2_\rmw(X;\mr)
\cong L^2(X)\otimes\mr$, viewed as a space of operators $\ce\to
L^2(X)\otimes\cf$, is more natural in our context.

\section{Preliminaries on groups and crossed products}
\label{s:hlca}
\protect\setcounter{equation}{0}

In this section we review notations and describe some preliminary
results concerning the locally compact abelian (lca) groups and
their crossed products with $C^*$-algebras.

\PAR\label{ss:group} Let us consider a lca group $X$ (with operation
denoted additively) and a closed subgroup $Y\subset X$ equipped with
Haar measures $\rmd x$ and $\rmd y$.  We shall write $X=Y\oplus Z$
if $X$ is the direct sum of the two closed subgroups $Y,Z$ equipped
with compatible Haar measures, in the sense that $\rmd x=\rmd
y\otimes \rmd z$. We set $\rl_X\equiv L(L^2(X))$ and $\rk_X\equiv
K(L^2(X))$ and note that these are $C^*$-algebras independent of the
choice of the measure on $X$.  If $X=Y\oplus Z$ then
$L^2(X)=L^2(Y)\otimes L^2(Z)$ as Hilbert spaces and
$\rk_X=\rk_Y\otimes \rk_Z$ as $C^*$-algebras. It will also be
convenient to use the abbreviations
$$
\rl_{XY}=L(L^2(Y),L^2(X)) \text{ and } \rk_{XY}=K(L^2(Y),L^2(X)).
$$ 
The bounded uniformly continuous functions on $X$ form a
$C^*$-algebra $\Cbu(X)$ which contains the algebras $\Cc(X)$,
$\Co(X)$ of functions which have compact support or tend to zero at
infinity.  We embed $\Cbu(X/Y)\subset\Cbu(X)$ with the help of the
injective morphism $\varphi\mapsto\varphi\circ\pi_Y$ where
$\pi_Y:X\to X/Y$ is the canonical surjection. So $\Cbu(X/Y)$ is
identified with the set of functions $\varphi\in\Cbu(X)$ such that
$\varphi(x+y)=\varphi(x)$ for all $x\in X$ and $y\in Y$.

In particular, $\Co(X/Y)$ is identified with the set of continuous
functions $\varphi$ on $X$ such that $\varphi(x+y)=\varphi(x)$ for
all $x\in X$ and $y\in Y$ and such that for each $\varepsilon>0$
there is a compact $K\subset X$ such that $|\varphi(x)|<\varepsilon$
if $x\nin K+Y$. By $x/Y\to\infty$ we mean $\pi_Y(x)\to\infty$, so
the last condition is equivalent to $\varphi(x)\to0$ if
$x/Y\to\infty$.  To avoid cumbersome expressions like $\Co(X/(Y\cap
Z))$ and also for coherence in later notations we set
\begin{equation}\label{eq:coxy}
\cc_X(Y)=\Co(X/Y)
\end{equation}
If $X=Y\oplus Z$ then $\cc_X(Y)=1\otimes\Co(Z)$ relatively to the
tensor factorization 
$L^2(X)=L^2(Y)\otimes L^2(Z)$.

We denote by $\varphi(Q)$ the operator in $L^2(X)$ of multiplication
by a function $\varphi$ and if $X$ has to be explicitly specified we
set $Q=Q_X$.  The map $\varphi\mapsto\varphi(Q)$ is an embedding
$\Cbu(X)\subset\rl_X$.

The translation operator $U_x$ on $L^2(X)$ associated to $x\in X$ is
defined by $(U_xu)(y)=u(y+x)$. We set
$\tau_xS\equiv\tau_x(S)=U_xSU_x^{*}$ for $S\in \rl_X$ and also
$(\tau_x\varphi)(y)=\varphi(y+x)$ for an arbitrary function
$\varphi$ on $X$, so that $\tau_x(\varphi(Q))=(\tau_x\varphi)(Q)$.
To an element $y\in Y$ we may associate a translation operator $U_y$
in $L^2(X)$ and another translation operator in $L^2(Y)$. However,
in order not to overcharge the writing we shall denote the second
operator also by $U_y$.

Let $X^*$ be the group dual to $X$ with operation denoted
additively\symbolfootnote[2]{\ Then $(k+p)(x)=k(x)p(x)$, $0(x)=1$,
and the element $-k$ of $X^*$ represents the function $\bar{k}$. In
order to avoid such strange looking expressions  one might use the
notation $k(x)=[x,k]$.  }.   
If $k\in X^*$ we define a 
unitary operator $V_k$ on $L^2(X)$ by $(V_ku)(x)=k(x) u(x)$.  The
restriction map $k\mapsto k|_Y$ is a continuous surjective group
morphism $X^*\to Y^*$ with kernel equal to $Y^\perp=\{k\in X^*\mid
k(y)=1\hspace{1mm}\forall y\in Y\}$ which defines the canonical
identification $Y^*\cong X^*/Y^\perp$. We denote by the same symbol
$V_k$ the operator of multiplication by the character $k\in X^*$ in
$L^2(X)$ and by the character $k|_Y\in Y^*$ in $L^2(Y)$.

Let $\cc^*(X)$ be the group $\cc^*$-algebra of $X$: this is the closed
linear subspace of $\rl_X$ generated by the convolution operators
of the form $(\varphi*u)(x)=\int_X \varphi(x-y)u(y)\rmd y$ with
$\varphi\in\Cc(X)$. We recall the notation
$\varphi^*(x)=\bar\varphi(-x)$.  Note that if we set
$C(\varphi)u\equiv\varphi*u$, then
$C(\varphi)=\int_X\varphi(-x)U_x\rmd x$.

The Fourier transform of an integrable measure $\mu$ on $X$ is
defined by $(F\mu)(k)=\int \bar{k}(x)\mu(\rmd x)$.  Then $F$ induces
a bijective map $L^2(X)\rarrow L^2(X^*)$ hence a canonical
isomorphism $S\mapsto F^{-1}S F$ of $\rl_{X^*}$ onto $\rl_X$.  If
$\psi$ is a function on $X^*$ we set $\psi(P)=F^{-1}M_\psi F$, where
$M_\psi$ is the operator of multiplication by $\psi$ on $L^2(X^*)$.
The map $\psi\mapsto\psi(P)$ gives an isomorphism $\Co(X^*)\cong
\cc^*(X)$.  If the group has to be specified, we set $P=P_X$.

\PAR\label{ss:nbcrp} A $C^*$-subalgebra stable under translations of
$\Cbu(X)$ will be called \emph{$X$-algebra}. The operation of
restriction of functions allows us to associate to each $X$-algebra
$\ca$ a $Y$-algebra $\ca|_Y=\{\varphi|_Y\mid\varphi\in\ca\}$. The
map $\ca\mapsto\ca|_Y$ from the set of $X$-algebras to the set of
$Y$-algebras is surjective.

If $\ca$ is an $X$-algebra then the \emph{crossed product of $\ca$ by
  the action of $X$} is an abstractly defined $C^*$-algebra
$\ca\rtimes X$ but we shall always identify it with the $C^*$-algebra
of operators on $L^2(X)$ given by
\begin{equation}\label{eq:crp}
\ca\rtimes X\equiv\ca\cdot \cc^*(X)=\cc^*(X)\cdot\ca \subset \rl_X,
\end{equation}
see, for example, Theorem 4.1 in \cite{GI1}.  The next result, due
to Landstad \cite{Ld}, gives an ``intrinsic'' characterization of
crossed products. We follow the presentation from \cite[Theorem
  3.7]{GI4} which takes advantage of the fact that $X$ is abelian.

\begin{theorem}\label{th:land}
A $C^*$-algebra $\ra\subset \rl_X$ is a crossed product
if and only for each $A\in\ra$ we have:
\begin{itemize}
\vspace{-2mm} 
\item
if $k\in X^*$ then $V_k^*AV_k\in\ra$ and
$\lim_{k\rarrow0}\|V_k^*AV_k-A\|=0$,
\item
if $x\in X$ then $U_xA\in\ra$ and $\lim_{x\rarrow0}\|(U_x-1)A\|=0$.
\vspace{-2mm}
\end{itemize} 
In this case one has $\ra=\ca\rtimes X$ for a unique $X$-algebra
$\ca\subset\Cbu(X)$  and this algebra is given by
\begin{equation}\label{eq:land}
\ca =\{\varphi\in\Cbu(X)\mid 
\varphi(Q)S \in {\ra} \hspace{1mm}\text{and}\hspace{1mm}
\bar\varphi(Q)S \in {\ra}
\hspace{1mm}\text{for all}\hspace{1mm} S \in \cc^*(X)\}.
\end{equation} 
\end{theorem}
Note that the second condition of Landstad's theorem is equivalent
to $\cc^*(X)\cdot\ra=\ra$, cf. Lemma \ref{lm:help}.

We discuss now crossed products of the form $\cc_X(Y)\rtimes X$
which play an important role in the $N$-body problem. To simplify
notations we set
\begin{equation}\label{eq:Cxy}
\rc_X(Y)\equiv\cc_X(Y)\rtimes X=\cc_X(Y)\cdot\cc^*(X)=
\cc^*(X)\cdot\cc_X(Y).
\end{equation}
If $X=Y\oplus Z$ and if we identify $L^2(X)=L^2(Y)\otimes L^2(Z)$ then
$\cc^*(X)=\cc^*(Y)\otimes\cc^*(Z)$ hence
\begin{equation}\label{eq:crxyz}
\rc_X(Y)=\cc^*(Y)\otimes \rk_Z.
\end{equation}
A useful ``symmetric'' description of $\rc_X(Y)$ is contained in the
next lemma. Let $Y^{(2)}$ be the closed subgroup of $X^2\equiv X\oplus
X$ consisting of elements of the form $(y,y)$ with $y\in Y$.

\begin{lemma}\label{lm:sym}
$\rc_X(Y)$ is the closure of the
set of integral operators with kernels $\theta\in\Cc(X^2/Y^{(2)})$.
\end{lemma}
\proof Let $\rc$ be the norm closure of the set of integral
operators with kernels $\theta\in\Cbu(X^2)$ having the properties:
(1) $\theta(x+y,x'+y)=\theta(x,x')$ for all $x,x'\in X$ and $y\in
Y$; (2) $\supp\theta\subset K_\theta+Y$ for some compact
$K_\theta\subset X^2$. We show $\rc=\rc_X(Y)$.  Observe that the map
in $X^2$ defined by $(x,x')\mapsto(x-x',x')$ is a topological group
isomorphism with inverse $(x_1,x_2)\mapsto(x_1+x_2,x_2)$ and sends
the subgroup $Y^{(2)}$ onto the subgroup $\{0\}\oplus Y$. This map
induces an isomorphism $X^2/Y^{(2)}\simeq X\oplus(X/Y)$. Thus any
$\theta\in\Cc(X^2/Y^{(2)})$ is of the form
$\theta(x,x')=\wtilde\theta(x-x',x')$ for some
$\wtilde\theta\in\Cc(X\oplus(X/Y))$. Thus $\rc$ is the closure in
$\rl_X$ of the set of operators of the form
$(Tu)(x)=\int_X\wtilde\theta(x-x',x')u(x') \rmd x'$. Since we may
approximate $\wtilde\theta$ with linear combinations of functions of
the form $a\otimes b$ with $a\in\Cc(X), b\in\Cc(X/Y)$ we see that
$\rc$ is the clspan of the set of operators of the form
$(Tu)(x)=\int_X a(x-x')b(x')u(x') \rmd x'$. But this clspan is
$\cc^*(X)\cdot\cc_X(Y)=\rc_X(Y)$.  \qed

Our purpose now is to give an intrinsic description of $\rc_X(Y)$. We
need the following result, which will be useful in other contexts
too. Let $\{T_g\}$ be a strongly continuous unitary representation of
a lca group $G$ on a Hilbert space $\ch$ and let $\psi\mapsto T(\psi)$
be the morphism $\Co(G^*)\to L(\ch)$ associated to it.

\begin{lemma}\label{lm:help}
If $A\in L(\ch)$ then $\lim_{g\to0}\|(T_g-1)A\|=0$ if and only if
$A=T(\psi)B$ for some $\psi\in\Co(G^*)$ and $B\in L(\ch)$.
\end{lemma}

This is an easy consequence of the Cohen-Hewitt factorization theorem,
see Lemma 3.8 from \cite{GI4}.

\begin{theorem}\label{th:cxy} 
$\rc_X(Y)$ is the set of $A\in \rl_X$ such that $U_y^*AU_y=A$ for all
$y\in Y$ and:
\begin{enumerate}
\item
$\|U_x^*AU_x-A\|\to 0$ if $x\to 0$ in $X$ and 
$\|V_k^*AV_k-A\|\to 0$ if $k\to 0$ in $X^*$,
\item
$\|(U_x-1)A\|\to 0$ if $x\to 0$ in $X$ and $\|(V_k-1)A\|\to 0$ if
$k\to 0$ in $Y^\perp$. 
\end{enumerate}   
\end{theorem}

By ``$k\to 0$ in $Y^\perp$'' we mean: $k\in Y^\perp$ and $k\to 0$.
Note that the second condition above is equivalent to:
\begin{equation}\label{eq:cxy}
\text{there are } \theta\in\cc^*(X),\ \psi\in\cc_X(Y)
\text{ and }  B,C\in\rl_X \text{ such that } 
A=\theta(P)B=\psi(Q)C.
\end{equation}
For the proof, use $Y^\perp\cong (X/Y)^*$ and apply Lemma
\ref{lm:help}. In particular, the last factorization shows that for
each $\varepsilon>$ there is a compact set $M\subset X$ such that
$\|\cchi_V(Q)A\|<\varepsilon$, where $V=X\setminus(M+Y)$.

\noindent{\bf Proof of Theorem \ref{th:cxy}:} This has been proved by
direct means for $X$ a finite dimensional real vector space in
\cite{DG2}. Here we use Theorem \ref{th:land} which allows us to treat
arbitrary groups.  Let $\ra\subset \rl_X$ be the set of operators $A$
satisfying the conditions from the statement of the theorem.  We first
prove that $\ra$ satisfies the two conditions of Theorem
\ref{th:land}. Let $A\in\ra$.  We have to show that $A_p\equiv
V_p^*AV_p\in\ra$ and $\|V_p^*AV_p-A\|\to0$ as $p\to0$.  From the
commutation relations $U_xV_p=p(x)V_pU_x$ we get
$\|(U_x-1)A_p\|=\|(U_x-p(x))A\|\to0$ if $x\to0$ and the second part
of condition 1 of the theorem is obviously satisfied by $A_p$. Then
for $y\in Y$
$$
U_y^*A_pU_y=U_y^*V_p^*AV_pU_y=V_p^*U_y^*AU_yV_p=V_p^*AV_p=A_p.
$$ Condition 2 is clear so we have $A_p\in\ra$ and the fact that
$\|V_p^*AV_p-A\|\to0$ as $p\to0$ is obvious. That $A$ satisfies the
second Landstad condition, namely that for each $a\in X$ we have
$U_aA\in\ra$ and $\|(U_a-1)A\|\to0$ as $a\to0$, is also clear because
$\|[U_a,V_k]\|\to0$ as $k\to0$.

Now we have to find the algebra $\ca$ defined by \eqref{eq:land}.
Assume that $\varphi\in\Cbu(X)$ satisfies $\varphi(Q)S\in\ra$ for all
$S\in \cc^*(X)$. Since $U_y^*\varphi(Q)U_y=\varphi(Q-y)$ we get
$(\varphi(Q)-\varphi(Q-y))S=0$ for all such $S$ and all $y\in Y$,
hence $\varphi(Q)-\varphi(Q-y)=0$ which means $\varphi\in\Cbu(X/Y)$.
We shall prove that $\varphi\in\cc_X(Y)$ by reductio ad absurdum. 

If $\varphi\nin\cc_X(Y)$ then there is $\mu>0$ and there is a
sequence of points $x_n\in X$ such that $x_n/Y\to\infty$ and
$|\varphi(x_n)|>2\mu$. From the uniform continuity of $\varphi$ we
see that there is a compact neighborhood $K$ of zero in $X$ such
that $|\varphi|>\mu$ on $\bigcup_n(x_n+K)$.  Let $K'$ be a compact
neighborhood of zero such that $K'+K'\subset K$ and let us choose
two positive not zero functions $\psi,f\in\Cc(K')$. We define $S\in
\cc^*(X)$ by $Su=\psi*u$ and recall that $\supp Su\subset\supp
\psi+\supp u$. Thus $\supp SU_{x_n}^*f\subset K'+x_n+K'\subset
x_n+K$. Now let $V$ be as in the remarks after \eqref{eq:cxy}. Since
$\pi_Y(x_n)\to\infty$ we have $x_n+K\subset V$ for $n$ large enough,
hence
$$
\|\cchi_V(Q)\varphi(Q)SU_{x_n}^*f\|\geq\mu\|SU_{x_n}^*f\|=
\mu\|Sf\| >0.
$$
On the other hand, for each $\varepsilon>0$ one can choose $V$ such
that $\|\cchi_V(Q)\varphi(Q)S\|<\varepsilon$.  Then we shall have
$\|\cchi_V(Q)\varphi(Q)SU_{x_n}^*f\|\leq\varepsilon\|f\|$ so
$\mu\|Sf\|\leq\varepsilon\|f\|$ for all $\varepsilon>0$ which is
absurd.  \qed

\section{Compatible groups and associated Hilbert $C^*$-modules}
\label{s:pair}
\protect\setcounter{equation}{0}

\PAR\label{ss:compat} If $X,Y$ is an arbitrary pair of lca groups
then $X\oplus Y$ is the set $X\times Y$ equipped with the product
topology and group structure, so that $X\oplus Y$ is a lca group.
Assume that $X,Y$ are closed subgroups (equipped with the induced
lca group structure) of a lca group $G$.  
Let us identify $X\cap Y$ with the closed subgroup of $X\oplus Y$
consisting of the elements of the form $(z,z)$ with $z\in X\cap Y$.
Then we may construct the lca quotient group

\begin{equation}\label{eq:boxplus}
 X\uplus Y \equiv (X\oplus Y)/(X\cap Y).
\end{equation}
On the other hand, we may also consider the subgroup $X+Y$ of $G$
generated by $X\cup Y$ equipped with the topology induced by
$G$. Note that if $H$ is a closed subgroup of $G$ such that $X\cup
Y\subset H$ and if we construct $X+Y$ by using $H$ instead of $G$
then we get the same topological group: thus the group $G$ does not
play a fundamental role in what follows. We have a natural map
\begin{equation}\label{eq:nat}
\phi:X\oplus Y \to X+Y \hspace{2mm}\text{defined by}\hspace{2mm}
 \phi(x,y)=x-y
\end{equation}
which is a continuous surjective group morphism $X\oplus Y\to X+Y$
with $X\cap Y$ as kernel hence it induces a continuous bijective group
morphism $\phi^\circ:X\uplus Y\to X+Y$. Clearly $\phi$ is an open
map if and only if $\phi^\circ$ is a homeomorphism and then $X+Y$ is a
locally compact group hence\symbolfootnote[2]{\ 
We recall that a subgroup $H$ of a locally compact group
$G$ is closed if and only if $H$ is locally compact for the induced
topology; see Theorem 5.11 in \cite{HR}.} \label{footnote}
a closed subgroup of $\rx$.

\begin{definition}\label{df:reg}
Two closed subgroups $X,Y$ of a lca group are \emph{compatible} if the
map \eqref{eq:nat} is open.
\end{definition}

\begin{remark}\label{re:reg2}
If $G$ is $\sigma$-compact then $X,Y$ are compatible if and only if
$X+Y$ is closed. Indeed, a continuous surjective morphism between
two locally compact $\sigma$-compact groups is open
 (see Theorem 5.29 in \cite{HR}; we thank Lo\"ic Dubois and Benoit
Pausader for enlightening discussions on this matter).
\end{remark}

Other useful descriptions of the compatibility condition may be found
in Lemma 6.1.1 from \cite{Ma} (or Lemma 3.1 from \cite{Ma3}), we quote
now two of them. Let $X/Y$ be the image of $X$ in $G/Y$ considered as
a subgroup of $G/Y$ equipped with the induced topology. On the other
hand, the group $X/(X\cap Y)$ is equipped with the locally compact
quotient topology and we have a natural map $X/(X\cap Y)\to X/Y$ which
is a bijective continuous group morphism.  Then $X,Y$ are compatible
if and only if the following equivalent conditions are satisfied:
\begin{align}
& \text{the natural map} \hspace{2mm} X/(X\cap Y)\to X/Y \hspace{2mm}
\text{is a homeomorphism}, \label{eq:ma1} \\
&
\text{the natural map }
G/(X\cap Y)\to G/X\times G/Y \hspace{1mm} \text{is closed}.
\label{eq:ma2}
\end{align}

The next three lemmas will be needed later on.

\begin{lemma}\label{lm:reg}
If $X,Y$ are compatible then
\begin{align}
& \cc_G(X)\cdot\cc_G(Y) = \cc_G(X\cap Y)  \label{eq:reg1}\\
& \cc_G(Y)|_X =  \cc_X(X\cap Y).
\label{eq:reg2}
\end{align}
The second relation remains valid for the subalgebras $\Cc$.
\end{lemma}
\proof The fact that the inclusion $\subset$ in \eqref{eq:reg1} is
equivalent to the compatibility of $X$ and $Y$ is shown in Lemma 6.1.1
from \cite{Ma}, so we only have to prove that the equality holds. Let
$E = (G/X) \times (G/Y)$. If $\varphi \in \Co(G/X)$ and $\psi \in
\Co(G/Y)$ then $\varphi \otimes \psi$ denotes the function $(s, t)
\longmapsto \varphi(s) \psi(t)$, which belongs to $\Co(E)$. The
subspace generated by the functions of the form $\varphi \otimes \psi$
is dense in $\Co(E)$ by the Stone-Weierstrass theorem. If $F$ is a
closed subset of $E$ then, by the Tietze extension theorem, each
function in $\Cc(F)$ extends to a function in $\Cc(E)$, so the
restrictions $(\varphi \otimes \psi)|_F$ generate a dense linear
subspace of $\Co(F)$.  Let us denote by $\pi$ the map $x \mapsto
(\pi_X(x), \pi_Y(x))$, so $\pi$ is a group morphism from $G$ to $E$
with kernel $V=X\cap Y$.  Then by \eqref{eq:ma2} the range $F$ of
$\pi$ is closed and the quotient map $\wtilde\pi : G/V \to F$ is a
continuous and closed bijection, hence is a homeomorphism. So $\theta
\mapsto \theta \circ \tilde \pi$ is an isometric isomorphism of
$\Co(F)$ onto $\Co(G/V)$. Hence for $\varphi \in \Co(G/X)$ and $\psi
\in \Co(G/Y)$ the function $\theta = (\varphi \otimes \psi) \circ
\tilde \pi$ belongs to $\Co(G/V)$, it has the property $\theta \circ
\pi_V = \varphi \circ \pi_X \cdot \psi \circ \pi_Y$, and the functions
of this form generate a dense linear subspace of $\Co(G/V)$.

Now we prove \eqref{eq:reg2}. Recall that we identify $\cc_G(Y)$
with a subset of $\Cbu(G)$ by using $\varphi\mapsto\varphi\circ\pi_Y$
so in terms of $\varphi$ the restriction map which defines
$\cc_G(Y)|_X$ is just $\varphi\mapsto\varphi|_{X/Y}$. Thus we have a
canonical embedding $\cc_G(Y)|_X\subset\Cbu(X/Y)$ for an arbitrary
pair $X,Y$ . Then the continuous bijective group morphism
$\theta:X/(X\cap Y)\to X/Y$ allows us to embed
$\cc_G(Y)|_X\subset\Cbu(X/(X\cap Y))$. That the range of this map is
not $\cc_X(X\cap Y)$ in general is clear from the example $G=\mbr,
X=\pi\mbz,Y=\mbz$. But if $X,Y$ are compatible then $X/Y$ is closed
in $G/Y$, so $\cc_G(Y)|_X=\Co(X/Y)$ by the Tietze extension theorem,
and $\theta$ is a homeomorphism, hence we get \eqref{eq:reg2}.  \qed

\begin{lemma}\label{lm:double}
If $X,Y$ are compatible then $X^2=X\oplus X$ and
$Y^{(2)}=\{(y,y)\mid y\in Y\}$ is a compatible pair of closed
subgroups of $G^2=G\oplus G$.
\end{lemma}
\proof Let $D=X^2\cap Y^{(2)}=\{(x,x)\mid x\in X\cap Y\}$. Due to to
\eqref{eq:ma1} it suffices to show that the natural map $Y^{(2)}/D\to
Y^{(2)}/X^2$ is a homeomorphism.  Here $Y^{(2)}/X^2$ is the image of
$Y^{(2)}$ in $G^2/X^2\cong (G/X)\oplus(G/X)$, more precisely it
is the subset of pairs $(a,a)$ with $a=\pi_X(z)$ and $z\in Y$,
equipped with the topology induced by $(G/X)\oplus(G/X)$. Thus the
natural map $Y/X\to Y^{(2)}/X^2$ is a homeomorphism. On the other
hand, the natural map $Y/(X\cap Y)\to Y^{(2)}/D$ is clearly a
homeomorphism.  To finish the proof note that $Y/(X\cap Y)\to Y/X$ is
a homeomorphism because $X,Y$ is a regular pair.  \qed

\begin{lemma}\label{lm:regp}
If the closed subgroups $X,Y$ of $G$ are compatible then $(X\cap
Y)^\perp=X^\perp+Y^\perp$ and the closed subgroups $X^\perp,Y^\perp$
of $G^*$ are compatible.
\end{lemma}
\proof $X+Y$ is closed and, since $(x,y)\mapsto(x,-y)$ is a
homeomorphism, the map $S:X\oplus Y\to X+Y$ defined by $S(x,y)=x+y$
is an open surjective morphism. Then from the Theorem 9.5, Chapter 2
of \cite{Gu} it follows that the adjoint map $S^*$ is a
homeomorphism between $(X+Y)^*$ and its range. In particular its
range is a locally compact subgroup for the topology induced by
$X^*\oplus Y^*$ hence is a closed subgroup of $X^*\oplus Y^*$,
see the footnote on page \pageref{footnote}.
We have $(X+Y)^\perp=X^\perp\cap Y^\perp$, cf. 23.29 in
\cite{HR}. Thus from $X^*\cong G^*/X^\perp$ and similar
representations for $Y^*$ and $(X+Y)^*$ we see that
$$
S^*:G^*/(X^\perp \cap Y^\perp)\to G^*/X^\perp\oplus G^*/Y^\perp
$$ is a closed map. But $S^*$ is clearly the natural map involved in
\eqref{eq:ma2}, hence the pair $X^\perp,Y^\perp$ is
regular. Finally, note that $(X\cap Y)^\perp$ is always equal to the
closure of the subgroup $X^\perp+Y^\perp$, cf. 23.29 and 24.10 in
\cite{HR}, and in our case $X^\perp+Y^\perp$ is closed.
\qed

\PAR\label{ss:hilbert} 
The lca group $X\uplus Y$ as defined in
\eqref{eq:boxplus} is a quotient of $X\oplus Y$ hence, according to
our general conventions, we have an embedding $\Cc(X\uplus Y)\subset
\Cbu(X\oplus Y)$. Then the elements $\theta\in\Cc(X\uplus Y)$ are
functions $\theta:X\times Y\to\mbc$ and we may think of them as
kernels of integral operators.

\begin{lemma}\label{lm:bound}
If $\theta\in\Cc(X\uplus Y)$ then $(T_\theta)(y)=\int_Y\theta(y,z)u(z)
\rmd z$ defines an operator in $\rl_{XY}$ with norm $\|T_\theta\|\leq
C\sup|\theta|$ where $C$ depends only on a compact which contains the
support of $\theta$.
\end{lemma}
\proof 
By the Schur test
$$
\|T_\theta\|^2\leq 
{\textstyle\sup_{y\in X}}\int_Y|\theta(y,z)\rmd  z \cdot
{\textstyle\sup_{z\in Y}}\int_X|\theta(y,z)\rmd y.
$$ 
Let $K\subset X$ and $L\subset Y$ be compact sets such that
$K\times L + D$ contains the support of $\theta$. Thus if
$\theta(y,z)\neq0$ then $y\in x+K$ and $z\in x+L$ for some $k\in K$
and $x\in X\cap Y$ hence $ \int_Y|\theta(y,z)\rmd z \leq
\sup|\theta| \lambda_Y(L).  $ Similarly $\int_X|\theta(y,z)\rmd y
\leq \sup|\theta| \lambda_X(K)$.  \qed

\begin{definition}\label{df:ryz}
$\rt_{XY}$ is the norm closure in $\rl_{XY}$ of the set of operators
$T_\theta$ as in Lemma \ref{lm:bound}.
\end{definition}

We give now an alternative definition of $\rt_{XY}$. If
$\varphi\in\Cc(G)$ we define $T_{XY}(\varphi):\Cc(Y)\to\Cc(X)$ by
\begin{equation}\label{eq:ryz}
(T_{XY}(\varphi)u)(x)=\int_Y\varphi(x-y)u(y)\rmd y.
\end{equation}
This operator depends only the restriction $\varphi|_{X+Y}$ hence,
by the Tietze extension theorem, we could take $\varphi\in\Cc(Z)$
instead of $\varphi\in\Cc(G)$, where $Z$ is any closed subgroup of
$G$ containing $X\cup Y$.

\begin{proposition}\label{pr:def2}
$T_{XY}(\varphi)$ extends to a bounded operator $L^2(Y)\to L^2(X)$,
also denoted $T_{XY}(\varphi)$, and for each compact $K\subset G$
there is a constant $C$ such that if $\supp\varphi\subset K$
\begin{equation}\label{eq:nyz}
\|T_{XY}(\varphi)\|\leq C \sup\nolimits_{x\in G}|\varphi(x)|.
\end{equation}
The adjoint operator is given by $T_{XY}(\varphi)^*=
T_{YX}(\varphi^*)$ where $\varphi^*(x)=\bar\varphi(-x)$.  The space
$\rt_{XY}$ coincides with the closure in $\rl_{XY}$ of the set of
operators of the from $T_{XY}(\varphi)$.
\end{proposition}
\proof The set $X+Y$ is closed in $G$ hence the restriction map
$\Cc(G)\to\Cc(X+Y)$ is surjective. On the other hand, the map
$\phi^\circ:X\uplus Y\to X+Y$, defined after \eqref{eq:nat}, is a
homeomorphism so it induces an isomorphism
$\varphi\to\varphi\circ\phi^\circ$ of $\Cc(X+Y)$ onto $\Cc(X\uplus
Y)$. Clearly $T_{XY}(\varphi)=T_\theta$ if $\theta=\varphi\circ\phi$,
so the proposition follows from Lemma \ref{lm:bound}.  
\qed

We discuss now some properties of the spaces $\rt_{XY}$.
We set $\rt_{XY}^*\equiv(\rt_{XY})^*\subset\rl_{YX}$.

\begin{proposition}\label{pr:nyza}
We have  $\rt_{XX}=\cc^*(X)$ and:
\begin{align}
& \rt_{XY}^* =\rt_{YX} \label{eq:rad} \\
& \rt_{XY}   =\rt_{XY}\cdot \cc^*(Y)=\cc^*(X)\cdot\rt_{XY}
\label{eq:cyzc}  \\
& \ca|_X\cdot\rt_{XY} =\rt_{XY}\cdot\ca|_Y \label{eq:ayza}
\end{align}
where $\ca$ is an arbitrary  $G$-algebra.
\end{proposition}
\proof The relations $\rt_{XX}=\cc^*(X)$ and \eqref{eq:rad} are
obvious.  Now we prove the first equality in \eqref{eq:cyzc} (then
the second one follows by taking adjoints). If $C(\eta)$ is the
operator of convolution in $L^2(Y)$ with $\eta\in \Cc(Y)$ then a
short computation gives
\begin{equation}\label{eq:yzc}
T_{XY}(\varphi)C(\eta)=T_{XY}(T_{G Y}(\varphi)\eta)
\end{equation}
for $\varphi\in\Cc(G)$. Since $T_{G Y}(\varphi)\eta\in\Cc(G)$ we get
$T_{XY}(\varphi)C(\eta)\in\rt_{GX}$, so $\rt_{XY}\cdot
\cc^*(Y)\subset\rt_{XY}$. The converse follows by a standard
approximation argument.

Let $\varphi\in\Cc(G)$ and $\theta\in\ca$. We shall denote by
$\theta(Q_X)$ the operator of multiplication by $\theta|_X$ in
$L^2(X)$ and by $\theta(Q_Y)$ that of multiplication by $\theta|_Y$
in $L^2(Y)$.  Choose some $\varepsilon>0$ and let $V$ be a compact
neighborhood of the origin in $G$ such that
$|\theta(z)-\theta(z')|<\varepsilon$ if $z-z'\in V$. There are
functions $\alpha_k\in\Cc(G)$ with $0\leq\alpha_k\leq1$ such that
$\sum_k\alpha_k=1$ on the support of $\varphi$ and
$\supp\alpha_k\subset z_k+V$ for some points $z_k$.  Below we shall
prove:
\begin{equation}\label{eq:byza}
\|T_{XY}(\varphi)\theta(Q_Y)- 
{\textstyle\sum_k}\theta(Q_X-z_k)T_{XY}(\varphi\alpha_k)\| 
\leq
\varepsilon\|T_{XY}(|\varphi|)\|.
\end{equation}
This implies $\rt_{XY}\cdot\ca|_Y\subset\ca|_X\cdot\rt_{XY}$. If we
take adjoints, use \eqref{eq:rad} and interchange $X$ and $Y$ in the
final relation, we obtain $\ca|_X\cdot\rt_{XY}=\rt_{XY}\cdot\ca|_Y$
hence the proposition is proved.  For $u\in\Cc(X)$ we have:
\begin{align*}
(T_{XY}(\varphi)\theta(Q_Y)u)(x) &=
\int_Y\varphi(x-y)\theta(y)u(y)\rmd y 
=\sum_k\int_Y\varphi(x-y)\alpha_k(x-y)\theta(y)u(y)\rmd y \\
&=
\sum_k\int_Y\varphi(x-y)\alpha_k(x-y)\theta(x-z_k)u(y)\rmd y
+(Ru)(x)\\ 
&= 
\sum_k\left(\theta(Q_X-z_k)T_{XY}(\varphi\alpha_k)u\right)(x) 
+(Ru)(x).
\end{align*}
We can estimate the remainder as follows
$$
|(Ru)(x)|=\left|\sum_k\int_Y\varphi(x-y)\alpha_k(x-y)
[\theta(y)-\theta(x-z_k)]u(y)\rmd y \right|\leq
\varepsilon\int_Y|\varphi(x-y)u(y)|\rmd y.
$$ 
because $x-z_k-y\in V$.  This proves \eqref{eq:byza}. 
\qed

\begin{proposition}\label{pr:ryz}
 $\rt_{XY}$ is a Hilbert $C^*$-submodule of $\rl_{XY}$ and
\begin{equation}\label{eq:hyz}
\rt_{XY}^*\cdot\rt_{XY}=\rc_Y(X\cap Y), \hspace{2mm}
\rt_{XY}\cdot\rt_{XY}^*=\rc_X(X\cap Y).
\end{equation}
Thus $\rt_{XY}$ is a $(\rc_X(X\cap Y),\rc_Y(X\cap Y))$-imprimitivity
bimodule.
\end{proposition}
\proof Due to \eqref{eq:rad}, to prove the first relation in
\eqref{eq:hyz} we have to compute the clspan $\rc$ of the operators
$T_{XY}(\varphi)T_{YX}(\psi)$ with $\varphi,\psi$ in $\Cc(G)$.  We
recall the notation $G^2=G\oplus G$, this is a locally compact
abelian group and $X^2=X\oplus X$ is a closed subgroup. Let us
choose functions $\varphi_k,\psi_k\in\Cc(G)$ and let
$\Phi=\sum_k\varphi_k\otimes\psi_k\in\Cc(G^2)$. If
$\psi_k^\dag(x)=\psi_k(-x)$, then $\sum_k
T_{XY}(\varphi_k)T_{YX}(\psi_k^\dag)$ is an integral operator on
$L^2(X)$ with kernel $\theta_X=\theta|_{X^2}$ where $\theta:G^2\to
\mbc$ is given by
$$ 
\theta(x,x')= \int_Y\Phi(x+y,x'+y)\rmd y.
$$ Since the set of decomposable functions is dense in $\Cc(G^2)$ in
the inductive limit topology, an easy approximation argument shows
that $\rc$ contains all integral operators with kernels of the same
form as $\theta_X$ but with arbitrary $\Phi\in\Cc(G^2)$.  Let
$Y^{(2)}$ be the closed subgroup of $G^2\equiv G\oplus G$ consisting
of the elements $(y,y)$ with $y\in Y$. Then $K=\supp\Phi\subset G^2$
is a compact, $\theta$ is zero outside $K+Y^{(2)}$, and
$\theta(a+b)=\theta(a)$ for all $a\in G^2,b\in Y^{(2)}$. Thus
$\theta\in\Cc(G^2/Y^{(2)})$, with the usual identification
$\Cc(G^2/Y^{(2)})\subset\Cbu(G^2)$. From Proposition 2.48 in
\cite{Fo} it follows that reciprocally, any function $\theta$ in
$\Cc(G^2/Y^{(2)})$ can be represented in terms of some $\Phi$ in
$\Cc(G^2)$ as above. Thus $\rc$ is the closure of the set of
integral operators on $L^2(X)$ with kernels of the form $\theta_X$
with $\theta\in\Cc(G^2/Y^{(2)})$. According to Lemma
\ref{lm:double}, the pair of subgroups $X^2,Y^{(2)}$ is regular, so
we may apply Lemma \ref{lm:reg} to get
$\Cc(G^2/Y^{(2)})|_{X^2}=\Cc(X^2/D)$ where $D=X^2\cap
Y^{(2)}=\{(x,x)\mid x\in X\cap Y\}$.  But by Lemma \ref{lm:sym} the
norm closure in $\rl_X$ of the set of integral operators with
kernel in $\Cc(X^2/D)$ is $\rc_X/(X\cap Y)$. This proves
\eqref{eq:hyz}.

It remains to prove that $\rt_{XY}$ is a Hilbert $C^*$-submodule of
$\rl_{XY}$, i.e. that we have
\begin{equation}\label{eq:hyz1}
\rt_{XY}\cdot\rt_{XY}^*\cdot\rt_{XY}=\rt_{XY}.
\end{equation}
The first identity in \eqref{eq:hyz} and \eqref{eq:cyzc} imply
\begin{equation*}
\rt_{XY}\cdot\rt_{XY}^*\cdot\rt_{XY} = 
\rt_{XY}\cdot \cc^*(Y)\cdot\cc_Y(X\cap Y)=
\rt_{XY}\cdot\cc_Y(X\cap Y).
\end{equation*}
From Lemma \ref{lm:reg} we get 
$$
\cc_Y(X\cap Y)=\cc_G(X\cap Y)|_Y =
\cc_G(X)|_Y \cdot \cc_G(Y)|_Y = \cc_G(X)|_Y 
$$ because $\cc_G(Y)|_Y=\mbc$.  Then by using Proposition
\ref{pr:nyza} we obtain
$$
\rt_{XY}\cdot\cc_Y(X\cap Y) = \rt_{XY}\cdot\cc_G(X)|_Y =
\cc_G(X)|_X \cdot\rt_{XY} =\rt_{XY}
$$
because $\cc_G(X)|_X=\mbc$. 
\qed

\begin{corollary}\label{co:txy}
We have
\begin{align}
\rt_{XY} &= \rt_{XY}\cc^*(Y)=\rt_{XY}\cc_Y(X\cap Y) 
\label{eq:txy1}
\\
&= \cc^*(X)\rt_{XY}=\cc_X(X\cap Y)\rt_{XY}.
\label{eq:txy2}
\end{align}
\end{corollary}
\proof If $\mr$ is a Hilbert $\ra$-module then $\mr=\mr\ra$ by
Proposition 2.31 in \cite{RW} for example, hence Proposition
\ref{pr:ryz} implies $\rt_{XY}=\rt_{XY}\rc_Y(X\cap Y)$. The space
$\rc_Y(X\cap Y)$ is a $\cc^*(Y)$-bimodule and $\rc_Y(X\cap
Y)=\rc_Y(X\cap Y)\cdot \cc^*(Y)$ by \eqref{eq:Cxy} hence we get
$\rc_Y(X\cap Y)=\rc_Y(X\cap Y)\cc^*(Y)$ by the Cohen-Hewitt theorem.
This proves the first equality in \eqref{eq:txy1} and the other ones
are proved similarly.
\qed

If $\cg$ is a set of closed subgroups of $G$ then the
\emph{semilattice generated by $\cg$} is the set of finite
intersections of elements of $\cg$.

\begin{proposition}\label{pr:product}
Let $X,Y,Z$ be closed subgroups of $G$ such that any two subgroups
from the semilattice generated by the family $\{X,Y,Z\}$ are
compatible. Then:
\begin{align}\label{eq:product}
\rt_{XZ}\cdot\rt_{ZY} &=
\rt_{XY}\cdot\cc_Y(Y\cap Z)= \cc_X(X\cap Z)\cdot\rt_{XY} \\
&= \rt_{XY}\cdot\cc_Y(X\cap Y\cap Z)= 
\cc_X(X\cap Y\cap Z)\cdot\rt_{XY}.
\end{align} 
In particular, if $Z\supset X\cap Y$ then
\begin{equation}\label{eq:factor}
\rt_{XZ}\cdot\rt_{ZY}=\rt_{XY}.
\end{equation}
\end{proposition}
\proof We first prove \eqref{eq:factor} in the particular case
$Z=G$. As in the proof of Proposition \ref{pr:ryz} we see that
$\rt_{XG}\cdot\rt_{G Y}$ is the the closure in $\rl_{XY}$ of the
set of integral operators with kernels 
$\theta_{XY}=\theta|_{X\times Y}$ where $\theta:G^2\to \mbc$ is
given by 
$$ 
\theta(x,y)= \int_G\sum_k\varphi_k(x-z)\psi_k(z-y)\rmd z=
\int_G\sum_k\varphi_k(x-y-z)\psi_k(z)\rmd z\equiv\xi(x-y)
$$ where $\varphi_k,\psi_k\in\Cc(G)$ and $\xi=\sum_k\varphi_k*\psi_k$
convolution product on $G$. Since $\Cc(G)*\Cc(G)$ is dense in
$\Cc(G)$ in the inductive limit topology, the space
$\rt_{XG}\cdot\rt_{G Y}$ is the the closure of the set of integral
operators with kernels $\theta(x,y)=\xi(x-y)$ with $\xi\in\Cc(G)$.
By Proposition \ref{pr:def2} this is  $\rt_{XY}$. 

Now we prove \eqref{eq:product}. From \eqref{eq:factor} with $Z=G$
and \eqref{eq:hyz} we get:
\begin{align*}
\rt_{XZ}\cdot\rt_{ZY} &=
\rt_{XG}\cdot\rt_{G Z}\cdot\rt_{ZG}\cdot\rt_{G Y} \\
&= \rt_{XG}\cdot\rt_{G Z}\cdot\rt_{Z G}\cdot\rt_{G Y} \\
&= \rt_{XG}\cdot\cc_G(Z)\cdot \cc^*(G)\cdot\rt_{GY}.
\end{align*}
Then from Proposition \eqref{pr:nyza} and Lemma \ref{lm:reg} we get:
$$
\cc_G(Z)\cdot \cc^*(G)\cdot\rt_{GY}=\cc_G(Z)\cdot\rt_{G Y}=
\rt_{G Y}\cdot\cc_G(Z)|_Y= \rt_{G Y}\cdot\cc_Y(Y\cap Z).
$$ 
We obtain \eqref{eq:product} by using once again
\eqref{eq:factor} with $Z=G$ and taking adjoints. On the other hand,
the relation $\rt_{XY}=\rt_{XY}\cdot\cc_Y(X\cap Y)$ holds because
of \eqref{eq:txy1}, so we have
$$
\rt_{XY}\cdot\cc_Y(Y\cap Z)=
\rt_{XY}\cdot\cc_Y(X\cap Y)\cdot\cc_Y(Y\cap Z)=
\rt_{XY}\cdot\cc_Y(X\cap Y\cap Z)
$$ where we also used \eqref{eq:reg1} and the fact that $X\cap Y$,
$Z\cap Y$ are compatible. Finally, to get \eqref{eq:factor} for
$Z\supset X\cap Y$ we use once again \eqref{eq:hyz}.  \qed

\begin{definition}\label{df:nxyz}
If $X,Y$ are compatible subgroups and $Z$ is a closed subgroup of
$X\cap Y$ then we set
\begin{equation}\label{eq:nxyz}
\rc_{XY}(Z)\equiv\rt_{XY}\cdot\cc_Y(Z)=\cc_X(Z)\cdot\rt_{XY}.
\end{equation} 
\end{definition}
The equality above follows from \eqref{eq:ayza} with $\ca=\cc_G(Z)$.
We clearly have $\rc_{XY}(X\cap Y)=\rt_{XY}$ and
$\rc_{XX}(Y)=\rc_X(Y)$ if $X\supset Y$. Moreover
\begin{equation}\label{eq:nadj}
\rc_{XY}^*(Z)\equiv\rc_{XY}(Z)^*=\rc_{YX}(Z)
\end{equation}
because of \eqref{eq:rad}.

\begin{theorem}\label{th:nxyz}
$\rc_{XY}(Z)$ is a Hilbert $C^*$-submodule of $\rl_{XY}$ such that
\begin{equation}\label{eq:nnz}
\rc_{XY}^*(Z)\cdot\rc_{XY}(Z)=\rc_Y(Z)
\hspace{2mm}\text{and}\hspace{2mm}
\rc_{XY}(Z)\cdot\rc_{XY}^*(Z)=\rc_X(Z).
\end{equation} 
In particular, $\rc_{XY}(Z)$ is a
$(\rc_X(Z),\rc_Y(Z))$-imprimitivity bimodule.
\end{theorem}
\proof 
By using \eqref{eq:nadj}, the definition \eqref{eq:nxyz}, and
\eqref{eq:reg1} we get
\begin{align*}
\rc_{XY}(Z)\cdot\rc_{YX}(Z) &=
\cc_X(Z)\cdot\rt_{XY}\cdot \rt_{YX}\cdot\cc_X(Z)\\
&=
\cc_X(Z)\cdot\cc_X(X\cap Y)\cdot \cc^*(X)\cdot\cc_X(Z)\\
&=
\cc_X(Z)\cdot \cc^*(X)\cdot\cc_X(Z)= \cc_X(Z)\cdot \cc^*(X)
\end{align*}
which proves the second equality in \eqref{eq:nnz}.  The first one
follows by interchanging $X$ and $Y$.
\qed

Below we give an intrinsic characterization of $\rc_{XY}(Z)$.  We
recall that for $k\in G^*$ the operator $V_k$ acts in $L^2(X)$ as
multiplication by $k|_X$ and in $L^2(Y)$ as multiplication by
$k|_Y$. Moreover, by Lemma \ref{lm:regp} and since $X,Y$ are
compatible, we have $(X\cap Y)^\perp=X^\perp+Y^\perp$ and the
natural map $X^\perp\oplus Y^\perp\to X^\perp+Y^\perp$ is an open
surjection. The orthogonals are taken relatively to $G$ unless
otherwise specified.

The following fact should be noted. Let $H,K,L$ be topological
spaces and let $\theta:H\to K$ be a continuous open surjection. If
$f:K\to L$ and $\theta(h_0)=k_0$ then $\lim_{k\to k_0}f(k)$ exists
if and only if $\lim_{h\to h_0}f(\theta(h))$ exists and then the
limits are equal. For example, in condition 2 of Theorem
\ref{th:yzintr} one may replace $G^*$ by $(X+Y)^*$ because the later
is a quotient of the first.

\begin{theorem}\label{th:yzintr}
$\rc_{XY}(Z)$ is the set of $T\in\rl_{XY}$ satisfying $U_z^*T U_z=T$
  if $z\in Z$ and such that
\begin{enumerate} \vspace{-2mm}
\item 
$\|(U_x-1)T\|\to 0$ if $x\to 0$ in $X$ and 
$\|T(U_y-1)\|\to 0$ if $y\to 0$ in $Y$, 
\item
$\|V^*_k T V_k-T\|\to 0$ if $k\to 0$ in $G^*$ and 
$\|(V_k-1)T\|\to 0$ if $k\to 0$ in $Z^\perp$.
\end{enumerate}
\end{theorem}

\begin{remark}\label{re:nxy1}
Observe that from condition 2 we also get $\|T(V_k-1)\|\to 0$ so we
may replace the second part of this condition by the apparently
stronger ``$\|(V_k-1)T^{(*)}\|\to 0$ if $k\to 0$ in $Z^\perp$''.
Most of the assumptions of Theorem \ref{th:yzintr} are decay
conditions in certain directions in $P$ or $Q$ space. Indeed, by
Lemma \ref{lm:help} condition 1 is equivalent to:
\begin{equation}\label{eq:cond1}
\text{there are } S_1\in \cc^*(X), S_2\in \cc^*(Y) \text{ and }
R_1,R_2\in\rl_{XY} \text{ such that } T=S_1R_1=R_2S_2.
\end{equation}
Recall that $\cc^*(X)\cong\Co(X^*)$ for example. Then the full
version $\|(V_k-1)T^{(*)}\|\to 0$ of the second part of condition 2
is equivalent to:
\begin{equation}\label{eq:cond2}
\text{there are } S_1\in \cc_X(Z), S_2\in \cc_Y(Z) \text{ and }
R_1,R_2\in\rl_{XY} \text{ such that } T=S_1R_1=R_2S_2.
\end{equation}
\end{remark}

\noindent{\bf Proof of Theorem \ref{th:yzintr}:} The set $\rc$ of all
the operators satisfying the conditions of the theorem is clearly a
closed subspace of $\rl_{XY}$.  We have $\rc_{X,Y}(Z)\subset\rc$
because \eqref{eq:cond1}, \eqref{eq:cond2} are satisfied by any
$T\in\rc_{XY}(Z)$ as a consequence of Theorem \ref{th:nxyz}.  Then we
get:
$$
\rc_Y(Z)= 
\rc_{XY}^*(Z)\cdot\rc_{XY}(Z)\subset \rc^*\cdot\rc, 
\hspace{1mm}
\rc_X(Z)=
\rc_{XY}(Z)\cdot\rc_{XY}^*(Z)\subset \rc\cdot\rc^*.
$$ 
We prove that equality holds in both these relations. We show, for
example, that $A\equiv TT^*$ belongs to $\rc_X(Z)$ if $T\in\rc$ and
for this we shall use Theorem \ref{th:cxy} with $Y$ replaced by
$Z$. That $U_z^*AU_z=A$ for $z\in Z$ is clear. From \eqref{eq:cond1}
we get $A=S_1R_1R_1^*S_1^*$ with $S_1\in\cc^*(X)$ hence
$\|(U_x-1)A\|\to 0$ and $\|A(U_x-1)\|\to 0$ as $x\to0$ in $X$ are
obvious and imply $\|U_x^*AU_x-A\|\to 0$. Then \eqref{eq:cond2}
implies $A=\psi(Q)C$ with $\psi\in\cc_X(Z)$ and bounded $C$ hence
\eqref{eq:cxy} is satisfied.

That $\rc\rc_Y(Z)\subset\rc$ is easily proven because $T=SA$ has the
properties \eqref{eq:cond1} and \eqref{eq:cond2} if $S$ belongs to
$\rc$ and $A$ to $\rc_Y(Z)$, cf. Theorem \ref{th:cxy}. From what we
have shown above we get $\rc\rc^*\rc\subset\rc\rc_Y(Z)\subset\rc$ so
$\rc$ is a Hilbert $C^*$-submodule of $\rl_{XY}$. On the other hand,
$\rc_{XY}(Z)$ is a Hilbert $C^*$-submodule of $\rl_{XY}$ such that
$\rc_{XY}^*(Z)\cdot\rc_{XY}(Z)=\rc^*\cdot\rc$ and
$\rc_{XY}(Z)\cdot\rc_{XY}^*(Z)=\rc\cdot\rc^*$. Since
$\rc_{XY}(Z)\subset\rc$ we get $\rc=\rc_{XY}(Z)$ from Proposition
\ref{pr:clsubmod}.  \qed

\begin{remark}\label{re:nxy2}
We shall make several more comments on the conditions of Theorem
\ref{th:yzintr}. All the convergences below are norm convergences.
First, it is clear that the condition 1 is equivalent to 
\begin{equation}\label{eq:cond1a}
U_x T U_y \to T \hspace{2mm} \text{if}\ (x,y)\to (0,0)
\ \text{in}\ X\oplus Y. 
\end{equation}
Let $Z^\perp_X$ be the orthogonal of $Z$ relatively to $X$, so that
$(X/Z)^*\cong Z^\perp_X \subset X^*$. We similarly have
$(Y/Z)^*\cong Z^\perp_Y \subset Y^*$. Then the condition
$(V_k-1)T^{(*)}\to 0$ if $k\to 0$ in $Z^\perp$ means
\begin{equation}\label{eq:cond2a}
\|(V_k-1)T\|\to 0 \hspace{2mm}\text{if}\ k\to 0 \ \text{in}\ (X/Z)^*
\quad\text{and}\quad 
\|T(V_k-1)\|\to 0 \ \text{if}\ k\to 0 \ \text{in}\ (Y/Z)^*
\end{equation}
which may also be written as
\begin{equation}\label{eq:cond2aa}
V_k T V_p\to T \hspace{2mm}\text{if}\hspace{2mm}
(k,p)\to (0,0) \ \text{in}\ (X/Z)^*\oplus(Y/Z)^*.
\end{equation}
Now we shall prove that condition 2 of Theorem \ref{th:yzintr} can
be re-expressed as follows:
\begin{equation}\label{eq:cond2b}
V_k T- T V_p\to 0 \ \text{if}\hspace{2mm} k\in X^*,\ p\in Y^*,\
k|_{X\cap Y}=p|_{X\cap Y},\
k|_{Z}=p|_{Z}=1,\ \text{and}\ (k,p)\to (0,0).
\end{equation}
For this we note that the map $\phi$ defined in \eqref{eq:nat}
induces an embedding $\phi^*(k)=(k|_X,\bar{k}|_Y)$ of $(X+Y)^*$ into 
$X^*\oplus Y^*$ whose range is the set of $(k,p)\in X^*\oplus Y^*$
such that $k|_{X\cap Y}=p|_{X\cap Y}$.

\end{remark}

If $Z=X\cap Y$ then Theorem \ref{th:yzintr} gives an intrinsic
description of the space $\rt_{XY}$. The case $X\supset Y$ is
particularly simple. 

\begin{corollary}\label{co:txyintr}
If $X\supset Y$ then $\rt_{XY}$ is the set of $T\in\rl_{XY}$
satisfying $U_y^*T U_y=T$    if $y\in Y$ and such that:
$U_xT\to T$ if $x\to 0$ in $X$, 
$V^*_k T V_k\to T$ if $k\to 0$ in $X^*$ and 
$V_kT\to T$ if $k\to 0$ in $Y^\perp$.
\end{corollary}

We say that \emph{$Z$ is complemented in $X$} if $X=Z\oplus E$ for
some closed subgroup $E$ of $X$. If $X,Z$ are equipped with Haar
measures then $X/Z$ is equipped with the quotient Haar measure and we
have $E\simeq X/Z$.  If $Z$ is complemented in $X$ and $Y$ then
$\rc_{XY}(Z)$ can be expressed as a tensor product.

\begin{proposition}\label{pr:def3}
If $Z$ is complemented in $X$ and $Y$ then
\begin{equation}\label{eq:ryzsum}
\rc_{XY}(Z)\simeq \cc^*(Z)\otimes \rk_{X/Z,Y/Z}.
\end{equation}
If $Y\subset X$ then $\rt_{XY}\simeq \cc^*(Y)\otimes L^2(X/Y)$ tensor
product of Hilbert $C^*$-modules.
\end{proposition} 
\proof Note first that the tensor product in \eqref{eq:ryzsum} is
interpreted as the exterior tensor product of the Hilbert
$C^*$-modules $\cc^*(Z)$ and $\rk_{X/Z,Y/Z}$. Let $X=Z\oplus E$ and
$Y=Z\oplus F$ for some closed subgroups $E,F$. Then, as explained in
\S\ref{ss:ha}, we may also view the tensor product as the norm closure
in the space of continuous operators from $L^2(Y)\simeq L^2(Z)\otimes
L^2(F)$ to $L^2(X)\simeq L^2(Z)\otimes L^2(E)$ of the linear space
generated by the operators of the form $T\otimes K$ with $T\in
\cc^*(Z)$ and $K\in \rk_{EF}$.

We now show that under the conditions of the proposition $X+Y\simeq
Z\oplus E\oplus F$ algebraically and topologically. The natural map
$\theta:Z\oplus E\oplus F\to Z+E+F=X+Y$ is a continuous bijective
morphism, we have to prove that it is open.  Since $X,Y$ are
compatible, the  map \eqref{eq:nat} is a continuous open
surjection.  If we represent $X\oplus Y\simeq Z\oplus Z\oplus E\oplus
F$ then this map becomes $\phi(a,b,c,d)=(a-b)+c+d$.  Let
$\psi=\xi\oplus{\rm id}_E\oplus{\rm id}_F$ where $\xi:Z\oplus Z\to Z$
is given by $\xi(a,b)=a-b$. Then $\xi$ is continuous surjective and
open because if $U$ is an open neighborhood of zero in $Z$ then
$U-U$ is also an open neighborhood of zero. Thus
$\psi:(Z\oplus Z)\oplus E\oplus F \to Z\oplus E\oplus F$ is a
continuous open surjection and $\phi=\theta\circ\psi$. So if $V$ is
open in $Z\oplus E\oplus F$ then there is an open 
$U\subset Z\oplus Z\oplus E\oplus F$ such that $V=\psi(U)$ and then
$\theta(V)=\theta\circ\psi(U)=\phi(U)$ is open in $Z+E+F$.

Thus we may identify $L^2(Y)\simeq L^2(Z)\otimes L^2(F)$ and
$L^2(X)\simeq L^2(Z)\otimes L^2(E)$ and we must describe the norm
closure of the set of operators $T_{XY}(\varphi)\psi(Q)$ with
$\varphi\in\Cc(X+Y)$ (cf. the remark after \eqref{eq:ryz} and the fact
that $X+Y$ is closed) and $\psi\in\Co(Y/Z)$. Since $X+Y\simeq Z\oplus
E\oplus F$ and $Y=Z\oplus F$ it suffices to describe the clspan of the
operators $T_{XY}(\varphi)\psi(Q)$ with
$\varphi=\varphi_Z\otimes\varphi_E\otimes\varphi_F$ and
$\varphi_Z,\varphi_E,\varphi_F$ continuous functions with compact
support on $Z,E,F$ respectively and $\psi=1\otimes\eta$ where $1$ is
the function identically equal to $1$ on $Z$ and
$\eta\in\Co(F)$. Then, if $x=(a,c)\in Z\times E$ and $y=(b,d)\in
Z\times F$, we get:
$$
(T_{XY}(\varphi)\psi(Q)u)(a,c)=\int_{Z\times F} \varphi_Z(a-b)
\varphi_E(c)
\varphi_F(d)  \eta(d) u(b,d) \rmd b \rmd d. 
$$ But this is just
$C(\varphi_Z)\otimes\ket{\varphi_E}\bra{\bar\eta\bar\varphi_F}$ where
$\ket{\varphi_E}\bra{\bar\eta\bar\varphi_F}$ is a rank one operator
$L^2(F)\to L^2(E)$ and $C(\varphi_Z)$ is the operator of convolution
by $\varphi_Z$ on $L^2(Z)$.  \qed

\section{Graded Hilbert $C^*$-modules}
\label{s:grad}
\protect\setcounter{equation}{0}

\PAR\label{ss:grca}
The natural framework for the systems considered
in this paper is that of $C^*$-algebras graded by semilattices. We
recall below their definition and a result which plays an important
role in our arguments.
Let $\cs$ be a \emph{semilattice}, i.e.\ $\cs$ is a set equipped
with an order relation $\leq$ such that the lower bound
$\sigma\wedge\tau$ of each couple of elements $\sigma,\tau$ exists.
We say that $\cs$ is \emph{atomic} if $\cs$ has a smallest element
$o\equiv\min \cs$ and if each $\sigma\neq o$ is minorated by an atom,
i.e. by some $\alpha\in\cs$ with $\alpha\neq o$ and such that $o\leq
\tau\leq\alpha\Rightarrow \tau=o \text{ or } \tau=\alpha$.  In this
case we denote by $\cp(\cs)$ the set of atoms of $\cs$.  

\begin{definition}\label{df:ga}
A $C^*$-algebra $\ra$ is called \emph{$\cs$-graded} if a linearly
independent family of $C^*$-subalgebras 
$\{\ra(\sigma)\}_{\sigma\in \cs}$ of $\ra$ has been given such that 
$\sum^\rmc_{\sigma\in \cs}\ra(\sigma)=\ra$ and
$\ra(\sigma)\ra(\tau)\subset\ra(\sigma\wedge\tau)$ for all
$\sigma,\tau$. The algebras $\ra(\sigma)$ are the \emph{components
  of $\ra$}.
\end{definition}

This notion has been introduced in \cite{BG1,DG1} but with the
supplementary assumption that the sum of a finite number of
$\ra(\sigma)$ be closed. That this condition is automatically
satisfied has been shown in \cite{Ma} where one may also find a
detailed study of this class of algebras.  The following has been
proved in \cite{DG1} (see also \cite[Sec. 3]{DG3}).  Let
$\ra_{\geq\sigma}\equiv\sum^\rmc_{\tau\geq\sigma}\ra(\tau)$, this is
clearly a $C^*$-subalgebra of $\ra$.

\begin{theorem}\label{th:ga}  
For each $\sigma\in \cs$ there is a unique linear continuous map
$\rp_{\geq\sigma}:\ra\rarrow\ra$ such that $\rp_{\geq\sigma}A=A$ if
$A\in\ra(\tau)$ for some $\tau\geq\sigma$ and $\rp_{\geq\sigma}A=0$
otherwise. The map $\rp_{\geq\sigma}$ is an idempotent morphism of the
algebra $\ra$ onto the subalgebra $\ra_{\geq\sigma}$. If
$\cs$ is atomic then $\rp A=(\rp_{\geq\alpha}A)_{\alpha\in\cp(\cs)}$
defines a morphism
$\rp:\ra\to\prod_{\alpha\in\cp(\cs)}\ra_{\geq\alpha}$ with $\ra(o)$
as kernel. This gives us a canonical embedding
\begin{equation}\label{eq:quot}
\ra/\ra(o)\subset\pprod_{\alpha\in\cp(\cs)}\ra_{\geq\alpha}.
\end{equation}
\end{theorem}

This result has important consequences in the spectral theory of the
operators of interest to us: it allows one to compute their
essential spectrum and to prove the Mourre estimate.  For the case
of finite $\cs$ this has been pointed out in \cite{BG1,BG2} (see
Theorems 3.1 and 4.4 in \cite{BG2} for example) and then extended to
the general case in \cite{DG1,DG2}. We shall recall here an
abstract version of the HVZ theorem which follows from
\eqref{eq:quot}.

We assume that $\cs$ is atomic so that $\ra$ comes equipped with a
remarkable ideal $\ra(o)$. Then for $A\in\ra$ we define its
\emph{essential spectrum} (relatively to $\ra(o)$) by the formula
\begin{equation}\label{eq:eso}
\spe(A)\equiv\sp(\rp A).
\end{equation}
In our concrete examples $\ra$ is represented on a Hilbert space
$\ch$ and $\ra(o)= K(\ch)$, so we get the usual Hilbertian notion of
essential spectrum. 

In order to extend this to unbounded operators it is convenient to
define an \emph{observable affiliated to $\ra$} as a morphism
$H:\Co(\mbr)\to\ra$.  We set $\varphi(H)\equiv H(\varphi)$.  If
$\ra$ is realized on $\ch$ then a self-adjoint operator on $\ch$
such that $(H+i)^{-1}\in\ra$ is said to be affiliated to $\ra$; then
$H(\varphi)=\varphi(H)$ defines an observable affiliated to $\ra$
(see Appendix A in \cite{DG3} for a precise description of the
relation between observables and self-adjoint operators affiliated
to $\ra$). The spectrum of an observable is by definition the
support of the morphism $H$:
\begin{equation}\label{eq:sp}
\sp(H)=\{\lambda\in\mbr \mid
\varphi\in\Co(\mbr),\varphi(\lambda)\neq 0 \Rightarrow
\varphi(H)\neq0\}. 
\end{equation}
Now note that $\rp H\equiv\rp\circ H$ is an observable affiliated to
the quotient algebra $\ra/\ra(o)$ so we may define the essential
spectrum of $H$ as the spectrum of $\rp H$. Explicitly, we get:
\begin{equation}\label{eq:es1}
\spe(H)=\{\lambda\in\mbr \mid 
\varphi\in\Co(\mbr),\varphi(\lambda)\neq 0 \Rightarrow
\varphi(H)\notin \ra(o)\}. 
\end{equation}
Now the first assertion of the next theorem follows immediately from
\ref{th:ga}. For the second assertion, see the proof of Theorem 2.10
in \cite{DG2}. By $\overline{\ccup}$ we denote the closure of the
union.

\begin{theorem}\label{th:gas}
Let $\cs$ be atomic. If $H$ is an observable affiliated to $\ra$
then $H_{\geq\alpha}=\rp_{\geq\alpha}H$ is an observable affiliated
to $\ra_{\geq\alpha}$ and we have:
\begin{equation}\label{eq:es2}
\spe(H)=\overline{\ccup}_{\alpha\in\cp(\cs)}\sp(H_{\geq\alpha}).
\end{equation}
If for each $A\in\ra$ the set of $\rp_{\geq\alpha}A$ with
$\alpha\in\cp(\cs)$ is compact in $\ra$ then the union in
\eqref{eq:es2} is closed.
\end{theorem}

\PAR\label{ss:ideal} A subset $\ct$ of a semilattice $\cs$ is called
a \emph{sub-semilattice} if $\sigma,\tau\in\ct \Rightarrow
\sigma\wedge\tau\in\ct$. We say that $\ct$ is an \emph{ideal of
  $\cs$} if $\sigma\leq\tau\in\ct \Rightarrow \sigma\in\ct$.  If
$\sigma\in \cs$ then we denote
\begin{equation}\label{eq:Aa}
\cs_{\geq\sigma}=\{\tau\in \cs\mid \tau\geq\sigma\},
\hspace{2mm}
\cs_{\leq\sigma}=\{\tau\in \cs\mid \tau\leq\sigma\}, 
\hspace{2mm}
\cs_{\not\geq\sigma}=\{\tau\in\cs\mid\tau\not\geq\sigma\}.
\end{equation}
Then $\cs_{\geq\sigma}$ is a sub-semilattice while the sets
$\cs_{\leq\sigma}$ and $\cs_{\not\geq\sigma}$ are ideals.  If $\ct$
is an ideal of $\cs$ and $\cs$ is atomic then $\ct$ is atomic, we
have $\min\ct=\min \cs$ and $\cp(\ct)=\cp(\cs)\cap\ct$.

An $\cs$-graded $C^*$-algebra $\ra$ is \emph{supported by a
sub-semilattice $\ct$} if $\ra(\sigma)=\{0\}$ for
$\sigma\nin\ct$. Then clearly $\ra$ is also $\ct$-graded.  The
smallest sub-semilattice with this property will be called
\emph{support of $\ra$}.  On the other hand, if $\ct$ is a
sub-semilattice of $\cs$ and $\ra$ is a $\ct$-graded algebra then
$\ra$ is canonically $\cs$-graded: we set $\ra(\sigma)=\{0\}$ for
$\sigma\in\cs\setminus\ct$.

For each $\ct\subset\cs$ let 
$\ra(\ct)=\sum^\rmc_{\sigma\in\ct}\ra(\sigma)$ 
(if $\ct$ is finite the sum is already closed).  If $\ct$ is a 
sub-semilattice then $\ra(\ct)$ is a $C^*$-subalgebra of $\ra$ and
if $\ct$ is an ideal then $\ra(\ct)$ is an ideal of $\ra$. 

Following \cite{Ma,Ma2} we say that $\rb\subset\ra$ is a \emph{graded
  $C^*$-subalgebra} if $\rb$ is a $C^*$-subalgebra of $\ra$ and it is
equal to the closure of $\sum_\sigma\rb\cap\ra(\sigma)$. Then $\rb$
has a natural structure of graded $C^*$-algebra:
$\rb(\sigma)=\rb\cap\ra(\sigma)$.  If $\rb$ is also an ideal of $\ra$
we shall say \emph{graded ideal}.  For example,
$\ra_{\geq\sigma}=\ra(\cs_{\geq\sigma})$ is a graded $C^*$-subalgebra
of $\ra$ supported by $\cs_{\geq\sigma}$ while $\ra(\cs_{\leq\sigma})$
and $\ra(\cs_{\not\geq\sigma})$ are graded ideals supported by
$\cs_{\leq\sigma}$ and $\cs_{\not\geq\sigma}$ respectively.

\PAR\label{ss:gf} The notion of graded Hilbert $C^*$-module that we
use is due to George Skandalis \cite{Sk}.

\begin{definition}\label{df:grm}
Let $\cs$ be a semilattice and $\ra$ an $\cs$-graded
$C^*$-algebra. A Hilbert $\ra$-module $\mr$ is an \emph{$\cs$-graded
  Hilbert $\ra$-module} if a linearly independent family
$\{\mr(\sigma)\}_{\sigma\in \cs}$ of closed subspaces of $\mr$ is
given such that $\sum_\sigma\mr(\sigma)$ is dense in $\mr$ and:
\begin{equation}\label{eq:grm}
\mr(\sigma)\ra(\tau)\subset\mr(\sigma\wedge\tau) 
\hspace{2mm}\text{and}\hspace{2mm}
\braket{\mr(\sigma)}{\mr(\tau)}\subset\ra(\sigma\wedge\tau)
\hspace{2mm} \text{for all } \sigma,\tau\in \cs.
\end{equation}
\end{definition}
Observe that $\ra$ equipped with its canonical Hilbert $\ra$-module
structure is an $\cs$-graded Hilbert $\ra$-module. Note that from
\eqref{eq:grm} it follows that each $\mr(\sigma)$ is a Hilbert
$\ra(\sigma)$-module and if $\sigma\leq\tau$ then $\mr(\sigma)$ is an
$\ra(\tau)$-module.

From \eqref{eq:grm} and the discussion in \S\ref{ss:not} we see that
\emph{the imprimitivity algebra $\ck(\mr(\sigma))$ of the Hilbert
  $\ra(\sigma)$-module $\mr(\sigma)$ is naturally identified with
  the clspan in $\ck(\mr)$ of the elements $MM^*$ with
  $M\in\mr(\sigma)$}. Thus $\ck(\mr(\sigma))$ is identified with a
$C^*$-subalgebra of $\ck(\mr)$. We use this identification below.

\begin{theorem}\label{th:kghm}
If $\mr$ is a graded Hilbert $\ra$-module then $\ck(\mr)$ becomes a
graded $C^*$-algebra if we define
$\ck(\mr)(\sigma)=\ck(\mr(\sigma))$. If $M\in\mr(\sigma)$ and
$N\in\mr(\tau)$ then there are elements $M'$ and $N'$ in
$\mr(\sigma\wedge\tau)$ such that $MN^*=M'N'^*$;
in particular $MN^*\in\ck(\mr)(\sigma\wedge\tau)$.
\end{theorem}
\proof As explained before, $\ck(\mr)(\sigma)$ are $C^*$-subalgebras
of $\ck(\mr)$. To show that they are linearly independent, let
$T(\sigma)\in\ck(\mr)(\sigma)$ such that $T(\sigma)=0$ but for a
finite number of $\sigma$ and assume $\sum_\sigma T(\sigma)=0$. Then
for each $M\in\mr$ we have $\sum_\sigma T(\sigma)M=0$. Note that the
range of $T(\sigma)$ is included in $\mr(\sigma)$. Since the linear
spaces $\mr(\sigma)$ are linearly independent we get $T(\sigma)M=0$
for all $\sigma$ and $M$ hence $T(\sigma)=0$ for all $\sigma$.

We now prove the second assertion of the proposition. Since
$\mr(\sigma)$ is a Hilbert $\ra(\sigma)$-module there are
$M_1\in\mr(\sigma)$ and $S\in\ra(\sigma)$ such that $M=M_1S$, cf. the
Cohen-Hewitt theorem or Lemma 4.4 in \cite{La}. Similarly, $N=N_1T$
with $N_1\in\mr(\tau)$ and $T\in\ra(\tau)$. Then $MN^*=M_1(S
T^*)N_1^*$ and $S T^*\in \ra(\sigma\wedge\tau)$ so we may factorize it
as $S T^*=UV^*$ with $U,V\in \ra(\sigma\wedge\tau)$, hence
$MN^*=(M_1U)(N_1V)^*$.  By using \eqref{eq:grm} we see that $M'=M_1U$
and $N'=N_1V$ belong to $\mr(\sigma\wedge\tau)$.  In particular, we
have $MN^*\in\ck(\mr)(\sigma\wedge\tau)$ if $M\in\mr(\sigma)$ and
$N\in\mr(\tau)$.

Observe that the assertion we just proved implies that
$\sum_\sigma\ck(\mr)(\sigma)$ is dense in $\ck(\mr)$.  
It remains to see that
$\ck(\mr)(\sigma)\ck(\mr)(\tau)\subset\ck(\mr)(\sigma\wedge\tau)$.
For this it suffices that $M\braket{M}{N}N^*$ be in
$\ck(\mr)(\sigma\wedge\tau)$ if $M\in\mr(\sigma)$ and
$N\in\mr(\tau)$. Since $\braket{M}{N}\in\ra(\sigma\wedge\tau)$ we
may write $\braket{M}{N}=S T^*$ with $S,T\in\ra(\sigma\wedge\tau)$
so $M\braket{M}{N}N^*=(MS)(NT)^*\in\ck(\mr)(\sigma\wedge\tau)$ by
\eqref{eq:grm}. 
\qed

We recall that the direct sum of a family $\{\rM_i\}$ of Hilbert
$\ra$-modules is defined as follows: $\oplus_i\rM_i$ is the space of
elements $(M_i)_i\in\prod_i\rM_i$ such that the series
$\sum_i\braket{M_i}{M_i}$ converges in $\ra$ equipped with the natural
$\ra$-module structure and with the $\ra$-valued inner product defined
by
\begin{equation}\label{eq:sum}
\braket{(M_i)_i}{(N_i)_i} 
=\textstyle\sum_i\braket{M_i}{N_i}.
\end{equation} 
The algebraic direct sum of the $\ra$-modules $\rM_i$ is dense in
$\oplus_i\rM_i$.

It is easy to check that if each $\mr_i$ is graded and if we set
$\mr(\sigma)=\oplus_i\mr_i(\sigma)$ then $\mr$ becomes a graded
Hilbert $\ra$-module.  For example, if $\rn$ is a graded Hilbert
$\ra$-module then $\rn\oplus\ra$ is a graded Hilbert $\ra$-module and
so the \emph{linking algebra $\ck(\rn\oplus\ra)$ is equipped with a
graded algebra structure}. We recall \cite[p. 50-52]{RW} that we
have a natural identification
\begin{equation}\label{eq:link}
\ck(\rn\oplus\ra)=
\begin{pmatrix}
\ck(\rn)& \rn\\
\rn^*& \ra
\end{pmatrix}
\end{equation}
and by Theorem \ref{th:kghm} this is a graded algebra whose
$\sigma$-component is equal to
\begin{equation}\label{eq:links}
\ck(\rn(\sigma)\oplus\ra(\sigma))=
\begin{pmatrix}
\ck(\rn(\sigma))& \rn(\sigma)\\
\rn(\sigma)^*& \ra(\sigma)
\end{pmatrix}.
\end{equation}
If $\rn$ is a $C^*$-submodule of $L(\ce,\cf)$ and if we set
$\rn^*\cdot\rn=\ra,\rn\cdot\rn^*=\rb$ then the linking algebra
$\begin{pmatrix}\rb& \mr\\\mr^*& \ra\end{pmatrix}$ of $\mr$ is a
$C^*$-algebra of operators on $\cf\oplus\ce$.

Some of the graded Hilbert $C^*$-modules which we shall use later on
will be constructed as follows.

\begin{proposition}\label{pr:rhm}
Let $\ce,\cf$ be Hilbert spaces and let $\mr\subset L(\ce,\cf)$ be a
Hilbert $C^*$-submodule, so that $\ra\equiv\mr^*\cdot\mr\subset
L(\ce)$ is a $C^*$-algebra and $\mr$ is a full Hilbert
$\ra$-module. Let $\cc$ be a $C^*$-algebra of operators on $\ce$
graded by the family of $C^*$-subalgebras
$\{\cc(\sigma)\}_{\sigma\in\cs}$. Assume that we have
\begin{equation}\label{eq:tas}
\ra\cdot\cc(\sigma)=\cc(\sigma)\cdot\ra
\equiv\rc(\sigma)
\hspace{2mm} \text{for all } \sigma\in\cs
\end{equation}
and that the family $\{\rc(\sigma)\}$ of subspaces of $L(\cf)$ is
linearly independent. Then the $\rc(\sigma)$ are $C^*$-algebras of
operators on $\ce$ and $\rc=\sum^\rmc_\sigma\rc(\sigma)$ is a
$C^*$-algebra graded by the family $\{\rc(\sigma)\}$. If
$\rn(\sigma)\equiv\mr\cdot\cc(\sigma)$ then
$\rn=\sum^\rmc_\sigma\rn(\sigma)$ is a full Hilbert $\rc$-module
graded by $\{\rn(\sigma)\}$.
\end{proposition}
\proof
We have
$$
\rc(\sigma)\cdot\rc(\tau)=\ra\cdot\cc(\sigma)\cdot\ra\cdot\cc(\tau)
=\ra\cdot\ra\cdot\cc(\sigma)\cdot\cc(\tau)\subset
\ra\cdot\cc(\sigma\wedge\tau)=\rc(\sigma\wedge\tau).
$$ 
This proves that the $\rc(\sigma)$ are $C^*$-algebras and that
$\rc$ is $\cs$-graded. Then:
$$
\rn(\sigma)\cdot\rc(\tau)=\mr\cdot\cc(\sigma)\cdot\cc(\tau)\cdot\ra
\subset\mr\cdot\cc(\sigma\wedge\tau)\cdot\ra=
\mr\cdot\ra\cdot\cc(\sigma\wedge\tau)=
\mr\cdot\cc(\sigma\wedge\tau)=\rn(\sigma\wedge\tau)
$$
and
$$
\rn(\sigma)^*\cdot\rn(\tau)=
\cc(\sigma)\cdot\mr^*\cdot\mr\cdot\cc(\tau)=
\cc(\sigma)\cdot\ra\cdot\cc(\tau)=
\ra\cdot\cc(\sigma)\cdot\cc(\tau)\subset
\ra\cdot\cc(\sigma\wedge\tau)=\rc(\sigma\wedge\tau).
$$
Observe that this computation also gives
$\rn(\sigma)^*\cdot\rn(\sigma)=\rc(\sigma)$.  Then
$$
\left(\sum\nolimits_\sigma\rn(\sigma)^*\right)
\left(\sum\nolimits_\sigma\rn(\sigma)\right)=
\sum\nolimits_{\sigma,\tau}\rn(\sigma)^*\rn(\tau)\subset
\sum\nolimits_{\sigma,\tau}\rc(\sigma\wedge\tau)\subset
\sum\nolimits_{\sigma}\rc(\sigma)
$$ and by the preceding remark we get $\rn^*\cdot\rn=\rc$ so $\rn$
is a full Hilbert $\rc$-module. To show the grading property it
suffices to prove that the family of subspaces $\rn(\sigma)$ is
linearly independent. Assume that $\sum N(\sigma)=0$ with
$N(\sigma)\in\rn(\sigma)$ and $N(\sigma)=0$ for all but a finite
number of $\sigma$. Assuming that there are non-zero elements in
this sum, let $\tau$ be a maximal element of the set of $\sigma$
such that $N(\sigma)\neq0$. From
$\sum_{\sigma_1,\sigma_2}N(\sigma_1)^*N(\sigma_2)=0$ and since
$N(\sigma_1)^*N(\sigma_2)\in\rc(\sigma_1\wedge\sigma_2)$ we get
$\sum_{\sigma_1\wedge\sigma_2=\sigma}N(\sigma_1)^*N(\sigma_2)=0$ for
each $\sigma$. Take here $\sigma=\tau$ and observe that if
$\sigma_1\wedge\sigma_2=\tau$ and $\sigma_1>\tau$ or $\sigma_2>\tau$
then $N(\sigma_1)^*N(\sigma_2)=0$. Thus $N(\tau)^*N(\tau)=0$ so
$N(\tau)=0$. But this contradicts the choice of $\tau$, so
$N(\sigma)=0$ for all $\sigma$.  \qed

\section{Graded $C^*$-algebras associated to semilattices of groups}
\label{s:grass}
\protect\setcounter{equation}{0}

In this section we construct $C^*$-algebras graded by semilattices of
the following type.

\begin{definition}\label{df:iscg}
An \emph{inductive semilattice $\cs$ of compatible lca groups} is a
set $\cs$ of lca groups (equipped with Haar measures) such that for
all $X,Y\in\cs$ the following three conditions are satisfied:
\begin{compactenum}
\item[(i)]
if $X\supset Y$ then the topology and the group structure of $Y$
coincide with those induced by $X$,
\item[(ii)]
$X\cap Y \in \cs$,
\item[(iii)]
there is $Z\in\cs$ such that $X,Y$ are compatible subgroups of $Z$. 
\end{compactenum}
\end{definition}

According to the Remark \ref{re:reg2}, if all $X\in\cs$ are
$\sigma$-compact then the condition (iii) is equivalent to:
\begin{compactenum}
\item[(iii$'$)]
there is $Z\in\cs$ with $X\cup Y\subset Z$ such that
the subgroup of $Z$ generated by $X\cup Y$ in $Z$ be closed.
\end{compactenum}

One may realize $\cs$ as a set of subgroups of the inductive limit
group $\cx=\lim_{X\in\cs}X$ equipped with the final topology defined
by the embeddings $X\hookrightarrow\cx$ but note that this is not a
group topology in general.  

In our main result we shall have to assume that $\cs$ satisfies one
more condition: 

\begin{definition}\label{df:ncq}
We say that \emph{$\cs$ has non-compact quotients} if:
$X\supsetneq Y \Rightarrow X/Y$ is not compact.
\end{definition}

The following notations are convenient. Since each $X\in\cs$ comes
with a Haar measure the Hilbert spaces
\begin{equation}\label{eq:hx}
\ch(X)\equiv L^2(X)
\end{equation}
are well defined. If $Y\subset X$ are groups in $\cs$ then their
quotient $X/Y$ is equipped with the quotient measure so $\ch(X/Y)=
L^2(X/Y)$ is also well defined.

We make now some comments in connection with the preceding conditions
and then give examples.

\begin{remarks}\label{re:cpt}{\rm
Since a subgroup of a locally compact group is closed if
and only if it is locally compact for the induced topology, condition
(i) can be restated as: if $X\supset Y$ then $Y$ is a closed subgroup
of $X$ equipped with the induced lca group structure. In particular,
$X/Y$ will then be a lca group hence Definition \ref{df:ncq} makes
sense.  By condition (ii) the set $X\cap Y$ is equipped with a lca
group structure. But $X\cap Y\subset X$ hence by using (i) we see that
$X\cap Y$ is a closed subgroup of $X$ and its lca group structure
coincides with that induced by $X$. Of course, we may replace here $X$
by $Y$.  If $Z$ is a lca group which contains $X,Y$ as closed
subgroups then the subgroup $X+Y$ of $Z$ generated by $X\cup Y$ is
closed and the map \eqref{eq:nat} is open. If the condition (iii) is
fulfilled by some $Z$ then it will hold for an arbitrary $Z\in\cs$
containing $X\cup Y$.  Indeed, if $Z'\in\cs$ is such that $X\cup
Y\subset Z'$ then $Z\cap Z'$ is a closed subgroup of $Z$ and of $Z'$
equipped with the induced lca group structure and so we get the same
topological group $X+Y$ if we use $Z$, $Z\cap Z'$, or $Z'$ for its
definition.  }\end{remarks}

\begin{remark}\label{re:fmax}
If $\ct$ is a finite part of $\cs$ then there is $X\in\cs$ such that
$Y\subset X$ for all $Y\in\ct$. This follows by induction from
condition (iii). Moreover, if $\cs$ has a maximal element $X$, then
$X$ is the largest element of $\cs$. Thus, \emph{if $\cs$ is finite
  then there is a largest element $\cx$ in $\cs$ and $\cs$ is a set of
  closed subgroups of $\cx$}.
\end{remark}

\begin{remark}\label{re:haar}
The $C^*$-algebras that we construct depend on the choice of Haar
measures $\lambda_X$ (or simply $\rmd x$ when there is no ambiguity)
on the groups $X\in\cs$ but different choices lead to isomorphic
algebras. Note that if an open relatively compact neighborhood
$\Omega$ of zero is given on some $X$ then one can fix the Haar
measure of the subgroups $Y\subset X$ by requiring
$\lambda_Y(\Omega\cap Y)=1$.
\end{remark}

\begin{example}
The simplest and most important example one should have in mind is
the following: $\cx$ is a $\sigma$-compact lca group and $\cs$ is a
set of closed subgroups of $\cx$ with $\cx\in\cs$ and such that: if
$X,Y\in\cs$ then $X\cap Y\in\cs$, $X+Y$ is closed, and $X/Y$ is not
compact if $X\supsetneq Y$.
\end{example}

\begin{example}\label{ex:vect}
One may take $\cs$ \emph{equal to the set of all finite dimensional
  vector subspaces of a vector space over an infinite locally
  compact field} (such a field is not compact): this is the main
example in the context of the many-body problem.  Of course,
subgroups which are not vector subspaces may be considered. We
recall (see Theorem 9.11 in \cite{HR}) that the closed additive
subgroups of a finite dimensional real vector space $X$ are of the
form $Y=E+L$ where $E$ is a vector subspace of $X$ and $L$ is a
lattice in a vector subspace $F$ of $X$ such that $E\cap F=\{0\}$.
More precisely, $L=\sum_k\mbz f_k$ where $\{f_k\}$ is a basis in
$F$. Thus $F/L$ is a torus and if $G$ is a third vector subspace
such that $X=E\oplus F\oplus G$ then the space $X/Y\simeq
(F/L)\oplus G$ is a cylinder with $F/L$ as basis.
\end{example}

\begin{example}\label{ex:hilbert}
This is a version of the preceding example and is the natural
framework for the nonrelativistic many-body problem. Let $\cx$ be a
real prehilbert space and let $\cs$ be a set of finite dimensional
subspaces of $\cx$ such that if $X,Y\in\cs$ then $X\cap Y\in\cs$ and
$X+Y$ is included in some subspace of $\cs$ (there is a canonical
choice, namely the set of all finite dimensional subspaces of
$\cx$). Then each $X\in\cs$ is an Euclidean space and so is equipped
with a canonical Haar measure and there is a canonical self-adjoint
operator in $\ch(X)$, the (positive) Laplacian $\Delta_X$ associated
to the Euclidean structure.
\end{example}

In what follows we fix $\cs$ as in Definition \ref{df:iscg}.  For each
$X\in\cs$ let $\cs(X)$ be the set of $Y\in\cs$ such that $Y\subset
X$. Then by Lemma \ref{lm:reg} the space
\begin{equation}\label{eq:sax}
\cc_X \equiv {\textstyle\sum^\rmc_{Y\in\cs(X)}}\cc_X(Y)
\end{equation}
is an $X$-algebra so $\cc_X\rtimes X$ is well defined and we clearly
have
\begin{equation}\label{eq:crsax}
\rc_X\equiv\cc_X\rtimes X =
{\textstyle\sum^\rmc_{Y\in\cs(X)}}\rc_X(Y).
\end{equation}
For each pair $X,Y\in\cs$ with $X\supset Y$ we set 
\begin{equation}\label{eq:saX}
\cc^Y_X\equiv 
{\textstyle\sum^\rmc_{Z\in\cs(Y)}}\cc_X(Z).
\end{equation}
This is also an $X$-algebra so we may define $\rc^Y_X=\cc^Y_X\rtimes
X$ and we have
\begin{equation}\label{eq:saX1}
\rc^Y_X\equiv\cc^Y_X\rtimes X=
{\textstyle\sum^\rmc_{Z\in\cs(Y)}}\rc_X(Z).
\end{equation}
If $X=Y\oplus Z$ then $\cc_X^Y\simeq\cc_Y\otimes 1$ and 
$\rc_X^Y\simeq\rc_Y\otimes \cc^*(Z)$. 

\begin{lemma}\label{lm:xyprod}
Let $X\in\cs$ and $Y\in\cs(X)$. Then
\begin{equation}\label{eq:xy1}
\cc_X^{Y}=\cc_X(Y)\cdot\cc_X \text{ and }
\rc_X^{Y}=\cc_X(Y)\cdot\rc_X=\rc_X\cdot\cc_X(Y).
\end{equation}
Moreover, if $Z\in\cs(X)$ then
\begin{equation}\label{eq:xy2}
\cc_X^Y\cdot\cc_X^Z=\cc_X^{Y\cap Z} \text{ and }
\rc_X^Y\cdot\rc_X^Z=\rc_X^{Y\cap Z}.
\end{equation}
\end{lemma}
\proof The abelian case is a consequence of \eqref{eq:reg1} and a
straightforward computation. For the crossed product algebras we use
$\cc_X(Y)\cdot\rc_X=\cc_X(Y)\cdot\cc_X\cdot \cc^*(X)$ and the first
relation in \eqref{eq:xy1} for example.  \qed

\begin{lemma}\label{lm:nxy}
For arbitrary $X,Y\in\cs$ we have
\begin{equation}\label{eq:main}
\cc_X\cdot\rt_{XY}=\rt_{XY}\cdot\cc_Y
=\rt_{XY}\cdot\cc_Y^{X\cap Y}=
\cc_X^{X\cap Y}\cdot\rt_{XY}.
\end{equation}
\end{lemma}
\proof
If $G\in\cs$ contains $X\cup Y$ then clearly
$$
\cc_X\cdot\rt_{XY}=
{\textstyle\sum^\rmc_{Z\in\cs(X)} }\cc_X(Z)\cdot\rt_{XY}=
{\textstyle\sum^\rmc_{Z\in\cs(X)} }\cc_G(Z)|_X\cdot\rt_{XY}.
$$
From \eqref{eq:ayza} and \eqref{eq:reg2} we get
$$
\cc_G(Z)|_X\cdot\rt_{XY}=\rt_{XY}\cdot\cc_Y(Y\cap Z).
$$ 
Since $Y\cap Z$ runs over $\cs(X\cap Y)$ when $Z$ runs over
$\cs(X)$ we obtain $\cc_X\cdot\rt_{XY}=\rt_{XY}\cdot\cc_Y^{X\cap Y}$.
Similarly $\rt_{XY}\cdot\cc_Y=\cc_X^{X\cap Y}\cdot\rt_{XY}$. 
On the other hand $\cc_X^{X\cap Y}=\cc_G^{X\cap Y}|_X$ and similarly
with $X,Y$ interchanged, hence 
$\cc_X^{X\cap Y}\cdot\rt_{XY}=\rt_{XY}\cdot\cc_Y^{X\cap Y}$
because of \eqref{eq:ayza}. 
\qed

\begin{definition}\label{df:main}
If $X,Y\in\cs$ then
$\rc_{XY}\equiv\rt_{XY}\cdot\cc_Y=\cc_X\cdot\rt_{XY}$.
In particular $\rc_{XX}=\rc_X$. 
\end{definition}

The $C^*$-algebra $\rc_X$ is realized on the Hilbert space $\ch(X)$
and we think of it as the algebra of energy observables of a system
with $X$ as configuration space. For $X\neq Y$ the space $\rc_{XY}$
is a closed linear space of operators $\ch(Y)\to \ch(X)$ canonically
associated to the semilattice of groups $\cs(X\cap Y)$. We call
these $\rc_{XY}$ \emph{coupling spaces} because they will determine
the way the systems corresponding to $X$ and $Y$ are allowed to
interact.

\begin{proposition}\label{pr:mxyz}
Let $X,Y,Z\in\cs$. Then $\rc_{XY}^*=\rc_{YX}$ and
\begin{equation}\label{eq:mxyz}
\rc_{XZ}\cdot\rc_{ZY}=\rc_{XY}\cdot\cc_Y^{X\cap Y\cap Z}=
\cc_X^{X\cap Y\cap Z}\cdot\rc_{XY} \subset \rc_{XY}. 
\end{equation}
In particular $\rc_{XZ}\cdot\rc_{ZY}=\rc_{XY}$ if $Z\supset X\cap Y$.
\end{proposition}
\proof 
The first assertion follows from \eqref{eq:rad}.  From the
Definition \ref{df:main} and Proposition \ref{pr:product} we then get
\begin{align*}
\rc_{XZ}\cdot\rc_{ZY} &=
\cc_X\cdot\rt_{XZ}\cdot\rt_{ZY}\cdot\cc_Y =
\cc_X\cdot\rt_{XY}\cdot\cc_Y(X\cap Y\cap Z)\cdot\cc_Y \\ &=
\rt_{XY}\cdot\cc_Y\cdot\cc_Y(X\cap Y\cap Z)\cdot\cc_Y =
\rt_{XY}\cdot\cc_Y(X\cap Y\cap Z)\cdot\cc_Y.
\end{align*}
But $\cc_Y(X\cap Y\cap Z)\cdot\cc_Y=\cc_Y^{X\cap Y\cap Z}$ by Lemma
\ref{lm:xyprod}. For the last inclusion in \eqref{eq:mxyz} we use
the obvious relation $\cc_Y^{X\cap Y\cap Z}\cdot\cc_Y\subset\cc_Y$.
The last assertion of the proposition follows from \eqref{eq:main}. 
\qed

The following theorem is a consequence of the results obtained so
far.

\begin{theorem}\label{th:nmod}
$\rc_{XY}$ is a Hilbert $C^*$-submodule of $\rl_{XY}$ such that 
\begin{equation}\label{eq:nmod}
\rc_{XY}^*\cdot\rc_{XY}=\rc_Y^{X\cap Y} \text{ and }
\rc_{XY}\cdot\rc_{XY}^*=\rc_X^{X\cap Y}.
\end{equation}
In particular, $\rc_{XY}$ is a
$(\rc_X^{X\cap Y},\rc_Y^{X\cap Y})$-imprimitivity bimodule. 
\end{theorem}

If $X\cap Y$ is complemented in $X$ and $Y$ then $\rc_{XY}$ can be
expressed (non canonically) as a tensor product.

\begin{proposition}\label{pr:xytens}
If $X\cap Y$ is complemented in $X$ and $Y$ then
\[
\rc_{XY}\simeq \rc_{X\cap Y} \otimes \rk_{X/(X\cap Y),Y/(X\cap Y)}.
\] 
In particular,
if $X\supset Y$ then $\rc_{XY}\simeq \rc_{Y} \otimes \ch(X/Y)$.
\end{proposition}
\proof If $X=(X\cap Y)\oplus E$ and $Y=(X\cap Y)\oplus F$ then we have
to show that $\rc_{XY}\simeq \rc_{X\cap Y} \otimes \rk_{EF}$ where the
tensor product may be interpreted either as the exterior tensor
product of the Hilbert $C^*$-modules $\rc_{X\cap Y}$ and $\rk_{EF}$ or
as the norm closure in the space of continuous operators from
$L^2(Y)\simeq L^2(X\cap Y)\otimes L^2(F)$ to $L^2(X)\simeq L^2(X\cap
Y)\otimes L^2(E)$ of the algebraic tensor product of $\rc_{X\cap Y}$
and $\rk_{EF}$.  From Proposition \ref{pr:def3} with $Z=X\cap Y$ we
get $\rt_{XY}\simeq\cc^*(X\cap Y)\otimes\rk_{EF}$. The relations
\eqref{eq:main} and the Definition \ref{df:main} imply
$\rc_{XY}=\rt_{XY}\cdot\cc_Y^{X\cap Y}$ and we clearly have
\[
\cc_Y^{X\cap Y}={\textstyle\sum_{Z\in\cs(X\cap Y)}^\rmc} \cc_Y(Z)
\simeq {\textstyle\sum_{Z\in\cs(X\cap Y)}^\rmc} 
\cc_{X\cap Y}(Z)\otimes \Co(F)
\simeq \cc_{X\cap Y}\otimes \Co(F).
\]     
Then we get
\[
\rc_{XY}\simeq \cc^*(X\cap Y)\otimes\rk_{EF}\cdot 
\cc_{X\cap Y}\otimes \Co(F)=
\big(\cc^*(X\cap Y)\cdot\cc_{X\cap Y} \big)\otimes
\big(\rk_{EF}\cdot\Co(F)\big)
\]
and this is $\rc_{X\cap Y} \otimes \rk_{EF}$.
\qed

From now on \emph{we suppose that $\cs$ has non-compact quotients}.

\begin{theorem}\label{th:grsax}
The $C^*$-algebras $\cc_X$ and $\rc_X$ are $\cs(X)$-graded by 
the decompositions \eqref{eq:sax} and \eqref{eq:crsax}. 
\end{theorem}

This is a particular case of results due to A. Mageira
\cite[Propositions 6.1.2, 6.1.3 and 4.2.1]{Ma,Ma3} and is rather
difficult to prove in this generality. We mention that in
\cite{Ma,Ma3} the groups are allowed to be not commutative and the
treatment is so that condition (iv) is not needed. The case when $\cs$
consists of linear subspaces of a finite dimensional real vector space
(this is of interest in physical applications) has been considered in
\cite{BG1,DG1} and the corresponding version of Theorem \ref{th:grsax}
is proved there by elementary means.

The following conventions are natural for what follows:
\begin{align}
& X,Y\in\cs \text{ and } Y\nin\cs(X) \Rightarrow 
\cc_X(Y)= \rc_X(Y)=\{0\}, \label{eq:convn} \\
& X,Y,Z\in\cs \text{ and } Z\not\subset X\cap Y \Rightarrow
\rc_{XY}(Z)=\{0\}.\label{eq:convn1}
\end{align}
From now by ``graded'' we mean $\cs$-graded.  Then
$\rc_X=\sum^\rmc_{Y\in\cs}\rc_X(Y)$ is a graded
$C^*$-algebras supported by the ideal $\cs(X)$ of $\cs$, in
particular it is a graded ideal in $\rc_X$.  With the notations of
Subsection \ref{ss:ideal} the algebra $\rc^Y_X=\rc_X(\cs(Y))$ is a
graded ideal of $\rc_X$ supported by $\cs(Y)$. Similarly for $\cc_X$
and $\cc_X^Y$.

Since $\rc_X^{X\cap Y}$ and $\rc_Y^{X\cap Y}$ are ideals in $\rc_X$
and $\rc_Y$ respectively, Theorem \ref{th:nmod} allows us to equip
$\rc_{XY}$ with (right) Hilbert $\rc_Y$-module and left Hilbert
$\rc_X$-module structures (which are not full in general).

\begin{theorem}\label{th:nmain}
The Hilbert $\rc_Y$-module $\rc_{XY}$ is graded by the family of
$C^*$-submodules  $\{\rc_{XY}(Z)\}_{Z\in\cs}$.
\end{theorem}
\proof We use Proposition \ref{pr:rhm} with $\mr=\rt_{XY}$ and
$\cc_Y(Z)$ as algebras $\cc(\sigma)$. Then $\ra=\rc_Y(X\cap Y)$ by
\eqref{eq:hyz} hence $\ra\cdot\cc_Y(Z)=\rc_Y(Z)$ and the conditions
of the proposition are satisfied.  \qed

\begin{remark}\label{re:precise}
The following more precise statement is a consequence of the Theorem
\ref{th:nmain}: the Hilbert $\rc_Y^{X\cap Y}$-module $\rc_{XY}$ is
$\cs(X\cap Y)$-graded by the family of $C^*$-submodules
$\{\rc_{XY}(Z)\}_{Z\in\cs(X\cap Y)}$.
\end{remark}

Finally, we may construct the $C^*$-algebra $\rc$ which is of main
interest for us. We shall describe it as an algebra of operators on
the Hilbert space
\begin{equation}\label{eq:bigh}
\ch\equiv\ch_\cs=\bigoplus\nolimits_{X\in\cs} \ch(X)
\end{equation}
which is a kind of total Fock space (without symmetrization or
anti-symmetrization) determined by the semilattice $\cs$. Note that if
the zero group $O=\{0\}$ belongs to $\cs$ then $\ch$ contains
$\ch(O)=\mbc$ as a subspace, this is the vacuum sector.  Let $\Pi_{X}$
be the orthogonal projection of $\ch$ onto $\ch(X)$ and let us think
of its adjoint $\Pi_{X}^*$ as the natural embedding
$\ch(X)\subset\ch$. Then for any pair $X,Y\in\cs$ we identify
\begin{equation}\label{eq:identc}
\rc_{XY}\equiv\Pi^*_{X}\rc_{XY}\Pi_{Y} \subset L(\ch).
\end{equation}
Thus we realize $\{\rc_{XY}\}_{X,Y\in\cs}$ as a linearly independent
family of closed subspaces of $L(\ch)$ such that
$\rc_{XY}^*=\rc_{YX}$ and $\rc_{XZ}\rc_{Z'Y}\subset\rc_{XY}$ for all
$X,Y,Z,Z'\in\cs$. Then by what we proved before, especially
Proposition \ref{pr:mxyz}, the space
$\sum\nolimits_{X,Y\in\cs}\rc_{XY}$ is a $*$-subalgebra of $L(\ch)$
hence its closure
\begin{equation}\label{eq:bigco}
\rc\equiv\rc_\cs= {\textstyle\sum^\rmc_{X,Y\in\cs}}\rc_{XY}.
\end{equation}
is a $C^*$-algebra of operators on $\ch$.  Note that one may view
$\rc$ as a matrix $(\rc_{XY})_{X,Y\in\cs}$.  

In a similar way one may associate to the algebras $\rt_{XY}$ a
closed self-adjoint subspace $\rt\subset L(\ch)$.  It is also useful
to define a new subspace $\rt^\circ\subset L(\ch)$ by
$\rt^\circ_{XY}=\rt_{XY}$ if $X\sim Y$ and $\rt^\circ=\{0\}$ if
$X\not\sim Y$. Recall that $X\sim Y$ means $X\subset Y$ or $Y\subset
X$ . Clearly $\rt^\circ$ is a closed self-adjoint linear subspace of
$\rt$.  Finally, let $\cc$ be the diagonal $C^*$-algebra
$\cc\equiv\oplus_X\cc_X$ of operators on $\ch$.

\begin{proposition}\label{pr:tc}
We have $\rc=\rt\cdot\cc=\cc\cdot\rt=\rt\cdot\rt=
\rt^\circ\cdot\rt^\circ$.
\end{proposition}
\proof The first two equalities are an immediate consequence of the
Definition \ref{df:main}. To prove the third equality we use
Proposition \ref{pr:product}, more precisely the relation
\[
\rt_{XZ}\cdot\rt_{ZY}=\rt_{XY}\cdot\cc_Y(X\cap Y\cap Z)=
\rc_{XY}(X\cap Y\cap Z)
\]
which holds for any $X,Y,Z$. Then
\[
{\textstyle\sum^\rmc_Z}\rt_{XZ}\cdot\rt_{ZY}=
{\textstyle\sum^\rmc_Z}\rc_{XY}(X\cap Y\cap Z)=
{\textstyle\sum^\rmc_Z}\rc_{XY}(Z)=\rc_{XY}
\]
which is equivalent to $\rt\cdot\rt=\rc$. Now we prove the last
equality in the proposition. We have
\[
{\textstyle\sum^\rmc_Z} \rt^\circ_{XZ}\cdot\rt^\circ_{ZY}= 
\text{ closure of the sum }
{\textstyle\sum_{\substack{Z\sim X\\ Z\sim Y}}}
\rt_{XZ}\cdot\rt_{ZY}. 
\]
In the last sum we have four possibilities: $Z\supset X\cup Y$,
$X\supset Z\supset Y$, $Y\supset Z\supset X$, and $Z\subset X\cap
Y$. In the first three cases we have $Z\supset X\cap Y$ hence
$\rt_{XZ}\cdot\rt_{ZY}=\rt_{XY}$ by \eqref{eq:factor}. In the last
case we have $\rt_{XZ}\cdot\rt_{ZY}=\rt_{XY}\cdot\cc_Y(Z)$ by
\eqref{eq:product}. This proves $\rt^\circ\cdot\rt^\circ=\rc$.
\qed

Finally, we are able to equip $\rc$ with an $\cs$-graded
$C^*$-algebra structure.

\begin{theorem}\label{th:cgrad}
For each $Z\in\cs$ the space $\rc(Z)\equiv
\sum^\rmc_{X,Y\in\cs}\rc_{XY}(Z)$ is a $C^*$-subalgebra of
$\rc$. The family $\{\rc(Z)\}_{Z\in\cs}$ defines a graded
$C^*$-algebra structure on $\rc$.
\end{theorem}
\proof 
We first prove the following relation:
\begin{equation}\label{eq:xyzef}
\rc_{XZ}(E)\cdot\rc_{ZY}(F)=\rc_{XY}(E\cap F)
\quad \text{if } X,Y,Z\in\cs \text{ and } E\in\cs(X\cap Z),
F\in\cs(Y\cap Z).
\end{equation}
From Definition \ref{df:nxyz}, Proposition \ref{pr:product},
relations \eqref{eq:reg1} and \eqref{eq:ayza}, and 
$F\subset Y\cap Z$, we get
\begin{align*}
\rc_{XZ}(E)\cdot\rc_{ZY}(F) &=
\cc_X(E)\cdot\rt_{XZ}\cdot\rt_{ZY}\cdot\cc_Y(F) \\
&= \cc_X(E)\cdot\rt_{XY}\cdot\cc_{Y}(Y\cap Z)\cdot\cc_Y(F) \\
&= \cc_X(E)\cdot\rt_{XY}\cdot\cc_{Y}(F) \\
&= \rt_{XY}\cdot\cc_Y(Y\cap E)\cdot\cc_{Y}(F) \\
&= \rt_{XY}\cdot\cc_Y(Y\cap E\cap F).
\end{align*}
At the next to last step we used $\cc_X(E)=\cc_G(E)|_X$ for some
$G\in\cs$ containing both $X$ and $Y$ and then \eqref{eq:ayza},
\eqref{eq:reg2}. Finally, we use $\cc_Y(Y\cap E\cap F)=\cc_Y(E\cap
F)$ and the Definition \ref{df:nxyz}. This proves \eqref{eq:xyzef}.
Due to the conventions \eqref{eq:convn}, \eqref{eq:convn1} we now
get from \eqref{eq:xyzef} for $E,F\in\cs$
\[
{\textstyle\sum_{Z\in\cs}}\rc_{XZ}(E)\cdot\rc_{ZY}(F)=\rc_{XY}(E\cap F).
\]
Thus $\rc(E)\rc(F)\subset\rc(E\cap F)$, in particular $\rc(E)$ is a
$C^*$-algebra. It remains to be shown that the family of
$C^*$-algebras $\{\rc(E)\}_{E\in\cs}$ is linearly independent. Let
$A(E)\in\rc(E)$ such that $A(E)=0$ but for a finite number of $E$
and assume that $\sum_E A(E)=0$. Then for all $X,Y\in\cs$
we have $\sum_E \Pi_X A(E) \Pi_Y^* =0$. Clearly 
$\Pi_X A(E) \Pi_Y^*\in\rc_{XY}(E)$ hence from Theorem \ref{th:nmain}
we get $\Pi_X A(E) \Pi_Y^*=0$ for all $X,Y$ so $A(E)=0$ for all $E$.
\qed

We now point out some interesting subalgebras of $\rc$. If
$\ct\subset\cs$ is any subset let
\begin{equation}\label{eq:t}
\rc_\ct\equiv{\textstyle\sum_{X,Y\in\ct}^\rmc}\rc_{XY} \quad
\text{and} \quad \ch_\ct\equiv\oplus_{X\in\ct}\ch(X).
\end{equation}
Note that the sum defining $\rc_\ct$ is already closed if $\ct$ is
finite and that $\rc_\ct$ is a $C^*$-algebra which lives on the
subspace $\ch_\ct$ of $\ch$. In fact, if $\Pi_\ct$ is the orthogonal
projection of $\ch$ onto $\ch_\ct$ then
\begin{equation}\label{eq:tt}
\rc_\ct=\Pi_\ct\rc\Pi_\ct
\end{equation}
and this is a $C^*$-algebra because $\rc\Pi_\ct\rc\subset\rc$ by
Proposition \ref{pr:mxyz}. 
It is easy to check that $\rc_\ct$ is a graded $C^*$-subalgebra of
$\rc$ supported by the ideal $\ccup_{X\in\ct}\cs(X)$ generated by
$\ct$ in $\cs$. Indeed, we have
\[
\rc_\ct\,\ccap\,\rc(E)=
\left({\textstyle\sum_{X,Y\in\ct}^\rmc}\rc_{XY}\right) \ccap
\left({\textstyle\sum_{X,Y\in\cs}^\rmc}\rc_{XY}(E)\right) =
{\textstyle\sum_{X,Y\in\ct}^\rmc}\rc_{XY}(E).
\]
It is clear that $\rc$ is the inductive limit of the increasing
family of $C^*$-algebras $\rc_\ct$ with finite $\ct$.

If $\ct=\{X\}$ then the definitions \eqref{eq:t} give $\rc_X$ and
$\ch(X)$. If $\ct=\{X,Y\}$ with distinct $X,Y$ we get a simple but
nontrivial situation. Indeed, we shall have 
$\ch_\ct=\ch(X)\oplus\ch(Y)$  and $\rc_\ct$ may be thought as a
matrix 
\[
\rc_\ct=
\begin{pmatrix}
\rc_X & \rc_{XY}\\
\rc_{YX} & \rc_Y
\end{pmatrix}.
\]
The grading is now explicitly defined as follows: 
\begin{compactenum}
\item
If  $E\subset X\cap Y$ then
\[
\rc_\ct(E)=
\begin{pmatrix}
\rc_X(E) & \rc_{XY}(E)\\
\rc_{YX}(E) & \rc_Y(E)
\end{pmatrix}.
\]
\item   \label{p:2ex}
If  $E\subset X$ and $E\not\subset Y$ then
\[
\rc_\ct(E)=
\begin{pmatrix}
\rc_X(E) & 0\\
0 & 0
\end{pmatrix}.
\]
\item
If  $E\not\subset X$ and $E\subset Y$ then
\[
\rc_\ct(E)=
\begin{pmatrix}
0 & 0\\
0 & \rc_Y(E)
\end{pmatrix}.
\]
\end{compactenum}

The case when $\ct$ is of the form $\cs(X)$ for some $X\in\cs$ is
especially interesting.

\begin{definition}\label{df:sc}
If $X\in\cs$ then we say that the $\cs(X)$-graded $C^*$-algebra
$\rc_X^\#\equiv\rc_{\cs(X)}$ is the \emph{second quantization}, or
\emph{unfolding}, of the algebra $\rc_X$.  More explicitly
\begin{equation}\label{eq:xc}
\rc_X^\#\equiv{\textstyle\sum^\rmc_{Y,Z\in\cs(X)}}\rc_{YZ}.
\end{equation}
\end{definition}
To justify the terminology, observe that the self-adjoint operators
affiliated to $\rc_X$ live on the Hilbert space $\ch(X)$ and are (an
abstract version of) Hamiltonians of an $N$-particle system $\rs$
with a fixed $N$ (the configuration space is $X$ and $N$ is the
number of levels of the semilattice $\cs(X)$).  One obtains
$\rc_X^\#$ by adding interactions which couple the subsystems of
$\rs$ which have the $Y\in\cs(X)$ as configuration spaces and have
$\rc_Y$ as algebras of energy observables.

Observe that $\rc_X^\#$ lives in the subspace $\ch_X=\ch_{\cs(X)}$
of $\ch$.  We have $\rc_X^\#\subset\rc_Y^\#$ if $X\subset Y$ and
$\rc$ is the inductive limit of the algebras $\rc_X^\#$.  Below we
give an interesting alternative description of $\rc_X^\#$.

\begin{theorem}\label{th:mor}
Let $\rn_X=\oplus_{Y\in\cs(X)}\rc_{YX}$ be the direct sum of the
Hilbert $\rc_X$-modules $\rc_{YX}$ equipped with the direct sum graded
structure. Then $\ck(\rn_X) \cong \rc_X^\#$ the isomorphism being such
that the graded structure on $\ck(\rn_X)$ defined in Theorem
\ref{th:kghm} is transported into that of $\rc_X^\#$.  In other terms,
$\rc_X^\#$ is the imprimitivity algebra of the full Hilbert
$\rc_X$-module $\rn_X$ and $\rc_X$ and $\rc_X^\#$ are Morita
equivalent.
\end{theorem}
\proof If $Y\subset X$ then $\rc^*_{YX}\cdot\rc_{YX}=\rc_X^Y$ and
$\rc_{YX}$ is a full Hilbert $\rc_X^Y$-module. Since the $\rc_X^Y$
are ideals in $\rc_X$ and their sum over $Y\in\cs(X)$ is equal to
$\rc_X$ we see that $\rn_X$ becomes a full Hilbert graded
$\rc_X$-module supported by $\cs(X)$, cf.  Section \ref{s:grad}. By
Theorem \ref{th:kghm} the imprimitivity $C^*$-algebra $\ck(\rn_X)$
is equipped with a canonical $\cs(X)$-graded structure.

We shall make a comment on $\ck(\mr)$ in the more general the case
when $\mr=\oplus_i\mr_i$ is a direct sum of Hilbert $\ra$-modules
$\mr_i$, cf. \S\ref{ss:gf}. First, it is clear that we have 
\[
\ck(\mr)={\textstyle\sum^\rmc_{ij}}\ck(\mr_j,\mr_i)\cong
(\ck(\mr_j,\mr_i))_{ij}.
\]
Now assume that $\ce,\ce_i$ are Hilbert spaces such that $\ra$ is a
$C^*$-algebra of operators on $\ce$ and $\mr_i$ is a Hilbert
$C^*$-submodule of $L(\ce,\ce_i)$ such that
$\ra_i\equiv\mr_i^*\cdot\mr_i$ is an ideal of $\ra$. 
Then by Proposition \ref{pr:2ss} we have
$\ck(\mr_j,\mr_i)\cong\mr_i\cdot\mr_j^*\subset L(\ce_j,\ce_i)$. 

In our case we take 
\[
i=Y\in\cs(X),\quad \mr_i=\rc_{YX}, \quad \ra=\rc_X,\quad
\ce=\ch(X),\quad  \ce_i=\ch(Y),\quad   \ra_i=\rc_X^Y.
\]
Then we get
\[
\ck(\mr_j,\mr_i)\equiv\ck(\rc_{ZX},\rc_{YX})\cong
\rc_{YX}\cdot\rc_{ZX}^*=\rc_{YX}\cdot\rc_{XZ}=\rc_{YZ}
\]
by Proposition \ref{pr:mxyz}.
\qed

\section{Operators affiliated to $\rc$ and their essential
spectrum} 
\label{s:af}
\protect\setcounter{equation}{0}

In this section we give examples of self-adjoint operators
affiliated to the algebra $\rc$ constructed in Section \ref{s:grass}
and then we give a formula for their essential spectrum. We refer to
\S\ref{ss:grca} for terminology and basic results related to the
notion of affiliation that we use and to \cite{ABG,GI1,DG3} for
details.

We recall that a self-adjoint operator $H$ on a Hilbert space $\ch$
is \emph{strictly affiliated} to a $C^*$-algebra of operators $\ra$
on $\ch$ if $(H+i)^{-1}\in\ra$ (then $\varphi(H)\in\ra$ for all
$\varphi\in\Co(\mbr)$) and if $\ra$ is the clspan of the elements
$\varphi(H)A$ with $\varphi\in\Co(\mbr)$ and $A\in\ra$. This class
of operators has the advantage that each time $\ra$ is
non-degenerately represented on a Hilbert space $\ch'$ with the help
of a morphism $\rp:\ra\to L(\ch')$, the observable $\rp H$ is
represented by a usual densely defined self-adjoint operator on
$\ch'$.

The diagonal algebra
\begin{equation}\label{eq:d}
\cc^*(\cs)\equiv\oplus_{X\in\cs} \cc^*(X)
\end{equation}
has a simple physical interpretation: this is the $C^*$-algebra
generated by the kinetic energy operators.  Since
$\rc_{XX}=\rc_X\supset\rc_{X}(X)= \cc^*(X)$ we see that $\cc^*(\cs)$ is a
$C^*$-subalgebra of $\rc$. From \eqref{eq:nxyz}, \eqref{eq:txy1},
\eqref{eq:txy2} and the Cohen-Hewitt theorem we get
\begin{equation}\label{eq:ed}
\rc(Z)\cc^*(\cs)=\cc^*(\cs)\rc(Z)=\rc(Z)\quad \forall Z\in\cs \quad 
\text{and}\hspace{2mm} \rc \cc^*(\cs)=\cc^*(\cs)\rc=\rc.
\end{equation}
In other terms, $\cc^*(\cs)$ acts
non-degenerately\symbolfootnote[2]{\ Note that if $\cs$ has a
  largest element $\cx$ then the algebra $\rc(\cx)$ acts on each
  $\rc(Z)$ but this action is degenerate.} 
on each $\rc(Z)$ and on $\rc$. It follows that a self-adjoint
operator strictly affiliated to $\cc^*(\cs)$ is also strictly
affiliated to $\rc$.

For each $X\in\cs$ let $h_X:X^*\to\mbr$ be a continuous function
such that $|h_X(k)|\to\infty$ if $k\to\infty$ in $X^*$. Then the
self-adjoint operator $K_X\equiv h_X(P)$ on $\ch(X)$ is strictly
affiliated to $\cc^*(X)$ and the norm of $(K_X+i)^{-1}$ is equal to
$\sup_k(h^2_X(k)+1)^{-1/2}$. Let $K\equiv\bigoplus_{X\in\cs}K_X$,
this is a self-adjoint operator $\ch$. Clearly $K$ is affiliated to 
$\cc^*(\cs)$ if and only if 
\begin{equation}\label{eq:kin}
\lim_{X\to\infty}\sup\nolimits_{k}(h^2_X(k)+1)^{-1/2}=0
\end{equation}
and then $K$ is strictly affiliated to $\cc^*(\cs)$ (the set $\cs$ is
equipped with the discrete topology). If the functions $h_X$ are
positive this means that $\min h_X$ tends to infinity when
$X\to\infty$. One could avoid such a condition by considering an
algebra larger then $\rc$ such as to contain 
$\prod_{X\in\cs} \cc^*(X)$, but we shall not develop this idea here.

Now let $H=K+I$ with $I\in\rc$ (or in the multiplier algebra) a
symmetric element. Then
\begin{equation}\label{eq:res}
(\lambda-H)^{-1}=(\lambda-K)^{-1}\left(1-I(\lambda-K)^{-1}\right)^{-1}
\end{equation}
if $\lambda$ is sufficiently far from the spectrum of $K$ such as to
have $\|I(\lambda-K)^{-1}\|<1$. Thus $H$ is strictly affiliated to
$\rc$.  We interpret $H$ as the Hamiltonian of our system of
particles when the kinetic energy is $K$ and the interactions
between particles are described by $I$. Even in the simple case
$I\in\rc$ these interactions are of a very general nature being a
mixture of $N$-body and quantum field type interactions (which
involve creation and annihilation operators so the number of
particles is not preserved). 

We shall now use Theorem \ref{th:gas} in order to compute the
essential spectrum of an operator like $H$. The case of unbounded
interactions will be treated later on. Let $\rc_{\geq E}$ be the
$C^*$-subalgebra of $\rc$ determined by $E\in\cs$ according to the
rules of $\S\ref{ss:grca}$. More explicitly, we set
\begin{equation}\label{eq:geqe}
\rc_{\geq E}={\textstyle\sum^\rmc_{F\supset E}\rc(F)}\cong 
\big({\textstyle\sum^\rmc_{F\supset E}}
\rc_{XY}(F)\big)_{X\cap Y\supset E}
\end{equation}
and note that $\rc_{\geq E}$ lives on the subspace $\ch_{\geq
  E}=\bigoplus_{X\supset E}\ch(X)$ of $\ch$. Since in the second sum
from \eqref{eq:geqe} the group $F$ is such that $E\subset F\subset
X\cap Y$ the algebra $\rc_{\geq E}$ is strictly included in the
algebra $\rc_\ct$ obtained by taking $\ct=\{F\in\cs \mid F\supset
E\}$ in \eqref{eq:t}. 

Let $\rp_{\geq E}$ be the canonical idempotent morphism of $\rc$
onto $\rc_{\geq E}$ introduced in Theorem \ref{th:ga}. We consider
the self-adjoint operator on the Hilbert space $\ch_{\geq E}$
defined as follows:
\begin{equation}\label{eq:post}
 H_{\geq E}=K_{\geq E}+I_{\geq E} \quad \text{where} \quad 
K_{\geq E}= \oplus_{X\geq E} K_X \hspace{2mm} \text{and} 
\hspace{2mm} I_{\geq E}=\rp_{\geq E}I.
\end{equation}
Then $H_{\geq E}$ is strictly affiliated to $\rc_{\geq E}$ and it
follows easily from \eqref{eq:res} that
\begin{equation}\label{eq:pre}
\rp_{\geq E}\varphi(H)=\varphi(H_{\geq E}) \quad \forall
\varphi\in\Co(\mbr).
\end{equation} 
Now let us assume that the group $O=\{0\}$ belongs to $\cs$. Then we
have
\begin{equation}\label{eq:O}
\rc(O)=K(\ch).
\end{equation} 
Indeed, from \eqref{eq:nxyz} we get
$\rc_{XY}(O)=\rt_{XY}\cdot\Co(Y)=\rk_{XY}$ which implies the
preceding relation. If we also assume that $\cs$ is atomic and we
denote $\cp(\cs)$ its set of atoms, then from Theorem \ref{th:ga} we
get a canonical embedding
\begin{equation}\label{eq:quotc}
\rc/K(\ch)\subset\pprod\nolimits_{E\in\cp(\cs)}\rc_{\geq E}
\end{equation} 
defined by the morphism $\rp\equiv(\rp_{\geq E})_{E\in\cp(\cs)}$.
Then from \eqref{eq:es2} we obtain:
\begin{equation}\label{eq:ess1}
\spe(H)=\overline{\ccup}_{E\in\cp(\cs)}\sp(H_{\geq E}).
\end{equation}
Our next purpose is to prove a similar formula for a certain class
of unbounded interactions $I$.

Let $\cg\equiv\cg_\cs=D(|K|^{1/2})$ be the form domain of $K$
equipped with the graph topology. Then $\cg\subset\ch$ continuously
and densely so after the Riesz identification of $\ch$ with its
adjoint space $\ch^*$ we get the usual scale
$\cg\subset\ch\subset\cg^*$ with continuous and dense embeddings.
Let us denote
\begin{equation}\label{eq:jap}
\jap{K} =|K+i|=\sqrt{K^2+1}.
\end{equation}
Then $\jap{K}^{1/2}$ is a self-adjoint operator on $\ch$ with domain
$\cg$ and $\jap{K}$ induces an isomorphism $\cg\to\cg^*$.  The
following result is a straightforward consequence of Theorem 2.8 and
Lemma 2.9 from \cite{DG3}. 

\begin{theorem}\label{th:af}
Let $I:\cg\to\cg^*$ be a continuous symmetric operator and let us
assume that there are real numbers $\mu,a$ with $0<\mu<1$ such that
one of the following conditions is satisfied:
\begin{compactenum}[(i)]
\item
$\pm I \leq\mu|K+ia|,$
\item
$K$ is bounded from below and $ I \geq -\mu|K+ia|.$
\end{compactenum}
Let $H=K+I$ be the form sum of $K$ and $I$, so $H$ has as domain the
set  of $u\in\cg$ such that $Ku+Iu\in\ch$ and acts as $Hu=Ku+Iu$. 
Then $H$ is a self-adjoint operator on $\ch$. If there is
$\alpha>1/2$ such that 
$\langle K\rangle^{-\alpha}I\langle K\rangle^{-1/2}\in\rc$ then $H$
is strictly affiliated to $\rc$. 
If $O\in\cs$ and the semilattice $\cs$ is atomic then
\begin{equation}\label{eq:ess2}
\spe(H)=\overline{\ccup}_{E\in\cp(\cs)}\sp(H_{\geq E}).
\end{equation}
\end{theorem}

The last assertion of the theorem follows immediately from Theorem
\ref{th:gas} and is a general version of the HVZ theorem.  In order
to have a more explicit description of the observables $H_{\geq
  E}\equiv\rp_{\geq E}H$ we now prove an analog of Theorem 3.5 from
\cite{DG3}. We cannot use that theorem in our context for three
reasons: first we did not suppose that $\cs$ has a maximal element,
then even if $\cs$ has a maximal element $\cx$ the action of the
corresponding algebra $\rc(\cx)$ on the algebras $\rc(E)$ is
degenerate, and finally our ``free'' operator $K$ is not affiliated
to $\rc(\cx)$.

\begin{theorem}\label{th:afi}
For each $E\in\cs$ let $I(E)\in L(\cg,\cg^*)$ be a symmetric
operator such that:
\begin{compactenum}[(i)]
\item
$\jap{K}^{-\alpha}I(E)\jap{K}^{-1/2}\in\rc(E)$ for some
$\alpha\geq 1/2$ independent of $E$,
\item
there are real positive numbers $\mu_E,a$ such that either $\pm
I(E) \leq\mu_E|K+ia|$ for all $E$ or $K$ is bounded from below and
$ I(E) \geq -\mu_E|K+ia|$ for all $E$,
\item
we have $\sum_E\mu_E\equiv\mu<1$ and the series $\sum_E I(E)\equiv I$
is norm summable in $L(\cg,\cg^*)$.
\end{compactenum}
Let us set $I_{\geq E}=\sum_{F\geq E}I(F)$. Define the self-adjoint
operator $H=K+I$ on $\ch$ as in Theorem \ref{th:af} and define
similarly the self-adjoint operator $H_{\geq E}=K_{\geq E}+I_{\geq
  E}$ on $\ch_{\geq E}$.  Then the operator $H$ is strictly
affiliated to $\rc$, the operator $H_{\geq E}$ is strictly
affiliated to $\rc_{\geq E}$, and we have $\rp_{\geq E}H=H_{\geq E}$.
\end{theorem}
\proof  We shall consider only the case when $\pm I(E)
\leq\mu_E|K+ia|$ for all $E$. The more singular situation when $K$
is bounded from below but there is no restriction on the positive
part of the operators $I(E)$ (besides summability) is more difficult
but the main idea has been explained in \cite{DG3}.

We first make some comments to clarify the definition of the
operators $H$ and $H_{\geq E}$.  Observe that our assumptions imply
$\pm I\leq\mu|K+ia|$ hence if we set
\[
\Lambda\equiv|K+ia|^{-1/2}=(K^2+a^2)^{-1/4}\in \cc^*(\cs)
\]
then  we obtain
\[
\pm\braket{u}{Iu}\leq\mu\braket{u}{|K+ia|u}=
\mu\| |K+ia|^{1/2}u\| = \mu\| \Lambda^{-1} u\|
\] 
which is equivalent to $\pm\Lambda I\Lambda\leq\mu$ or $\|\Lambda
I\Lambda\|\leq\mu$. In particular we may use Theorem \ref{th:af}
in order to define the self-adjoint operator $H$. Moreover, we have
\[
\jap{K}^{-\alpha}I\jap{K}^{-1/2}={\textstyle\sum_E}
\jap{K}^{-\alpha}I(E)\jap{K}^{-1/2}\in\rc
\]
because the series is norm summable in $L(\ch)$. Thus $H$ is
strictly affiliated to $\rc$.  

In order to define $H_{\geq E}$ we first make a remark on 
$I_{\geq E}$.  If we set $\cg(X)=D(|K_X|^{-1/2})$ and if we equip
$\cg$ and $\cg(X)$ with the norms  \label{p:formd}
\[
\|u\|_\cg=\|\jap{K}^{1/2}u\|_\ch \quad \text{and} \quad
\|u\|_{\cg(X)}=\|\jap{K_X}^{1/2}u\|_{\ch(X)}
\]
respectively then clearly 
\[
\cg=\oplus_X\cg(X) \quad  \text{and} \quad \cg^*=\oplus_X\cg^*(X)
\]
where the sums are Hilbertian direct sums and $\cg^*$ and
$\cg^*(X)\equiv\cg(X)^*$ are equipped with the dual norms.
Then each $I(F)$ may be represented as a matrix 
$I(F)=(I_{XY}(F))_{X,Y\in\cs}$ of continuous operators 
$I_{XY}(E):\cg(Y)\to\cg^*(X)$. Clearly
\[
\jap{K}^{-\alpha}I(F)\jap{K}^{-1/2}=
\left(\jap{K_X}^{-\alpha}I_{XY}(F)\jap{K_Y}^{-1/2}\right)_{X,Y\in\cs}
\] 
and since by assumption (i) this belongs to $\rc(F)$ we see that
$I_{XY}(F)=0$ if $X\not\supset F$ or $Y\not\supset F$. Now fix $E$
and let $F\supset E$. Then, when viewed as a sesquilinear form,
$I(F)$ is supported by the subspace $\ch_{\geq E}$ and has domain
$\cg_{\geq E}= D(|K_{\geq E}|^{1/2}$.  It follows that $I_{\geq E}$ 
is a sesquilinear form with domain $\cg_{\geq E}$ supported by the
subspace $\ch_{\geq E}$ and may be thought as an element of
$L(\cg_{\geq E},\cg^*_{\geq E})$ such that
$\pm I_{\geq E}\leq \mu |K_{\geq E}+ia|$ because 
$\sum_{F\supset E}\mu_F\leq \mu$. To conclude, we may now define 
$H_{\geq E}=K_{\geq E}+I_{\geq E}$ exactly as in the case of $H$ and
get a self-adjoint operator on $\ch_{\geq E}$ strictly affiliated to
$\rc_{\geq E}$. Note that this argument also gives
\begin{equation}\label{eq:ek}
\jap{K}^{-1/2} I(F) \jap{K}^{-1/2}=
\jap{K_{\geq E}}^{-1/2} I(F) \jap{K_{\geq E}}^{-1/2}.
\end{equation}
It remains to be shown that $\rp_{\geq E}H=H_{\geq E}$. If we set
$R\equiv(ia-H)^{-1}$ and $R_{\geq E}\equiv(ia-H_{\geq E})^{-1}$ then
this is equivalent to $\rp_{\geq E}R=R_{\geq E}$.  Let us set
\[
U=|ia-K|(ia-K)^{-1}=\Lambda^{-2}(ia-K)^{-1}, \quad
J=\Lambda I\Lambda U.
\]
Then $U$ is a unitary operator and $\|J\|<1$, so we get a norm
convergent series expansion
\[
R=(ia-K-I)^{-1}=
\Lambda U(1-\Lambda I\Lambda U)^{-1}\Lambda =
{\textstyle\sum_{n\geq0}}\Lambda U J^n\Lambda
\]
which implies
\[
\rp_{\geq E} (R)=
{\textstyle\sum_{n\geq0}}
\rp_{\geq E}\big(\Lambda U J^{n}\Lambda\big)
\]
the series being norm convergent. Thus it suffices to prove that for
each $n\geq0$ 
\begin{equation}\label{eq:ekk}
\rp_{\geq E}\big(\Lambda U J^{n}\Lambda\big)=
\Lambda_{\geq E} (J_{\geq E})^{n}\Lambda_{\geq E}
\end{equation}
where $J_{\geq E}=\Lambda_{\geq E} I_{\geq E}\Lambda_{\geq E}
U_{\geq E}$.  Here $\Lambda_{\geq E}$ and $U_{\geq E}$ are
associated to $K_{\geq E}$ in the same way $\Lambda$ and $K$ are
associated to $K$. For $n=0$ this is obvious because 
$\rp_{\geq E}K=K_{\geq E}$. If $n=1$ this is easy because
\begin{align}\label{eq:e}
\Lambda U J\Lambda &= \Lambda U \Lambda I \Lambda U\Lambda=
(ia-K)^{-1} I (ia-K)^{-1} \\
&=
[(ia-K)^{-1}\jap{K}^\alpha] \cdot 
[\jap{K}^{-\alpha} I \jap{K}^{-1/2}] \cdot
[\jap{K}^{1/2} (ia-K)^{-1}] \nonumber
\end{align}
and it suffices to note that 
$\rp_{\geq E}(\jap{K}^{-\alpha} I(F) \jap{K}^{-1/2})=0$ if
$F\not\supset E$ and to use \eqref{eq:ek} for $F\supset E$. 

To treat the general case we make some preliminary remarks. 
If $J(F)=\Lambda I(F) \Lambda U$ then $J=\sum_F J(F)$ where the
convergence holds in norm on $\ch$ because of the condition (iii). 
Then we have a norm convergent expansion
\[
\Lambda U J^n \Lambda ={\textstyle\sum_{F_1,\dots,F_n\in\cs}}
\Lambda U J(F_1)\dots J(F_n) \Lambda.
\]
Assume that we have shown $\Lambda U J(F_1)\dots
J(F_n)\Lambda\in\rc(F_1\cap\dots\cap F_n)$.  Then we get
\begin{equation}\label{eq:ekkk}
\rp_{\geq E}(\Lambda U J^n \Lambda)=
{\textstyle\sum_{F_1\geq E,\dots,F_n\geq E}}
\Lambda U J(F_1)\dots J(F_n) \Lambda
\end{equation}
because if one $F_k$ does not contain $E$ then the intersection
$F_1\cap\dots\cap F_n$ does not contain $E$ hence $\rp_{\geq E}$
applied to the corresponding term gives $0$. Because of
\eqref{eq:ek} we have $J(F)=\Lambda_{\geq E} I(F) \Lambda_{\geq E}
U_{\geq E}$ if $F\supset E$ and we may replace everywhere in the
right hand side of \eqref{eq:ekkk} $\Lambda$ and $U$ by
$\Lambda_{\geq E}$ and $U_{\geq E}$. This clearly proves
\eqref{eq:ekk}. 

Now we prove the stronger fact 
$\Lambda U J(F_1)\dots J(F_n)\in\rc(F_1\cap\dots\cap F_n)$.  
If $n=1$ this follows from a slight modification of
\eqref{eq:e}: the last factor on the right hand side of \eqref{eq:e}
is missing but is not needed. Assume that the assertion holds for
some   $n$. Since $K$ is strictly affiliated to $\cc^*(\cs)$ and
$\cc^*(\cs)$ acts non-degenerately on each $\rc(F)$ we may use the
Cohen-Hewitt theorem to deduce that there is $\varphi\in\Co(\mbr)$
such that
$\Lambda U J(F_1)\dots J(F_n)=T\varphi(K)$ 
for some $T\in\rc(F_1\cap\dots\cap F_n)$. 
Then
\[
\Lambda U J(F_1)\dots J(F_n)J(F_{n+1})=T\varphi(K)J(F_{n+1})
\]
hence it suffices to prove that $\varphi(K)J(F)\in\rc(F)$ for any
$F\in\cs$ and any $\varphi\in\Co(\mbr)$. But the set of $\varphi$
which have this property is a closed subspace of $\Co(\mbr)$ which
clearly contains the functions $\varphi(\lambda)=(\lambda -z)^{-1}$
if $z$ is not real hence is equal to $\Co(\mbr)$.  \qed

\begin{remark}\label{re:alpha}
Choosing $\alpha>1/2$ allows one to consider perturbations of $K$
which are of the same order as $K$, e.g. in the $N$-body situations
one may add to the Laplacian $\Delta$ on operator like $\nabla^*
M\nabla$ where the function $M$ is bounded measurable and has the
structure of an $N$-body type potential, cf. \cite{DG3,DerI}.  
\end{remark}

The only assumption of Theorem \ref{th:afi} which is really relevant
is $\jap{K}^{-\alpha}I(E)\jap{K}^{-1/2}\in\rc(E)$.  We shall give
below more explicit conditions which imply it.  If we change
notation $E\to Z$ and use the formalism introduced in the proof of
Theorem \ref{th:afi} we have
\begin{equation}\label{eq:xye}
I(Z)=(I_{XY}(Z))_{X,Y\in\cs} \quad \text{with} \quad
I_{XY}(Z):\cg(Y)\to\cg^*(X) \text{ continuous}.
\end{equation}
We are interested in conditions on $I_{XY}(Z)$ which imply
\begin{equation}\label{eq:xye1}
\jap{K_X}^{-\alpha}I_{XY}(Z)\jap{K_X}^{-1/2} \in \rc_{XY}(Z).
\end{equation}
For this we shall use Theorem \ref{th:yzintr} which gives a simple
intrinsic characterization of $\rc_{XY}(Z)$.

The construction which follows is interesting only if $X$ is not a
discrete group, otherwise $X^*$ is compact and many conditions are
trivially satisfied. We shall use weights only in order to avoid
imposing on the functions $h_X$ regularity conditions stronger than
continuity.

A positive function on $X^*$ is a \emph{weight} if
$\lim_{k\to\infty} w(k)=\infty$ and $w(k+p)\leq\omega(k)w(p)$ for
some function $\omega$ on $X^*$ and all $k,p$. We say that $w$ is
\emph{regular} if one may choose $\omega$ such that
$\lim_{k\to0}\omega(k)=1$.  The example one should have in mind when
$X$ is an Euclidean space is $w(k)=\jap{k}^s$ for some $s>0$. Note
that we have $\omega(-k)^{-1}\leq w(k+p)w(p)^{-1} \leq \omega(k)$
hence if $w$ is a regular weight then \label{p:regw}
\begin{equation}\label{eq:regw}
\theta(k)\equiv \sup_{p\in X^*}\frac{|w(k+p)-w(p)|}{w(p)} 
\Longrightarrow
\lim_{k\to0}\theta(k)=0.
\end{equation}
It is clear that if $w$ is a regular weight and $\sigma\geq 0$ is a
real number then $w^\sigma$ is also a regular weight.

We say that two functions $f,g$ defined on a neighborhood of 
infinity of $X^*$ are \emph{equivalent} and we write $f\sim g$ if
there are numbers $a,b$ such that $a|f(k)|\leq|g(k)|\leq
b|f(k)|$. Then $|f|^\sigma\sim|g|^\sigma$ for all $\sigma>0$.

In the next theorem we shall use the spaces
\[
\cg^\sigma(X)=D(|K_X|^{\sigma/2}) \hspace{2mm} \text{and}\hspace{2mm}
\cg^{-\sigma}(X)\equiv\cg^\sigma(X)^*
\]
with $\sigma \geq 1$.  In particular $\cg^1(X)=\cg(X)$ and
$\cg^{-1}(X)=\cg^*(X)$. 

\begin{proposition}\label{pr:tex}
Assume that $h_X,h_Y$ are equivalent to regular weights.  For
$Z\subset X\cap Y$ let $I_{XY}(Z):\cg(Y)\to\cg^*(X)$ be a continuous
map such that
\begin{enumerate}
\item
$U_z I_{XY}(Z)=I_{XY}(Z) U_z$ if $z\in Z$ and
$V^*_k I_{XY}(Z) V_k\to I_{XY}(Z)$ if $k\to 0$ in $(X+Y)^*$,
\item 
$I_{XY}(Z)(U_y-1)\to 0$ if $y\to 0$ in $Y$ and
$I_{XY}(Z)(V_k-1)\to 0$ if $k\to 0$ in $(Y/Z)^*$,
\end{enumerate}
where the limits hold in norm in $L(\cg^1(Y),\cg^{-\sigma}(X))$ for
some $\sigma\geq1$. Then \eqref{eq:xye1} holds with
$\alpha=\sigma/2$.
\end{proposition}
\proof We begin with some general comments on weights. Let $w$ be a
regular weight and let $\cg(X)$ be the domain of the operator $w(P)$
in $\ch(X)$ equipped with the norm $\|w(P)u\|$. Then $\cg(X)$ is a
Hilbert space and if $\cg^*(X)$ is its adjoint space then we get a
scale of Hilbert spaces $\cg(X)\subset\ch(X)\subset\cg^*(X)$ with
continuous and dense embeddings. Since $U_x$ commutes with $w(P)$ it
is clear that $\{U_x\}_{x\in X}$ induces strongly continuous unitary
representation of $X$ on $\cg(X)$ and $\cg^*(X)$.  Then
\[
\|V_k u\|_{\cg(X)}=\|w(k+P)u\|\leq\omega(k)\|u\|_{\cg(X)}
\]
from which it follows that $\{V_k\}_{k\in X^*}$ induces by
restriction and extension strongly continuous representations of
$X^*$ in $\cg(X)$ and $\cg^*(X)$. Moreover, as operators on $\ch(X)$
we have \label{p:RK}
\begin{align}  
|V_k^*w(P)^{-1}V_k-w(P)^{-1}| 
&=|w(k+P)^{-1}-w(P)^{-1}| 
= |w(k+P)^{-1}(w(P)-w(k+P))w(P)^{-1}|  \nonumber \\
& \leq \omega(-k)|(w(P)-w(k+P))w(P)^{-2}|
\leq \omega(-k)\theta(k) w(P)^{-1}. \label{eq:refw} 
\end{align}
Now let $w_X,w_Y$ be regular weights equivalent to
$|h_X|^{1/2},|h_Y|^{1/2}$ and let us set $S=I_{XY}(Z)$. Then
\[
\jap{K_X}^{-\alpha}S\jap{K_X}^{-1/2}=
\jap{K_X}^{-\alpha}w_X(P)^{2\alpha}\cdot 
w_X(P)^{-2\alpha}S w_Y(P)^{-1} \cdot
w_Y(P)\jap{K_X}^{-1/2}
\]
and $\jap{h_X}^{-\alpha}w_X^{2\alpha}$, $\jap{h_Y}^{-1/2}w_Y$ and
their inverses are bounded continuous functions on $X,Y$. Since
$\rc_{XY}(Z)$ is a non-degenerate left $\cc^*(X)$-module and right
$\cc^*(Y)$-module we may use the Cohen-Hewitt theorem to deduce that
\eqref{eq:xye1} is equivalent to
\begin{equation}\label{eq:xye2}
w_X(P)^{-\sigma}I_{XY}(Z) w_Y(P)^{-1} \in \rc_{XY}(Z)
\end{equation}
where $\sigma=2\alpha$.  To simplify notations we set
$W_X=w^\sigma_X(P), W_Y=w_Y(P)$.  We also omit the index $X$ or $Y$
for the operators $W_X,W_Y$ since their value is obvious from the
context. In order to show $W^{-1}SW^{-1}\in \rc_{XY}(Z)$ we check
the conditions of Theorem \ref{th:yzintr} with $T=W^{-1}SW^{-1}$. We
may assume $\sigma>1$ and then we clearly have
\[
\|(U_x-1)T\|\leq \|(U_x-1)w_X^{1-\sigma}(P)\|
\|w_X^{-1}(P)I_{XY}(Z)W^{-1}\|\to 0 \quad \text{if } x\to0.
\]
so the first part of condition 1 from Theorem \ref{th:yzintr} is
satisfied. The second part of that condition is trivially verified.
Condition 2 there is not so obvious, but if we set $W_k=V_k^*WV_k$
and $V_k^* SV_k$ we have:
\begin{align*}
V_k^*TV_k-T 
&= W_k^{-1} S_k W_k^{-1}-W^{-1}SW^{-1}\\
&= (W_k^{-1}-W^{-1})S_kW_k^{-1} + W^{-1}S_kW_k^{-1} -W^{-1}SW^{-1}\\
&= (W_k^{-1}-W^{-1})S_kW_k^{-1} + W^{-1}(S_k-S)W_k^{-1}
+W^{-1}S(W_k^{-1} -W^{-1}).
\end{align*}
Now if we use \eqref{eq:refw} and set $\xi(k)=\omega(-k)\theta(k)$
we get:
\begin{align*}
\|V_k^*TV_k-T\| &\leq 
\xi(k)\|W^{-1}S_kW_k^{-1}\| + \|W^{-1}(S_k-S)W^{-1}\|\|W W_k^{-1}\|
+\xi(k)\|W^{-1}SW^{-1}\|
\end{align*}
which clearly tends to zero if $k\to0$. The second part of condition
2 of Theorem \ref{th:yzintr} follows by a similar argument.
\qed

Th following algorithm summarizes the preceding construction of
Hamiltonians affiliated to $\rc$. \label{p:algor}

\begin{compactenum}
\item[(a)] For each $X$ we choose a kinetic energy operator
  $K_X=h_X(P)$ for the system having $X$ as configuration space. The
  function $h_X:X^*\to\mbr$ must be continuous and equivalent to a
  regular weight, in particular $|h_X(x)|\to\infty$ if $k\to\infty$.
  The equivalence to a weight is not an important assumption, it
  just allows us to consider below quite singular interactions
  $I$. If $\cs$ is infinite, we also require
  $\lim_X\inf_k|h_X(k)|=\infty$. This assumption is similar to the
  non-zero mass condition in quantum field theory models.

\item[(b)] The total kinetic energy of the system will be
  $K=\oplus_X K_X$. We denote $\cg=D(|K|^{1/2})$ its form domain
  equipped with the norm $\|u\|_\cg=\|\jap{K}^{1/2}u\|$ and observe
  that $\cg=\oplus_X\cg(X)$ Hilbert direct sum, where
  $\cg(X)=D(|K_X|^{1/2})$ is similarly related to $K_X$.  It is
  convenient to introduce the following topological vector spaces:
\[
\cg_\rmo=\ooplus_X^{\text{alg}}\cg(X), \quad
\cg_\rmo^*=\pprod_X\cg^*(X).
\]
$\cg_\rmo$ is an algebraic direct sum equipped with the inductive
limit topology and $\cg_\rmo^*$ is its adjoint space, direct product
of the adjoint spaces. $\cg_\rmo$ is a dense subspace of $\cg$ and
it has the advantage that its topology does not change if we replace
the norms on $\cg(X)$ by equivalent norms. 

\item[(c)] For each $Z\in\cs$ and for each couple $X,Y\in\cs$ such
  that $X\cap Y\supset Z$ let $I_{XY}(Z)$ be a continuous map
  $\cg(Y)\to\cg^*(X)$ such that the conditions of Proposition
  \ref{pr:tex} are fulfilled.  We require $I_{XY}(Z)^*=I_{YX}(Z)$
  and set $I_{XY}(Z)=0$ if $Z\not\subset X\cap Y$.

\item[(d)] The matrix $I(Z)=(I_{XY}(Z))_{X,Y\in\cs}$ can be realized
  as a continuous linear operator $\cg_\rmo\to\cg_\rmo^*$. We shall
  require that this be the restriction of a continuous map
  $I(Z):\cg\to\cg^*$.  Equivalently, the sesquilinear form
  associated to $I(Z)$ should be continuous for the $\cg$
  topology. We also require that $I(Z)$ be norm limit in
  $L(\cg,\cg^*)$ of its finite sub-matrices $\Pi_\ct
  I(Z)\Pi_\ct=(I_{XY}(Z))_{X,Y\in\ct}$, with notations as in
  \eqref{eq:tt}.

\item[(e)] Finally, we assume that there are real positive numbers
  $\mu_Z$ and $a$ with $\sum_Z\mu_Z<1$ and such that either $\pm
  I(Z) \leq\mu_Z|K+ia|$ for all $Z$ or $K$ is bounded from below and
  $ I(Z) \geq -\mu_Z|K+ia|$ for all $Z$. Furthermore, the series
  $\sum_E I(E)\equiv I$ should be norm summable in $L(\cg,\cg^*)$.
\end{compactenum}

We note that condition (i) of Theorem \ref{th:afi} will be satisfied
for all $\alpha >1/2$. Indeed, from Proposition \ref{pr:tex} it
follows that $\jap{K}^{-\alpha}\Pi_\ct
I(Z)\Pi_\ct\jap{K}^{-1/2}\in\rc(Z)$ for any finite $\ct$ and this
operator converges in norm to $\jap{K}^{-\alpha}
I(Z)\jap{K}^{-1/2}$.

Thus all conditions of Theorem \ref{th:afi} are fulfilled by the
Hamiltonian $H=K+I$ and so $H$ is strictly affiliated to $\rc$ and
its essential spectrum is given by
\begin{equation}\label{eq:ess4}
\spe(H)=\overline{\ccup}_{E\in\cp(\cs)}\sp(H_{\geq E}), \quad
\text{where }
H_{\geq E}=K_{\geq E}+{\textstyle\sum_{F\geq E}} I(F).
\end{equation}

\section{The Euclidean case} 
\label{s:morre}
\protect\setcounter{equation}{0}

In this section $\cs$ will be a set of finite dimensional vector
subspaces of a real prehilbert space which is stable under finite
intersections and such that for each pair $X,Y\in\cs$ there is
$Z\in\cs$ which contains both $X$ and $Y$. The ``ambient space'',
i.e. the prehilbert space in which the elements of $\cs$ are
embedded, does not really play a role in what follows so we shall
not need a notation for it.

It is interesting however to note that if $\cx$ is a real prehilbert
space then by taking in our construction from \S\ref{s:grass} the
semilattice $\cs$ equal to the set of all finite dimensional
subspaces of $\cx$ we canonically associate to $\cx$ a $C^*$-algebra
$\rc$. But if $\cx$ is finite dimensional then we may naturally
associate to it two $C^*$-algebras, namely $\rc_\cx$ and its second
quantization $\rc=\rc^\#_\cx$, cf. Definition \ref{df:sc}.

Since each $X\in\cs$ is an Euclidean space we have a canonical
identification $X^*=X$. Note that if $Y\subset X$ the notation
$Y^\perp$ is slightly ambiguous because we did not indicate if the
orthogonal is taken in the ambient prehilbert space or relatively to
$X$. To be precise we shall denote $X/Y$ the orthogonal of $Y$ in
$X$, and this is coherent with our previous notations. Thus
\begin{equation}\label{eq:xort}
X/Y=X\ominus Y=X\cap Y^\perp \quad \text{for} \quad  Y\subset X,
\quad \text{hence} \quad X=Y\oplus (X/Y).
\end{equation}
We choose the Euclidean measures as Haar measures, so that
\begin{equation}\label{eq:euclidtens}
\ch(X)=\ch(Y)\otimes\ch(X/Y) \quad \text{if} \quad Y\subset X.
\end{equation}
For arbitrary $X,Y$ the relation \eqref{eq:ma1} holds and so we set
\begin{equation}\label{eq:xyet}
X/Y=X/(X\cap Y)= X\ominus(X\cap Y).
\end{equation}
Now let $X,Y,Z\in\cs$ with $Z\subset X\cap Y$. Then we have
$X=Z\oplus (X/Z)$ and $Y=Y\oplus (Y/Z)$ so
\begin{equation}\label{eq:xyzet}
\ch(X)=\ch(Z)\otimes\ch(X/Z) \quad\text{and}\quad 
\ch(Y)=\ch(Z)\otimes\ch(Y/Z).
\end{equation}
Proposition \ref{pr:def3} gives now relatively to these tensor
decompositions: 
\begin{equation}\label{eq:xyzetens}
\rc_{XY}(Z)=\cc^*(Z)\otimes \rk_{X/Z,Y/Z}\cong
\Co(Z^*;\rk_{X/Z,Y/Z}).
\end{equation}
We have written $Z^*$ above in spite of the canonical isomorphism
$Z^*\cong Z$ in order to stress that we have functions of momentum
not of position.  
Since 
\[
X/Z=X/(X\cap Y)\oplus (X\cap Y)/Z = X/Y\oplus (X\cap Y)/Z 
\]
and similarly for $Y/Z$ we get by using \eqref{eq:comtens} the finer
factorization:
\begin{equation}\label{eq:xyzeptens}
\rc_{XY}(Z)=\cc^*(Z)\otimes \rk_{(X\cap Y)/Z}
\otimes \rk_{X/Y,Y/X}. 
\end{equation}
Then from Proposition \ref{pr:xytens} we obtain
\begin{equation}\label{eq:xyetens}
\rc_{XY}=\rc_{X\cap Y}\otimes \rk_{X/Y,Y/X}
\end{equation}
tensor product of Hilbert modules or relatively to the tensor
factorizations 
\begin{equation}\label{eq:xyeht}
\ch(X)=\ch(X\cap Y)\otimes\ch(X/Y) \quad\text{and}\quad 
\ch(Y)=\ch(X\cap Y)\otimes\ch(Y/X).
\end{equation}
In the special cases $Y\subset X$ we have 
\begin{equation}\label{eq:xby}
\rc_{XY}=\rc_{Y}\otimes \rk_{X/Y,O}=\rc_{Y}\otimes\ch(X/Y)
\end{equation}
and if $Z\subset Y\subset X$ then
\begin{equation}\label{eq:xbybz}
\rc_{XY}(Z)=\cc^*(Z)\otimes \rk_{Y/Z} \otimes \ch(X/ Y)
\end{equation}
where all the tensor products are in the category of Hilbert
modules. 

Theorem \ref{th:yzintr} can be improved in the present context. Note
that $V_k$ is the operator of multiplication by the function
$x\mapsto\rme^{i\braket{x}{k}}$ where the scalar product
$\braket{x}{k}$ is well defined for any $x,k$ in the ambient space.

\begin{theorem}\label{th:xyzeintr}
$\rc_{XY}(Z)$ is the set of $T\in\rl_{XY}$ satisfying:
\begin{enumerate} 
\item 
$U_z^*T U_z=T$ for $z\in Z$ and
$\|V^*_z T V_z-T\|\to 0$ if $z\to 0$ in $Z$,
\item 
$\|T(U_y-1)\|\to 0$ if $y\to 0$ in $Y$ and $\|T(V_k-1)\|\to 0$ if
$k\to 0$ in $Y/Z$.
\end{enumerate}
\end{theorem}

\begin{remark}\label{re:xyzeintr}
Condition 2 may be replaced by
\begin{compactenum}
\item[3.] 
$\|(U_x-1)T\|\to 0$ if $x\to 0$ in $X$ and $\|(V_k-1)T\|\to 0$ if
$k\to 0$ in $X/Z$.
\end{compactenum}
This will be clear from the next proof.
\end{remark}
\proof Let $\cf\equiv\cf_Z$ be the Fourier transformation in the
space $Z$, this is a unitary operator in the space $L^2(Z)$ which
interchanges the position and momentum observables $Q_Z,P_Z$. We
denote also by $\cf$ the operators $\cf\otimes1_{\ch(X/Z)}$ and
$\cf\otimes1_{\ch(Y/Z)}$ which are unitary operators in the spaces
$\ch(X)$ and $\ch(Y)$ due to \eqref{eq:xyzet}.  If $S=\cf T
\cf^{-1}$ then $S$ satisfies the following conditions:
\begin{enumerate} 
\item[(i)]
$V_z^*S V_z=S$ for $z\in Z$, $\|S(V_z-1)\|\to 0$ if $z\to 0$ in $Z$,
and $\|U_z S U^*_z-S\|\to 0$ if $z\to 0$ in $Z$;
\item[(ii)] 
$\|S(U_y-1)\|\to 0$ and $\|S(V_y-1)\|\to 0$ if $y\to 0$ in $Y/Z$.
\end{enumerate}
For the proof, observe that the first part of condition 2 may be
written as the conjunction of the two relations $\|T(U_z-1)\|\to 0$
if $z\to 0$ in $Z$ and $\|T(U_y-1)\|\to 0$ if $y\to 0$ in $Y/Z$.  We
shall work in the representations
\begin{equation}\label{eq:fiber}
\ch(X)= L^2(Z;\ch(X/Z)) \quad \text{and} \quad
\ch(Y)= L^2(Z;\ch(Y/Z))
\end{equation} 
which are versions of \eqref{eq:xyzet}. Then from $V_z^*S V_z=S$ for
$z\in Z$ it follows that there is a bounded weakly measurable
function $S(\cdot):Z\to\rl_{X/Z,Y/Z}$ such that in the
representations \eqref{eq:fiber} $S$ is the operator of
multiplication by $S(\cdot)$. Then $\|U_z S U^*_z-S\|\to 0$ if $z\to
0$ in $Z$ means that the function $S(\cdot)$ is uniformly
continuous. And $\|S(V_z-1)\|\to 0$ if $z\to 0$ in $Z$ is equivalent
to the fact that $S(\cdot)$ tends to zero at infinity. Thus we see
that $S(\cdot)\in\Co(Z;\rl_{X/Z,Y/Z})$. 

The condition (ii) is clearly equivalent to
\[
\sup_{z\in Z}\big(\|S(z)(U_y-1)\|+ \|S(z)(V_y-1)\|\big)\to 0
\quad \text{if } y\to 0 \text{ in } Y/Z.
\]
From the Riesz-Kolmogorov theorem (cf. the presentation on
\cite{GI3}) it follows that each $S(z)$ is a compact operator. This
clearly implies 
\[
\|(U_x-1)S(z)\|+ \|(V_x-1)S(z)\|\to 0 
\quad \text{if } x\to 0 \text{ in } X/Z
\]
for each $z\in Z$. Since $S(\cdot)$ is continuous and tends to zero
at infinity, for each $\varepsilon>0$ there are points
$z_1,\dots,z_n\in Z$ and complex functions
$\varphi_1,\dots,\varphi_n\in\Cc(Z)$ such that
\[
\|S(z)-{\textstyle\sum_k}\varphi_k(z)S(z_k)\|\leq\varepsilon\quad
\forall z\in Z.
\]
This proves \eqref{eq:xyzetens} from which one may deduce our
initial description of $\rc_{XY}(Z)$. However, we prefer to get it
as a consequence of Theorem \ref{th:yzintr}. First, from the
preceding relation we obtain
\[
\sup_{z\in Z}\big(\|(U_x-1)S(z)\|+ \|(V_x-1)S(z)\|\big)\to 0 
\quad \text{if } x\to 0 \text{ in } X/Z.
\]
Now going back through this argument we see that if $T$ satisfies
the conditions of the theorem then it satisfies the stronger
conditions
\begin{enumerate} 
\item[(a)]
$U_z^*T U_z=T$ for $z\in Z$ and
$\|V^*_k T V_k-T\|\to 0$ if $k\to 0$ in $Z$,
\item[(b)]
$\|(U_x-1)T\|\to 0$ if $x\to 0$ in $X$ and 
$\|T(U_y-1)\|\to 0$ if $y\to 0$ in $Y$,
\item[(c)]
$\|(V_k-1)T\|\to 0$ if $k\to 0$ in $X/Z$ and
$\|T(V_k-1)\|\to 0$ if $k\to 0$ in $Y/Z$.
\end{enumerate}
Finally, we show that the conditions of Theorem \ref{th:yzintr} are
fulfilled.  Due to \eqref{eq:cond2a} we have only to discuss the
condition $\|V_k^*TV_k-T\|\to0$ as $k\to0$ in $G^*$.  We write this
as $V_kT\sim TV_k$ and use similar abbreviations below.  We may take
$G=X+Y$ and since $X+Y$ is a quotient of $X\oplus Y$ this condition
is equivalent to $V_{p+q}T\sim TV_{p+q}$ as $p\to0$ in $X$ and
$q\to0$ in $Y$. Since $X=Z\oplus X/Z$ and $Y=Z\oplus Y/Z$ we may
take $p=z+x$ and $q=z'+y$ with $z,z'\in Z$ and $x\in X/Z$, $y\in
Y/Z$ and make $x,y,z,z'$ tend to zero. Then $V_p=V_z V_x$ and
$V_q=V_{z'}V_y$ and since conditions (a) and (c) are satisfied we
have
\[
V_{p+q}T=V_xV_yV_{z+z'}T \sim V_yV_xTV_{z+z'}\sim V_yTV_{z+z'}.
\]
Let $\pi,\pi',\pi''$ be the orthogonal projections of $X+Y$ onto
$X,Z,X/Z$ respectively, so that $\pi=\pi'+\pi''$. Then for $y\in Y/Z$
we have $\pi'y=0$ hence for $x\in X$ we have
$\braket{x}{y}=\braket{\pi x}{y}=\braket{x}{\pi y}=\braket{x}{\pi''
  y}$. Since for $y\to0$ in $Y/Z$ we have $\pi''y\to0$ in $X/Z$ by
using again the first part of condition (c) we get
\[
V_yTV_{z+z'}=V_{\pi''y}TV_{z+z'}\sim TV_{z+z'}.
\]
A similar argument gives $TV_{z+z'}\sim TV_xV_yV_{z+z'}=TV_pV_q$ which
finishes the proof.
\qed

We shall present below a Sobolev space version of Proposition
\ref{pr:tex} which uses the class of weights $\jap{\cdot}^s$ and is
convenient in applications. For each real $s$ let $\ch^s(X)$ be the
Sobolev space defined by the norm
\[
\|u\|_{\ch^s}=\|\jap{P}^s u\| =\|(1+\Delta_X)^s/2u\|
\] 
where $\Delta_X$ is the (positive) Laplacian associated to the
Euclidean space $X$. The space $\ch^s(X)$ is equipped with two
continuous representations of $X$, a unitary one induced by
$\{U_x\}_{x\in X}$ and a non-unitary one induced by $\{V_x\}_{x\in
X}$. This gives us a weighted Sobolev-Besov scale $\ch^s_{t,p}$,
cf. Chapter 4 in \cite{ABG}.  Let
\begin{equation}\label{eq:sob}
\rl^{s,t}_{XY}=L(\ch^t(Y),\ch^{-s}(X)) \quad
\text{with norm} \quad \|\cdot\|_{s,t}.
\end{equation}
We mention a compactness criterion which follows from the
Riesz-Kolmogorov theorem and the argument page \pageref{p:RK}
involving the regularity of the weight.

\begin{proposition}\label{pr:stsob}
If $s,t\in\mbr$ and $T\in\rl^{s,t}_{XY}$ then $T$ is compact if and
only if one of the next two equivalent conditions is satisfied:
\begin{compactenum}
\item[(i)]
$\|(U_x-1)T\|_{s,t} + \|(V_x-1)T\|_{s,t}\to 0 \quad 
\text{if } x\to 0 \text{ in }X$,
\item[(ii)]
$\|T(U_y-1)\|_{s,t} + \|T(V_y-1)\|_{s,t}\to 0 \quad 
\text{if } y\to 0 \text{ in }Y$.
\end{compactenum}
\end{proposition}

The next result follows from Proposition \ref{pr:tex} or directly
from Theorem \ref{th:xyzeintr}.

\begin{proposition}\label{pr:etex}
Let $s,t>0$ and $Z\subset X\cap Y$. Let $I_{XY}(Z)\in\rl^{s,t}_{XY}$
such that the following relations hold in norm in
$\rl^{s,t+\varepsilon}_{XY}$ for some $\varepsilon\geq 0$:
\begin{enumerate}
\item
$U_z I_{XY}(Z)=I_{XY}(Z) U_z$ if $z\in Z$ and
$V^*_z I_{XY}(Z) V_z\to I_{XY}(Z)$ if $z\to 0$ in $Z$,
\item 
$I_{XY}(Z)(U_y-1)\to 0$ if $y\to 0$ in $Y$ and
$I_{XY}(Z)(V_k-1)\to 0$ if $k\to 0$ in $Y/Z$.
\end{enumerate}
If $h_X,h_Y$ are continuous real functions on $X,Y$ such that
$h_X(x)\sim\jap{x}^{2s}$ and $h_Y(y)\sim\jap{y}^{2t}$ and if we set
$K_X=h_X(P), K_Y=h_Y(P)$ then
$\jap{K_X}^{-\alpha}I_{XY}(Z)\jap{K_Y}^{-1/2}\in\rc_{XY}(Z)$ if
$\alpha>1/2$.
\end{proposition}

To give a more detailed description of $I_{XY}(Z)$ we make a Fourier
transformation $\cf_Z$ in the $Z$ variable as in the proof of
Theorem \ref{th:xyzeintr}. We have $X=Z\oplus (X/Z)$ so
$\ch(X)=\ch(Z)\otimes\ch(X/Z)$ and $\Delta_X=\Delta_Z\otimes 1 +
1\otimes \Delta_{X/Z}$. Thus if $t\geq0$
\begin{equation}\label{eq:stens}
\ch^t(X)=\ch(Z;\ch^t(X/Z))\cap \ch^t(Z;\ch(X/Z))=
\big(\ch(Z)\otimes\ch^t(X/Z)\big)\cap 
\big(\ch^t(Z)\otimes\ch(X/Z)\big)
\end{equation}
where our notations are extended to  vector-valued Sobolev spaces. 
Clearly
\begin{equation}\label{eq:lap}
\cf_Z \jap{P_Z}^t \cf_Z^{-1} = 
\int_Z^\oplus (1+|k|^2+|P_{X/Z}|^2)^{t/2} \rmd k.
\end{equation}
We introduce now a class of operators which tend weakly to zero as
$x\to\infty$:
\begin{equation}\label{eq:Lo}
\hat\rl^{s,t}_{XY}=\{T\in L(\ch^t(Y),\ch^{-s}(X)) \mid 
T:\ch^t(Y) \to \ch^{-s-\varepsilon }(X) 
\text{ is compact if } \varepsilon>0\}.
\end{equation} 
If $s=t$ we set $\hat\rl^{s,t}_{XY}=\hat\rl^s_{XY}$.
Note that if the compactness condition holds for one $\varepsilon>0$
then it holds for all $\varepsilon>0$. 
Thus the first part of condition (i) of Proposition \ref{pr:stsob} is
automatically satisfied, hence
\begin{equation}\label{eq:Loo}
\hat\rl^{s,t}_{XY}=\{T\in\rl^{s,t}_{XY} \mid 
\|(V_x-1)T\|_{s+\varepsilon,t}\to 0 \quad
\text{if } x\to 0 \text{ in }X \}.
\end{equation} 
Now we proceed as in the proof of Theorem \ref{th:xyzeintr} and work
in the representations \eqref{eq:fiber}. We define
\begin{equation}\label{eq:J}
\cf_Z I_{XY}(Z) \cf_Z^{-1} \equiv \int_Z^\oplus I_{XY}^Z(k) \rmd k
\end{equation}
where $I_{XY}^Z:Z\to \hat\rl^{s,t}_{X/Z,Y/Z}$ is a continuous
operator valued function satisfying
\begin{equation}\label{eq:Jest}
\sup\nolimits_k
\|(1+|k|+|P_{X/Z}|)^{-s}I_{XY}^Z(k)(1+|k|+|P_{Y/Z}|)^{-t}\| <\infty.
\end{equation} 
In $N$-body type situations such conditions have been introduced in
\cite{DG2} and in Section 4 of \cite{DG3} and we refer to these
papers for some examples of physical interest. We mention that if we
take $\varepsilon=0$ in \eqref{eq:Lo} then we obtain interactions
which have relatively compact fibers $J(k)$. But in \eqref{eq:Loo}
we may take $\varepsilon=0$ and still get a very large class of
singular interactions. For example, if $a_{jk}$ are bounded
measurable functions on $X$ such that $\int_{|x-y|<1}|a_{jk}(x)|
\rmd x\to0$ when $y\to\infty$ then $\sum\partial_j
a_{jk}\partial_k\in\hat\rl^1_{XX}$ will be an admissible
perturbation of $\Delta$.

In order to take advantage of the Euclidean setting the algorithm
for the construction of Hamiltonians affiliated to $\rc$ described
on page \pageref{p:algor} should be modified by adding to the first
three steps the following: 
\begin{compactenum}

\item[(a)] The $h_X$ are functions on $X$ and we assume that
  $a_X\jap{x}^{2s_X} \leq |h_X(x)| \leq b_X\jap{x}^{2s_X}$ for some
  strictly positive real numbers $s_X$ and all large $x$.

\item[(b)] We take $\cg(X)=\ch^{s_X}(X)$.

\item[(c)] The $I_{XY}(Z)$ are continuous maps
  $\ch^{s_Y}(Y)\to\ch^{-s_X}(X)$ such that the conditions of
  Proposition \ref{pr:etex} are fulfilled with $s=s_Y$ and $t=s_X$.

\end{compactenum}

\section{Non relativistic Hamiltonians and the Mourre estimate} 
\label{s:mou}
\protect\setcounter{equation}{0}

\PAR\label{ss:mou}
Assume that $\cs$ is an inductive semilattice of finite dimensional
vector subspaces of a real vector space (then $\cs$ has non-compact
quotients). This means that $\cs$ is a set of finite dimensional
vector subspaces of a real vector space which is stable under finite
intersections and such that for each pair $X,Y\in\cs$ there is
$Z\in\cs$ which contains both $X$ and $Y$. Then dilations implement
a group of automorphisms of the $C^*$-algebra $\rc$ which is
compatible with the grading, i.e. it leaves invariant each component
$\rc(E)$ of $\rc$. To be precise, for each real $\tau$ let $W_\tau$
be the unitary operator in $\ch(X)$ defined by
\begin{equation}\label{eq:dil}
(W_\tau u)(x)=\rme^{n\tau/4}u\big(\rme^{\tau/2} x\big)
\end{equation}
where $n$ is the dimension of $X$. The unusual normalization is
convenient for non-relativistic operators. As in the case of the
operators $U_x$ and $V_k$ we shall not specify the space $X$ in the
notation of $W_\tau$. Moreover, we denote by the same symbol the
unitary operator $\bigoplus_X W_\tau$ on the direct sum
$\ch=\bigoplus_X\ch(X)$. Then it is clear that
$W_\tau^*\rc_{XY}(Z)W_\tau=\rc_{XY}(Z)$ for all $X,Y,Z$,
cf. \eqref{eq:ryz}. Let $D$ be the infinitesimal generator of
$\{W_\tau\}$, so $D$ is a self-adjoint operator such that
$W_\tau=\rme^{i\tau D}$. Formally
\begin{equation}\label{eq:fedex}
2iD_X=x\cdot\nabla_x+n/2= \nabla_x\cdot x-n/2 \quad 
\text{if $n$ is the dimension of } X.
\end{equation}
This structure allows one to prove the Mourre estimate for operators
affiliated to $\rc$ in a systematic way as shown in \cite{ABG,BG2}
in an abstract setting under the assumption that $\cs$ is
finite. This procedure has been extended in \cite{DG2} to the case
when $\cs$ is infinite and applied there to a class of dispersive
$N$-body type systems: more precisely, $\cs$ is allowed to be
infinite but the ambient space is finite dimensional.

For simplicity and since here we are mainly interested in
non-relativistic many-body systems we shall restrict ourselves to the
case when $\cs$ is a finite semilattice of subspaces of a finite
dimensional real prehilbert space. In fact, the extension of the
techniques of \cite{DG2} to the case when both $\cs$ and the ambient
space are infinite is rather straightforward but the condition
\eqref{eq:kin} is quite annoying in the non-relativistic case: we
should replace $\Delta_X$ by $\Delta_X+E_X$ where $E_X$ is a number
which tends to infinity with $X$, which is a rather artificial
procedure. On the other hand, we do not have satisfactory results in
the general case due to the well-known problem of dispersive $N$-body
Hamiltonians \cite{De1,Ger1,DG2}. We note that the quantum field case
is much easier from this point of view because of the special nature
of the interactions.  This is especially clear from the treatments in
\cite{Ger2,Geo}, but see also \cite{DeG2}. 

\PAR\label{ss:hvz} Thus from now on in this section \emph{$\cs$ is a
  finite set of subspaces of an Euclidean space such that if
  $X,Y\in\cs$ then $X\cap Y\in\cs$ and there is $Z\in\cs$ such that
  $X\cup Y\subset Z$}. As we noticed in the Remark \ref{re:fmax},
$\cs$ will have a largest element, but this space will not play a
special role in our arguments so it does not deserve to be named. On
the other hand, $\cs$ has a least element and is atomic.

We first point out a particular case of our preceding results which
is of interest in this section. Let us fix $s>0$ and for each
$X\in\cs$ let $h_X:X\to\mbr$ be a \emph{positive} continuous
function such that $h_X(k)\sim\jap{k}^{2s}$.  Recall that we denote
$K_X=h_X(P)$ and that the kinetic energy operator is $K=\oplus_X
K_X$ with form domain $\cg=\oplus_X\ch^s(X)$.  
In the next proposition we use the the embeddings
\begin{equation}\label{eq:sotens}
\ch^s(X)\subset\ch(Z)\otimes\ch^s(X/Z)\subset \ch(X)
\subset\ch(Z)\otimes\ch^{-s}(X/Z)\subset \ch^{-s}(X)
\end{equation}
which follow from \eqref{eq:stens}. Then if
$I^Z_{XY}:\ch^s(Y/Z)\to\ch^{-s}(X/Z)$ is a continuous operator
we may define $I_{XY}(Z)=1\otimes I^Z_{XY}$ which induces
a continuous operator $\ch^s(Y)\to\ch^{-s}(X)$.

\begin{proposition}\label{pr:ehvz}
For each $X,Y,Z\in\cs$ such that $Z\subset X\cap Y$ let
$I^Z_{XY}\in\hat\rl^s_{X/Z,Y/Z}$ with $(I^Z_{XY})^*=I^Z_{YX}$
and let $I_{XY}(Z)=1\otimes I^Z_{XY}$. Let $I_{XY}(Z)=0$ if
$Z\not\subset X\cap Y$.  We set $I(Z)=(I_{XY}(Z))_{X,Y\in\cs}$ and
assume that there are positive numbers $\mu_Z$ and $a$ with
$\sum_Z\mu_Z<1$ and such that $ I(Z) \geq -\mu_Z|K+ia|$ for all
$Z$. Let $I=\sum I(Z)$ and $I_{\geq E}=\sum_{Z\supset E}I(Z)$.  Then
the form sum $H=K+I$ is a self-adjoint operator strictly affiliated
to $\rc$, we have $\rp_{\geq X}H=K+I_{\geq X}\equiv H_{\geq X}$, and
\begin{equation}\label{eq:ess5}
\spe(H)=\ccup_{X\in\cp(\cs)}\sp(H_{\geq X}).
\end{equation}
\end{proposition}

This follows immediately from Proposition \ref{pr:etex}, the
discussion after it, and Theorem \ref{th:afi} (see page
\pageref{p:algor}).

We shall now restrict ourselves to the non-relativistic case,
cf. Definition \ref{df:NR}. In particular, in Proposition
\ref{pr:ehvz} we must take $h_X=\|k\|^2$ and $s=1$.  Then $\Delta_X$
is the (positive) Laplacian associated to the Euclidean space $X$
with the convention $\Delta_O=0$.  In order to point out a special
structure that have the Hamiltonians $H_{\geq E}$ we need to revert
to the more precise notations $\rc=\rc_\cs$ and $\ch=\ch_\cs$. We
also set $\Delta_\cs\equiv K=\oplus_X \Delta_X$, denote $I_\cs(Z)$
and $I_\cs$ the interaction terms $I(Z)$ and $I$ constructed as in
Proposition \ref{pr:ehvz}, and set $H_\cs=H$.

Let us assume that $\cs$ has a smallest element $E$. Then
\eqref{eq:xyzetens} implies for all $Z\subset X\cap Y$
\begin{equation}\label{eq:facte1}
\rc_{XY}(Z)=\cc^*(Z)\otimes \rk_{X/Z,Y/Z}
=\cc^*(E)\otimes \cc^*(Z/E)\otimes \rk_{X/Z,Y/Z}.
\end{equation}
Moreover, we have $\ch(X)=\ch(E)\otimes\ch(X/E)$ for all $X\in\cs$
hence 
\begin{equation}\label{eq:facte}
\ch_\cs=\oplus_X\ch(X)=\ch(E)\otimes\big(\oplus_X\ch(X/E)\big).
\end{equation}
We denote by $\cs/E$ the set of subspaces $X/E=X\cap E^\perp$, this
is clearly an inductive semilattice of finite dimensional subspaces
of the ambient space which contains $O=\{0\}$. Thus we can associate
to $\cs/E$ an algebra $\rc_{\cs/E}$ which acts on the Hilbert space
$\ch_{\cs/E}=\oplus_X\ch(X/E)$. From \eqref{eq:facte1} and
\eqref{eq:facte} we get
\begin{equation}\label{eq:fact}
\rc_\cs=\cc^*(E)\otimes\rc_{\cs/E} \quad \text{and} \quad
\ch_\cs=\ch(E)\otimes\ch_{\cs/E}.
\end{equation}
Then we have
\begin{equation}\label{eq:dcs}
\Delta_X=\Delta_E\otimes 1 +1\otimes\Delta_{X/E} 
\quad \text{hence we get} \quad
\Delta_\cs=\Delta_E\otimes 1 +1\otimes\Delta_{\cs/E}.
\end{equation}
Since $Z\supset E$ for all $Z\in\cs$ we may
write\symbolfootnote[2]{\ 
We shall not use the natural but excessive notation
$I^{Z/E}_{X/E,Y/E}$. }
$I_{XY}(Z)=1_E \otimes 1_{Z/E} \otimes I^Z_{XY}$ where $1_E$
for example is the identity operator on $\ch(E)$. Hence we get 
$I_\cs(Z)=1\otimes I_{\cs/E}(Z)$ and $I_\cs=1\otimes I_{\cs/E}$
the tensor products being relative to the factorization
\eqref{eq:fact}. Finally we get
\begin{equation}\label{eq:tdec}
H_\cs=\Delta_E\otimes1+ 1\otimes H_{\cs/E} \quad
\text{if $E$ is the smallest element of $\cs$}. 
\end{equation}
We shall apply these remarks to the sub-semilattice $\cs_{\geq E}$
of $\cs$ for some $E\in\cs$. Then:
\[
\rc_{\cs_{\geq E}}=\rc_{\geq E}, \quad 
\ch_{\cs_{\geq E}}=\ch_{\geq E}, \quad
H_{\cs_{\geq E}}=H_{\geq E}
\]
with our old notations. We extend the preceding definition of
$\cs/E$ and for an arbitrary $E\in\cs$ we denote by $\cs/E$ the set
of subspaces $X/E$ where $X$ runs over $\cs$ with the condition
$X\supset E$.  Thus we get
\begin{equation}\label{eq:edec}
\ch_{\geq E}=\ch(E)\otimes\ch_{\cs/E}, \quad
\rc_{\geq E}=\cc^*(E)\otimes\rc_{\cs/E}, \quad
H_{\geq E}=\Delta_E\otimes1+ 1\otimes H_{\cs/E}.
\end{equation}
Let us denote $\tau_E=\min H_{\cs/E}$ the bottom of the spectrum of
$H_{\cs/E}$.  From the last relation we get
\begin{equation}\label{eq:spe}
\sp(H_{\geq E})=[0,\infty) + \sp(H_{\cs/E})=
[\tau_E,\infty) \quad
\text{if } E\neq O
\end{equation}
and then \eqref{eq:ess5} implies:

\begin{corollary}\label{co:hvznr}
Under the conditions of Proposition \ref{pr:ehvz} and if we are in
the non-relativistic case then we have $\spe(H)=[\tau,\infty)$ with
$\tau=\min_{E\in\cp(\cs)} \tau_E$ where $\tau_E=\min H_{\cs/E}$.
\end{corollary}

\PAR\label{ss:mest} We shall now define the threshold set and prove
the Mourre estimate outside it for $H=H_\cs$. The strategy of our
proof is that introduced in \cite{BG2} and further developed in
\cite{ABG,DG2}. The case of graded $C^*$-algebras over infinite
semilattices and of dispersive Hamiltonians is treated in Section 5
from \cite{DG2}.  We choose the generator $D$ of the dilation group
$W_\tau$ in $\ch$ as conjugate operator. For special type of
interactions, e.g. of quantum field type, which are allowed by our
formalism and are physically interesting, much better choices can be
made, but technically speaking there is nothing new in that with
respect to \cite{Geo}.

Form \eqref{eq:tdec} we see that we can restrict ourselves to the
case when $O\in\cs$ so we suppose this from now on. The properties
of the dilation group, cf. the beginning of \S\ref{ss:mou}, which
are important for us are: (i) $W^*_\tau\rc(Z)W_\tau\subset\rc(Z)$
for each $\tau$ and $Z$, and (ii) for each $T\in\rc$ the map
$\tau\mapsto W^*_\tau T W_\tau$ is norm continuous. The relation
\begin{equation}\label{eq:dlap}
W^*_\tau \Delta_X W_\tau = \rme^\tau\Delta \quad \text{or}\quad
[\Delta_X, i D]=\Delta_X 
\end{equation}
is not really important but it will allow us to make a very explicit
computation.

We say that a self-adjoint operator $H$ \emph{is of class $C^1(D)$}
or \emph{of class $C^1_\rmu(D)$} if $W^*_\tau RW_\tau$ as a function
of $\tau$ is of class $C^1$ strongly or in norm respectively.  Here
$R=(H-z)^{-1}$ for some $z$ outside the spectrum of $H$.  The formal
relation
\begin{equation}\label{eq:dres}
[D,R]= R[H,D] R 
\end{equation}
can be given a rigorous meaning as follows. If $H$ is of class
$C^1(D)$ then the intersection $\rd$ of the domains of the operators
$H$ and $D$ is dense in $D(H)$ and the sesquilinear form with domain
$\rd$ associated to the formal expression $HD-DH$ is continuous for
the topology of $D(H)$ so extends uniquely to a continuous
sesquilinear form on the domain of $H$ which is denoted
$[H,D]$. This defines the right hand side of \eqref{eq:dres}.  The
left hand side can be defined for example as 
$i\frac{d}{d\tau}W_\tau^*RW_\tau|_{\tau=0}$. 

For Hamiltonians as those considered here it is easy to decide that
$H$ is of class $C^1(D)$ in terms of properties of the commutator
$[H, D]$.  Moreover, the following is easy to prove: \emph{if $H$
  is affiliated to $\rc$ then $H$ is of class $C^1_\rmu(D)$ if and
  only if $H$ is of class $C^1(D)$ and $[R,D]\in\rc$}.

Let $H$ be of class $C^1(D)$ and $\lambda\in\mbr$. Then for each
$\theta\in\Cc(\mbr)$ with $\theta(\lambda)\neq0$ one may find a real
number $a$ and a compact operator $K$ such that 
\begin{equation}\label{eq:must}
\theta(H)^*[H,iD]\theta(H)\geq a|\theta(H)|^2+K.
\end{equation}

\begin{definition}\label{df:must}
The upper bound $\what\rho_H(\lambda)$ of the numbers $a$ for which
such an estimate holds is \emph{the best constant in the Mourre
  estimate for $H$ at $\lambda$}.  The \emph{threshold set} of $H$
(relative to $D$) is the closed real set
\begin{equation}\label{eq:thr0}
\tau(H)=\{\lambda \mid \what\rho_H(\lambda)\leq0\}
\end{equation}
One says that $D$ is \emph{conjugate to} $H$ at $\lambda$ if 
$\what\rho_H(\lambda)>0$. 
\end{definition}
The set $\tau(H)$ is closed because the function
$\what\rho_H:\mbr\to]-\infty,\infty]$ is lower semicontinuous.

The following notion will play an important role in our arguments: to
each closed real set $A$ we associate the function 
$N_A:\mbr\to[-\infty,\infty[$ defined by
\begin{equation}\label{eq:na}
N_A(\lambda)=\sup\{ x\in A \mid x\leq\lambda\}.
\end{equation}
We make the convention $\sup\emptyset=-\infty$. Thus $N_A$ may take
the value $-\infty$ if and only if $A$ is bounded from below and then
$N_A(\lambda)=-\infty$ if and only if $\lambda<\min A$. The function
$N_A$ is further discussed during the proof of Lemma \ref{lm:nab}.

The notion of non-relativistic many-body Hamiltonian has been
introduced in Definition \ref{df:NR}.  Recall that we assume $O\in\cs$
and that we denote $\mathrm{ev}(T)$ the set of eigenvalues of an
operator $T$.

\begin{theorem}\label{th:thr}
Let $H=H_\cs$ be a non-relativistic many-body Hamiltonian of class
$C^1_\rmu(D)$. Then
\begin{equation}\label{eq:thr}
\tau(H)=\ccup_{X\neq O}\mathrm{ev}(H_{\cs/X}).
\end{equation}
In particular $\tau(H)$ is a closed \emph{countable} real set. 
We have $\what\rho_H(\lambda)=\lambda-N_{\tau(H)}(\lambda)$ for all
real $\lambda$.
\end{theorem}
\proof We need a series of facts which are discussed in detail in
Sections 7.2, 8.3 and 8.4 from \cite{ABG} (see pages 51--61 in
\cite{BG2} for a shorter presentation).

\begin{compactenum}

\item[(i)] For each real $\lambda$ let $\rho_H(\lambda)$ be the
  upper bound of the numbers $a$ for which an estimate like
  \eqref{eq:must} but with $K=0$ holds. This defines a lower
  semicontinuous function $\rho_H:\mbr\to ]-\infty,\infty] $ hence
the set $\varkappa(H)=\{\lambda \mid \rho_H(\lambda)\leq 0\}$ is a
closed real set called \emph{critical set} of $H$ (relative to
$D$). We clearly have $\rho_H\leq\what\rho_H$ and so
$\tau(H)\subset\varkappa(H)$.

\item[(ii)] Let $\mu(H)$ be the set of eigenvalues of $H$ such that
  $\what\rho_H(\lambda)>0$. Then $\mu(H)$ is a discrete subset of
  $\mathrm{ev}(H)$ consisting of eigenvalues of finite
  multiplicity. This is essentially the virial theorem.

\item[(iii)] There is a simple and rather unexpected relation
  between the functions $\rho_H$ and $\what\rho_H$: they are
  ``almost'' equal. In fact, $\rho_H(\lambda)=0$ if
  $\lambda\in\mu(H)$ and $\rho_H(\lambda)=\what\rho_H(\lambda)$
  otherwise. In particular
\begin{equation}\label{eq:tev}
\varkappa(H)=\tau(H)\cup \mathrm{ev}(H)=\tau(H)\sqcup\mu(H)
\end{equation}
where $\sqcup$ denotes disjoint union. 

\item[(iv)] This step is easy but rather abstract and the
  $C^*$-algebra setting really comes into play. We assume that $H$
  is affiliated to our algebra $\rc$. The preceding arguments did
  not require more than the $C^1(D)$ class. Now we require $H$ to be
  of class $C^1_\rmu(D)$.  Then the operators $H_{\geq X}$ are also
  of class $C^1_\rmu(D)$ and we have the important relation (Theorem
  8.4.3 in \cite{ABG} or Theorem 4.4 in \cite{BG2})
\[
\what\rho_H=\min_{X\in\cp(\cs)}\rho_{H_{\geq X}}.
\]
To simplify notations we adopt the abbreviations $\rho_{H_{\geq
    X}}=\rho_{\geq X}$ and instead of $X\in\cp(\cs)$ we write
$X\gtrdot O$, which should be read ``$X$ covers $O$''. For coherence
with later notations we also set $\what\rho_H=\what\rho_\cs$.  So
\eqref{eq:mustq} may be written
\begin{equation}\label{eq:mustq}
\what\rho_\cs=\min_{X\gtrdot O}\rho_{\geq X}.
\end{equation}

\item[(v)] From \eqref{eq:dlap} and \eqref{eq:edec} we get
\[
H_{\geq X}=\Delta_X\otimes1+ 1\otimes H_{\cs/X}, \quad
[H_{\geq X},iD]=\Delta_X\otimes1+ 1\otimes [D,i H_{\cs/X}].
\]
Recall that we denote $D$ the generator of the dilation group
independently of the space in which it acts. We note that the formal
argument which gives the second relation above can easily be made
rigorous but this does not matter here. Indeed, since $H_{\geq X}$
is of class $C^1_\rmu(D)$ and by using the first relation above, one
can easily show that $H_{\cs/X}$ is also of class $C^1_\rmu(D)$ (see
the proof of Lemma 9.4.3 in \cite{ABG}). We do not enter into
details on this question because any reasonable conditions on the
interaction $I$ in Proposition \ref{pr:ehvz} which ensure that $H$
is of class $C^1_\rmu(D)$ will also imply that the $H_{\cs/X}$ are
of the same class. Anyway, we may use Theorem 8.3.6 from \cite{ABG}
to get
\[
\rho_{\geq X}(\lambda)=\inf_{\lambda_1+\lambda_2=\lambda}
\big(\rho_{\Delta_X}(\lambda_1) + \rho_{\cs/X}(\lambda_2) \big)
\]
where $\rho_{\cs/X}=\rho_{H_{\cs/X}}$. But clearly if $X\neq O$ we
have $\rho_{\Delta_X}(\lambda)=\infty$ if $\lambda<0$ and 
$\rho_{\Delta_X}(\lambda)=\lambda$ if $\lambda\geq0$. Thus we get
\begin{equation}\label{eq:mustt}
\rho_{\geq X}(\lambda)=\inf_{\mu\leq\lambda}
\big(\lambda-\mu + \rho_{\cs/X}(\mu) \big)
=\lambda-  \sup_{\mu\leq\lambda}\big(\mu - \rho_{\cs/X}(\mu) \big).
\end{equation}

\item[(vi)]
Now from \eqref{eq:mustq} and \eqref{eq:mustt} we get
\begin{equation}\label{eq:musr}
\lambda-\what\rho_\cs(\lambda)=
\max_{X\gtrdot O}\sup_{\mu\leq\lambda}\big(\mu - \rho_{\cs/X}(\mu)\big).
\end{equation}
\end{compactenum}

Finally, we prove the formula
$\what\rho_H(\lambda)=\lambda-N_{\tau(H)}(\lambda)$ from Theorem
\ref{th:thr} by induction over the semilattice $\cs$. In other
terms, we assume that the formula is correct if $H$ is replaced by
$H_{\cs/X}$ for all $X\neq O$ and we prove it for $H=H_{\cs/O}$. So
we have to show that the right hand side of \eqref{eq:musr} is equal
to $N_{\tau(H)}(\lambda)$.

According to step (iii) above we have $\rho_{\cs/X}(\mu)=0$ if
$\mu\in\mu(H_{\cs/X})$ and
$\rho_{\cs/X}(\mu)=\what\rho_{\cs/X}(\mu)$ otherwise. Since by the
explicit expression of $\what\rho_{\cs/X}$ this is a positive
function and since $\rho_H(\lambda)\leq0$ is always true if
$\lambda$ is an eigenvalue, we get $\mu-\rho_{\cs/X}(\mu)=\mu$ if
$\mu\in\mathrm{ev}(H_{\cs/X})$ and
\[
\mu-\rho_{\cs/X}(\mu)=\mu-\what\rho_{\cs/X}(\mu)=
N_{\tau(H_{\cs/X})}(\mu)
\]
otherwise. From the first part of Lemma \ref{lm:nab} below we get
\[
\sup_{\mu\leq\lambda}\big(\mu - \rho_{\cs/X}(\mu)\big)=
N_{\mathrm{ev}(H_{\cs/X}) \cup \tau(H_{\cs/X})}.
\]
If we use the second part of Lemma \ref{lm:nab} then we see that
\[
\max_{X\gtrdot O}\sup_{\mu\leq\lambda}\big(\mu -
\rho_{\cs/X}(\mu)\big)=
\max_{X\gtrdot O}N_{\mathrm{ev}(H_{\cs/X}) \cup \tau(H_{\cs/X})}
\]
is the $N$ function of the set 
\[
\bigcup_{X\gtrdot O}\big(\mathrm{ev}(H_{\cs/X}) 
\cup \tau(H_{\cs/X})\big)=
\bigcup_{X\gtrdot O}\left(\mathrm{ev}(H_{\cs/X}) 
\bigcup \bigcup_{Y>X}\mathrm{ev}(H_{\cs/Y})\right)=
\bigcup_{X>O}\mathrm{ev}(H_{\cs/X})
\]
which finishes the proof of
$\what\rho_H(\lambda)=\lambda-N_{\tau(H)}(\lambda)$ hence the proof
of the Theorem \ref{th:thr}. 
\qed

It remains, however, to show the following fact which was used above.

\begin{lemma}\label{lm:nab}
If $A$ and $A\cup B$ are closed and if $M$ is the function given by
$M(\mu)=N_A(\mu)$ for $\mu\nin B$ and $M(\mu)=\mu$ for $\mu\in B$
then $\sup_{\mu\leq\lambda}M(\mu)=N_{A\cup B}(\lambda)$.  If $A,B$
are closed then $\sup(N_A,N_B)=N_{A\cup B}$.
\end{lemma}
\proof The last assertion of the lemma is easy to check, we prove
the first one.  Observe first that the function $N_A$ has the
following properties:
\begin{compactenum}
\item[(i)]
$N_A$ is increasing and right-continuous,
\item[(ii)]
$N_A(\lambda)=\lambda$ if $\lambda\in A$,
\item[(iii)] $N_A$ is locally constant and $N_A(\lambda)<\lambda$ on
  $A^\rmc\equiv \mbr\setminus A$.
\end{compactenum} 
Indeed, the first assertion in (i) and assertion (ii) are obvious.
The second part of (i) follows from the more precise and easy to prove
fact 
\begin{equation}\label{eq:nae}
N_A(\lambda+\varepsilon)\leq N_A(\lambda)+\varepsilon \quad
\text{for all real } \lambda \text{ and } \varepsilon>0.
\end{equation}
A connected component of the open set $A^\rmc$ is necessarily an open
interval of one of the forms $]-\infty,y[$ or $]x,y[$ or $]x,\infty[$
with $x,y\in A$. On the first interval (if such an interval appears)
$N_A$ is equal to $-\infty$ and on the second one or the third one it
is clearly constant and equal to $N_A(x)$. We also note that the
function $N_A$ is characterized by the properties (i)--(iii).

Thus, if we denote $N(\lambda)=\sup_{\mu\leq\lambda}M(\mu)$, then it
will suffices to show that the function $N$ satisfies the conditions
(i)--(iii) with $A$ replace by $A\cup B$. Observe that $M(\mu)\leq\mu$
and the equality holds if and only if $\mu\in A\cup B$.  Thus $N$ is
increasing, $N(\lambda)\leq\lambda$, and $N(\lambda)=\lambda$ if
$\lambda\in A\cup B$.

Now assume that $\lambda$ belongs to a bounded connected component
$]x,y[ $ of $A\cup B$ (the unbounded case is easier to treat). If
$x<\mu<y$ then $\mu\nin B$ so $M(\mu)=N_A(\mu)$ and $]x,y[$ is
included in a connected component of $A^\rmc$ hence $M(\mu)=N_A(x)$.
Then $N(\lambda)= \max(\sup_{\nu\leq x}M(\nu),N_A(x))$ hence $N$ is 
constant on $]x,y[$. Here we have $M(\nu)\leq\nu\leq x$ so if $x\in A$
then $N_A(x)=x$ and we get $N(\lambda)=x$. If $x\in B\setminus A$ then
$M(x)=x$ so $\sup_{\nu\leq x}M(\nu)=x$ and $N_A(x)<x$ hence 
$N(\lambda)=x$. Since $x\in A\cup B$ one of these two cases is 
certainly realized and the same argument gives $N(x)=x$. Thus the 
value of $N$ on $]x,y[$ is $N(x)$ so $N$ is right continuous on 
$[x,y[$.  Thus we proved that $N$ is locally constant and right 
continuous on the complement of $A\cup B$ and also that 
$N(\lambda)<\lambda$ there.

It remains to be shown that $N$ is right continuous at each point of
$\lambda\in A\cup B$. We show that \eqref{eq:nae} hold with $N_A$
replaced by $N$. If $\mu\leq\lambda$ then 
$M(\mu)\leq\mu\leq\lambda=M(\lambda)$ hence we have 
\[
N(\lambda+\varepsilon)=
\sup_{\lambda\leq\mu\leq\lambda+\varepsilon}M(\mu). 
\]
But $M(\mu)$ above is either $N_A(\mu)$ either $\mu$. In the second
case $\mu\leq\lambda+\varepsilon$ and  in the first case
\[
N_A(\mu)\leq N_A(\lambda+\varepsilon)\leq N_A(\lambda)+\varepsilon
\leq \lambda +\varepsilon. 
\]
Thus we certainly have $N(\lambda+\varepsilon)\leq\lambda+\varepsilon$
and $\lambda=N(\lambda)$ because $\lambda\in A\cup B$.
\qed

\PAR\label{ss:scs} From Theorem \ref{th:thr} we shall deduce now an
optimal version of the limiting absorption principle. Optimality
refers both to the Besov spaces in which we establish the existence of
the boundary values of the resolvent and to the degree of regularity
of the Hamiltonian with respect to the conjugate operator $D$. This
regularity condition involves the following Besov type class of
operators.  An operator $T\in L(\ch)$ is of class $C^{1,1}(D)$ if
\begin{equation}\label{eq:c11}
\int_0^1\|W^*_{2\varepsilon}T W_{2\varepsilon}-
2W^*_{\varepsilon}T W_{\varepsilon} +T\|
\frac{\rmd\varepsilon}{\varepsilon^2} \equiv
\int_0^1\|(\cw_\varepsilon-1)^2
T\|\frac{\rmd\varepsilon}{\varepsilon^2} <\infty
\end{equation}
where $\cw_\varepsilon$ is the automorphism of $L(\ch)$ defined by
$\cw_\varepsilon T=W_\varepsilon^*T W_\varepsilon$. The condition
\eqref{eq:c11} implies $T$ is of class $C^1_\rmu(D)$ and is just
slightly more than this.  For example, if $T$ is of class $C^1(D)$, so
the commutator $[D,T]$ is a bounded operator, and if
\begin{equation}\label{eq:din}
\int_0^1\|W_\varepsilon^*[D,T]W_\varepsilon-T\|
\frac{\rmd\varepsilon}{\varepsilon} <\infty,
\end{equation}
then $T$ is  of class $C^{1,1}(D)$. A self-adjoint operator $H$ is
called of class $C^{1,1}(D)$ if its resolvent is of class
$C^{1,1}(D)$. We refer to \cite{ABG} for a more thorough discussion of
these matters.

The next result is a consequence of Theorem \ref{th:thr} and of
Theorem 7.4.1 from \cite{ABG}. We set $\ch_{s,p}=\oplus_X\ch_{s,p}(X)$
where the $\ch_{s,p}(X)$ are the Besov spaces associated to the
position observable on $X$ (these are obtained from the usual Besov
spaces associated to $L^2(X)$ by a Fourier transformation).  Let
$\mbc_+$ be the open upper half plane and
$\mbc^H_+=\mbc_+\cup(\mbr\setminus\tau(H))$. If we replace the upper
half plane by the lower one we similarly get the sets $\mbc_-$ and
$\mbc^H_-$.  

\begin{theorem}\label{th:c11}
If $H$ is of class $C^{1,1}(D)$ then its singular continuous spectrum
is empty. The holomorphic maps $\mbc_\pm\ni z\mapsto(H-z)^{-1}\in
L(\ch_{1/2,1},\ch_{-1/2,\infty})$ extend to weak$^*$ continuous
functions on $\mbc^H_\pm$.
\end{theorem}

\PAR\label{ss:mex} Here we describe an explicit class of
non-relativistic many-body Hamiltonians of class $C^1_\rmu(D)$ and
then make a comment on the class $C^{1,1}(D)$.  To simplify
notations we shall consider only interactions which are relatively
bounded in \emph{operator} sense with respect to the kinetic energy
and summarize all the conditions in this context below.

\begin{proposition}\label{pr:resume}
Under the following assumptions the conditions of Theorem
\ref{th:thr} are satisfied and the domain of $H$ is equal to
$\ch^2$.
\begin{compactenum}
\item[{\rm(i)}] $\cs$ is a finite set of subspaces of an Euclidean
  space $\cx$ with $\cx\in\cs$ and such that $X\cap Y\in\cs$ if
  $X,Y\in\cs$. The Hilbert space of the system is
  $\ch=\oplus_X\ch(X)$ and its kinetic energy is
  $K=\oplus_X\Delta_X$ with domain $\ch^2=\oplus_X\ch^2(X)$. The
  total Hamiltonian is $H=K+I$ where the interaction is an operator
  $I=(I_{XY})_{X,Y\in\cs}:\ch^2\to\ch$ with the properties described
  below.

\item[{\rm(ii)}] The operators $I_{XY}:\ch^2(Y)\to\ch(X)$ are of the
  form $I_{XY}=\sum_Z I_{XY}(Z)$ with $I_{XY}(Z)=0$ if $Z\not\subset
  X\cap Y$ and if $Z\subset X\cap Y$ then
\[
  I_{XY}(Z)=1\otimes I^Z_{XY} 
\hspace{2mm} \text{relatively to} \hspace{2mm}
\ch(Y)=\ch(Z)\otimes\ch(Y/Z), \hspace{2mm}
\ch(X)=\ch(Z)\otimes\ch(X/Z)
\]
where $I^Z_{XY}:\ch^2(Y/Z)\to\ch(X/Z)$ is a compact operator
satisfying $(I^{Z}_{XY})^*\supset I^Z_{YX}$. 
  
\item[{\rm(iii)}]\label{p:pmex} We require $[D,I^Z_{XY}]$ to be a
  compact operator $\ch^2(Y/Z)\to\ch^{-2}(X/Z)$.
\end{compactenum}
\end{proposition}

Note that under the assumption (ii) the operator
\begin{equation}\label{eq:D}
[D,I^Z_{XY}]\equiv 
D_{X/Z}I^Z_{XY}-I^Z_{XY}D_{Y/Z}:
\ch^2_{\mathrm{loc}}(Y/Z)\to\ch^{-1}_{\mathrm{loc}}(X/Z)
\end{equation}
is well defined.  We indicated by a subindex the space where the
operator $D$ acts and we used for example
\begin{equation}\label{eq:dfact}
D_X=D_Z\otimes 1 + 1\otimes D_{X/Z} \quad \text{relatively to }
\ch(X)=\ch(Z)\otimes\ch(X/Z).
\end{equation}

\begin{remark}\label{re:cod}
If condition (ii) is satisfied for \emph{all} $X,Y,Z$, and since
$I_{XY}^Z$ is a restriction of the adjoint of $I_{YX}^Z$, we get by
interpolation 
\begin{equation}\label{eq:interpol}
I^Z_{XY}:\ch^\theta(Y/Z)\to\ch^{\theta-2}(X/Z) \quad 
\text{is a compact operator for all } 0\leq \theta \leq 2.
\end{equation}
\end{remark}

We make a comment on the compactness assumption from condition (ii)
of Proposition \ref{pr:resume}.  If $E,F$ are Euclidean spaces let
us set
\begin{equation}\label{eq:kef}
\rk^2_{FE} =K(\ch^2(E),\ch(F)) \quad \text{and} \quad
\rk^{2}_E=\rk^2_{E,E}. 
\end{equation}
If we set $E=(X\cap Y)/Z$ then $Y/Z=E\oplus(Y/X)$ and
$X/Z=E\oplus(X/Y)$ hence
\begin{equation}\label{eq:ixyzc}
\ch(X/Z)=\ch(E)\otimes\ch(X/Y) \text{ and }
\ch^2(Y/Z)=\big(\ch^2(E)\otimes\ch(Y/X)\big)\cap
\big(\ch(E)\otimes\ch^2(Y/X)\big).
\end{equation}
From \eqref{eq:EFHEF} we then get
\begin{align*}
K(\ch^2(Y/Z),\ch(X/Z)) & =
K(\ch^2(E),\ch(E))\otimes K(\ch(Y/X),\ch(X/Y)) \\
& + K(\ch(E),\ch(E)) \otimes K(\ch^2(Y/X),\ch(X/Y)).
\end{align*}
With the abbreviations introduced before this may also be written
\begin{equation}\label{eq:kde3}
\rk^2_{X/Z,Y/Z} = \rk^2_E\otimes \rk_{X/Y,Y/X} + 
\rk_E\otimes\rk^2_{X/Y,Y/X}.
\end{equation}
Condition (ii) of Proposition \ref{pr:resume} requires
$I_{XY}^Z\in\rk^2_{X/Z,Y/Z}$. According to the preceding relation
this means
\begin{equation}\label{eq:kde4}
I_{XY}^Z=J+J' \text{ for some }
J\in \rk^2_E\otimes \rk_{X/Y,Y/X} \text{ and }
J'\in\rk_E\otimes\rk^2_{X/Y,Y/X}. 
\end{equation}
Some special cases of these conditions are worth to be mentioned, we
shall consider this only for $J$, the discussion for $J'$ is
similar. We recall the notation $X \boxplus Y= X/Y\times Y/X$ and that
we identify a Hilbert-Schmidt operator with its kernel. Thus we have
an embedding $L^2(X \boxplus Y)\subset\rk_{X/Y,Y/X}$ hence
\[
\rk^2_E\otimes \rk_{X/Y,Y/X}\supset
\rk^2_E\otimes L^2(X/Y\times Y/X) \supset
L^2(X/Y\times Y/X;\rk^2_E)
\]
cf. the discussion in \S\ref{ss:ha} and Definition \ref{df:ww}. The
condition $I_{XY}^Z \in L^2(X/Y\times Y/X;\rk^2_E)$ is very explicit
and seems to us already quite general. The action of $I_{XY}^Z$
under this condition may be described as follows.
Think of $u\in\ch^2(Y/Z)$ as an element of $L^2(Y/X;\ch^2(E))$. Then
we may represent $I_{XY}^Z u$ as element of 
$\ch(X/Z)=L^2(X/Y;\ch(E))$ as
\[
(I_{XY}^Z u)(x')={\textstyle\int_{Y/X}} I_{XY}^Z(x',y')u(y') \rmd y'.
\]
Observe that if we assume $I_{XY}^Z \in L^2(X \boxplus Y;\rk^2_E)$ for
\emph{all} $X,Y,Z$ then as in Remark \ref{re:cod} we get
\[
I_{XY}^Z\in L^2(X \boxplus Y;K(\ch(E)^\theta,\ch^{\theta-2}(E))
\quad\text{for all } 0\leq\theta\leq2.
\]

We now consider a Hamiltonian satisfying (i)--(iii) of Proposition
\ref{pr:resume} and discuss conditions which ensure that $H$ is of
class $\rc^{1,1}(D)$. It is important to observe that the domain
$\ch^2$ of $H$ is stable under the dilation group $W_\tau$.  Thus we
may use Theorem 6.3.4 from \cite{ABG} to see that $H$ is of class
$\rc^{1,1}(D)$ if and only if
\begin{equation}\label{eq:c11h}
\int_0^1\|(\cw_\varepsilon-1)^2H\|_{\ch^2\to\ch^{-2}}
\frac{\rmd\varepsilon}{\varepsilon^2} <\infty.
\end{equation}
Here $\cw_{\varepsilon} H=W^*_{\varepsilon}H W_{\varepsilon}$ hence
\[
(\cw_\varepsilon-1)^2H=W^*_{2\varepsilon}H W_{2\varepsilon}-
2W^*_{\varepsilon}H W_{\varepsilon} +H.
\]
The relation \eqref{eq:c11h} is trivially verified by the kinetic part
$\Delta$ of $H$ hence we need that \eqref{eq:c11h} be satisfied with
$H$ replaced by $I$. The condition we get will be satisfied if and
only if each coefficient $I_{XY}$ of $I$ satisfies a similar
relation. Thus it suffices to have
\begin{equation}\label{eq:c11xyz}
\int_0^1\|(\cw_\varepsilon-1)^2I_{XY}^Z\|_{\ch^2(Y/Z)\to\ch^{-2}(X/Z)}
\frac{\rmd\varepsilon}{\varepsilon^2} <\infty
\quad\text{for all } X,Y,Z.
\end{equation}
A similar argument may be used in the context of the Dini condition
\eqref{eq:dini} to get as sufficient conditions
\begin{equation}\label{eq:dinix}
\int_0^1\|W_\varepsilon^*[D,I_{XY}^Z]W_\varepsilon-[D,I_{XY}^Z]
\|_{\ch^2(Y/Z)\to\ch^{-2}(X/Z)}
\frac{\rmd\varepsilon}{\varepsilon} <\infty.
\end{equation}
In fact each of the three terms in the decomposition
\begin{equation}\label{eq:deco}
[D,I^Z_{XY}] =[D_E,I^Z_{XY}] +D_{X/Y}I^Z_{XY} -I^Z_{XY}D_{Y/X}
\end{equation}
(see \eqref{eq:dec}) should be treated separately.

The techniques developed in \S 7.5.3 and on pages 425--429 from
\cite{ABG} can be used to get optimal and more concrete conditions.
The only new fact with respect to the $N$-body situation as treated
in \cite{ABG} is that $\cw_\tau:T\mapsto W_{-\tau}T W_\tau$ when
considered as an operator on $L(\ch(Y/Z),\ch(X/Z))$ factorizes in a
product of three commuting operators. Indeed, if we write
\[
\ch(Y/Z)=\ch(E)\otimes\ch(Y/X), \quad
\ch(X/Z)=\ch(E)\otimes\ch(X/Y)
\]
then we get $\cw_\tau(T)=W^{X/Y}_{-\tau}\cw^E_\tau(T)W^{Y/X}_\tau$
where this time we indicated by an upper index the space to which
the operator is related and, for example, we identified
$W^{Y/X}_\tau=1_E\otimes W^{Y/X}_\tau$. Let $L_\tau$ be the operator
of left multiplication by $W^{X/Y}_{-\tau}$ and $N_\tau$ the
operator of right multiplication by $W^{Y/X}_{\tau}$ on
$L(\ch(Y/Z),\ch(X/Z))$. If we also set $M_\tau=\cw^E_\tau$ then we
get three commuting operators $L_\tau,M_\tau,N\tau$ on
$L(\ch(Y/Z),\ch(X/Z))$ such that $\cw_\tau=\cl_\tau \cm_\tau N_\tau$.
Then in order to check a Dini type condition as \eqref{eq:dinix} we
use 
\begin{equation}\label{eq:tmp}
\cw_\tau-1=(\cl_\tau-1)\cm_\tau N_\tau+(\cm_\tau -1)N_\tau+N_\tau-1 
\end{equation}
hence
\[
\|W^*_\tau T W_\tau-T\|\leq
\|(W^{X/Y}_{-\tau}-1) T\|+\|W^E_{-\tau} T W^E_\tau-T\| +
\|T(W^{Y/X}_{\tau}-1)\|.
\]
This relation remains true modulo a constant factor if the norms are
those of $L(\ch^2(Y/Z),\ch^{-2}(X/Z))$. An analog argument works for
the second order differences. Indeed, if $A,B,C$ are commuting
operators on a Banach space then starting from
\[
(AB-1)^2=(A-1)^2B^2 +2(A-1)(B-1)B+ (B-1)^2
\]
we obtain
\begin{align*}
(ABC-1)^2 &=(A-1)^2B^2C^2 +2(A-1)(B-1)BC^2+ 2(A-1)B(C-1)C^2 \\
& + (B-1)^2C^2 +2(B-1)(C-1)C +(C-1)^2.
\end{align*}
Thus in our case we get the estimate
\begin{align*}
\|(\cw_\tau-1)^2T\| & \leq 
\|(\cl_\tau-1)^2T\| + \|(\cm_\tau-1)^2T\| +\|(\cn_\tau-1)^2T\|
 + 2\|(\cl_\tau-1)(\cm_\tau-1)T\| \\ 
& + 2\|(\cl_\tau-1)(\cn_\tau-1)T\| + 2\|(\cm_\tau-1)(\cn_\tau-1)T\|
\end{align*}
which remains true modulo a constant factor if the norms are those
of $L(\ch^2(Y/Z),\ch^{-2}(X/Z))$. This relation is helpful in
checking the $C^{1,1}(D)$ property. However, it is possible to go
further and to get rid off the last three terms by interpreting
\eqref{eq:c11xyz} in terms of real interpolation theory.

\begin{lemma}\label{lm:inter}
If $T\in\rh\equiv L(\ch^2(Y/Z),\ch^{-2}(X/Z))$ then 
$\int_0^1\|(\cw_\varepsilon-1)^2T\|_\rh
\rmd\varepsilon/\varepsilon^2<\infty$ follows from
\begin{equation}\label{eq:inter}
\int_0^1\left(
\|(W^{X/Y}_{\varepsilon}-1)^2T\|_\rh+ 
\|(\cw^E_\varepsilon-1)^2T\|_\rh+
\|T(W^{Y/X}_{\varepsilon}-1)^2\|_\rh
\right)
\frac{\rmd\varepsilon}{\varepsilon^2} < \infty.
\end{equation}
\end{lemma}
\proof We use the notations and conventions from \cite{ABG}. Observe
that $\cw_\tau,\cl_\tau,\cm_\tau,\cn_\tau$ are one parameter groups
of operators on the Banach space $\rh=L(\ch^2(Y/Z),\ch^{-2}(X/Z))$.
These groups are not continuous in the ordinary sense but this does
not really matter, in fact we are in the setting of \cite[Chapter
  5]{ABG}. The main point is that the finiteness of the integral 
$\int_0^1\|(\cw_\varepsilon-1)^2T\|_\rh
\rmd\varepsilon/\varepsilon^2<\infty$ 
is equivalent to that of
$\int_0^1\|(\cw_\varepsilon-1)^6T\|_\rh
\rmd\varepsilon/\varepsilon^2<\infty$.
Now by taking the sixth power of \eqref{eq:tmp} and developing the
right hand side we easily get the result, cf. the formula on top of
page 132 of \cite{ABG}.
\qed

\PAR\label{ss:xsupy} To see the relation with the
creation-annihilation type interactions characteristic to quantum
field models we consider in detail the simplest situation when
$Y\subset X$ strictly. For any $X,Y$ we define
\[
\ri_{XY}= {\textstyle\sum_{Z\in\cs(X\cap Y)}} 
1_Z\otimes \rk^2_{X/Z,Y/Z}\subset \rl^{0,2}_{XY} \quad\text{and }
\ri_{X}\equiv\ri_{XX}.
\]
Note that the sum is direct and $\ri_{XY}$ is closed. A
non-relativistic $N$-body Hamiltonian associated to the semilattice
$\cs(X)$ of subspaces of $X$ is usually of the form $\Delta_X+V$ with
$V\in\ri_{X}$.

If $Y\subset X$ then, according to \eqref{eq:xby},
\begin{equation}\label{eq:ysux}
\rc_{XY}=\rc_Y\otimes\ch(X/Y), \quad 
\rc_{XY}(Z)=\rc_Y(Z)\otimes\ch(X/Y), \quad
\ch(X)=\ch(Y)\otimes\ch(X/Y)
\end{equation}
where the first two tensor product have to be interpreted as
explained in \S\ref{ss:ha}. In particular we have
\begin{equation}\label{eq:sux}
L^2(X/Y;\rc_Y)\subset\rc_{XY} \quad \text{and} \quad
L^2(X/Y;\rc_Y(Z))\subset\rc_{XY}(Z)
\quad \text{strictly}. 
\end{equation}
Note that for each $Z\subset Y$ we have $X=Z\oplus(Y/Z)\oplus(X/Y)$
and $X/Z=(Y/Z)\oplus(X/Y)$. Then $\ch(X/Z)=\ch(Y/Z)\otimes\ch(X/Y)$
and thus the operator $I^Z_{XY}$ from (ii) above is just a
compact operator
\begin{equation}\label{eq:ixyo}
I^Z_{XY} : \ch^2(Y/Z)\to\ch(Y/Z)\otimes\ch(X/Y).
\end{equation}
If $\ce,\cf,\cg$ are Hilbert spaces then $K(\ce,\cf\otimes\cg)\cong
K(\ce,\cf)\otimes\cg$, see \S\ref{ss:ha}. Hence \eqref{eq:ixyo} means
\begin{equation}\label{eq:ixyoo}
I^Z_{XY} \in \rk^2_{Y/Z}\otimes \ch(X/Y)
\end{equation}
so the interaction which couples the $X$ and $Y$ systems is
\begin{equation}\label{eq:xyinter}
I_{XY}={\textstyle\sum_{Z\in\cs(Y)}} 
1_Z\otimes I^Z_{XY} \in \ri_{Y}\otimes \ch(X/Y). 
\end{equation}
Now according to \eqref{eq:xyinter} we may view $I_{XY}$ as an
element of $L^2_\rmw(X/Y;\ri_Y)$ (see Definition \ref{df:ww}). This
``weakly square integrable'' function $I_{XY}:X/Y\to\ri_Y$
determines the operator $I_{XY}:\ch^2(Y)\to\ch(X)$ by the following
rule: it associates to $u\in\ch^2(Y)$ the function $y'\mapsto
I_{XY}(y')u$ which belongs to $L^2(X/Y;\ch(X/Y))=\ch(X)$.  We may
also write
\begin{equation}\label{eq:IV}
(I_{XY}u)(x)=(I_{XY}(y')u)(y) \quad \text{where }
x\in X=Y\oplus X/Y \text{ is written as } x=(y,y').
\end{equation}
We also say that the operator valued function $I_{XY}$ is the symbol
of the operator $I_{XY}$.

The particular case when the function $I_{XY}$ is factorizable gives
the connection with the quantum field type interactions: assume that
$I_{XY}$ is a finite sum $I_{XY}=\sum_i V_Y^i\otimes\phi_i$ where
$V^i_Y\in\ri_Y$ and $\phi_i\in \ch(X/Y)$, then
\begin{equation}\label{eq:qft}
I_{XY}u={\textstyle\sum_i} (V_Y^i u)\otimes\phi_i \quad
\text{as an operator }  
I_{XY}:\ch^2(Y)\to\ch(X)=\ch(Y)\otimes\ch(X/Y).
\end{equation}
This is a sum of $N$-body type interactions $V^i_Y$ tensorized with
operators which create particles in states $\phi_i$.  Note that this
type of interactions is more subtle than those usually considered in
quantum field theory.

We mention that the adjoint $I_{YX}=I_{XY}^*$ acts like an integral
operator in the $y'$ variable (like an annihilation operator). Indeed,
if $v\in\ch(X)$ is thought as a map $y'\mapsto v(y')\in\ch(Y)$ then we
have $I_{YX}v=\int_{X/Y} I^*_{XY}(y')v(y')\rmd y'$ at least formally.

Now the conditions on the ``commutator'' $[D,I_{XY}]$ may be written
in a quite explicit form in terms of the symbol $I_{XY}$. The
relation \eqref{eq:deco} becomes $
[D,I_{XY}]=[D_Y,I_{XY}]+D_{X/Y}I_{XY}$. The operator $D_Y$ acts only
on the variable $y$ and $D_{X/Y}$ acts only on the variable
$y'$. Thus $[D_Y,I_{XY}]$ and $D_{X/Y}I_{XY}$ are operators of the
same nature as $I_{XY}$ but more singular. Indeed, the symbol of
$[D_Y,I_{XY}]$ is the function $y'\mapsto [D_Y,I_{XY}(y')]$ and that
of $2iD_{X/Y}I_{XY}$ is the function $ y'\mapsto
(y'\cdot\nabla_{y'}+n/2) I_{XY}(y')$.  Thus we see that to get
condition (iii) of Proposition \ref{pr:resume} it suffices to
require two types of conditions on the symbol $I_{XY}$, one on
$[D_Y,I_{XY}(y')]$ and a second one on
$y'\cdot\nabla_{y'}I_{XY}(y')$.

To state more explicit conditions we need to decompose $I_{XY}$ as in
\eqref{eq:xyinter}. For this we assume given for each $Z\in\cs$ with
$Z\subset Y$ a function $I^Z_{XY}:X/Y\to \rk^2_{Y/Z}$ in
$L^2_\rmw(X/Y;\rk^2_{Y/Z})$. This is the symbol of an operator
$\ch^2(Y/Z)\to L^2(X/Y;\ch(Y/Z))=\ch(X/Z)$ that we also denote
$I^Z_{XY}$ and which is clearly compact. Then we take
$I_{XY}=\sum_{Z\in\cs}I^Z_{XY}$.

Now each ``commutator'' 
$[D,I_{XY}^Z]=[D_{Y/Z},I^Z_{XY}]+D_{X/Y}I^Z_{XY}$ should be a
compact operator from $\ch^2(Y/Z)$ to $\ch^{-2}(X/Z)$. For
simplicity we shall ask that each of the two components satisfies
this compactness condition.

As explained before the operator $[D_{Y/Z},I^Z_{XY}]$ is associated
to the symbol $y'\mapsto [D_{Y/Z},I^Z_{XY}(y')]$ and the main
contribution to the operator $2iD_{X/Y}I^Z_{XY}$ comes from the
operator associated to the symbol $y'\mapsto
y'\cdot\nabla_{y'}I^Z_{XY}(y')$. So we ask that these two symbols
induce compact operators $\ch^2(Y/Z)\to\ch^{-2}(X/Z)$.  On the other
hand, from \eqref{eq:stens} and $X/Z=(Y/Z)\oplus(X/Y)$ we get
\begin{align}
\ch^2(X/Z) &= \big(\ch(Y/Z)\otimes\ch^2(X/Y)\big)\cap
\big(\ch^2(Y/Z)\otimes\ch(X/Y)\big),
\label{eq:klean1}\\
\ch^{-2}(X/Z) &= \ch(Y/Z)\otimes\ch^{-2}(X/Y) +
\ch^{-2}(Y/Z)\otimes\ch(X/Y). \label{eq:klean2}
\end{align}  
This allows one to write down general and more or less explicit
conditions to ensure that that the operator $I_{XY}$ satisfies the
conditions (ii) and (iii) of Proposition \ref{pr:resume} in the case
$Y\subset X$. Without trying to go into any refinements we now state
a sufficient set of assumptions on the symbols $I^Z_{XY}$. We find
convenient to revert to the abstract tensor product notation.
\begin{compactenum}

\item[(a)] $I^Z_{XY}\in K(\ch^2(Y/Z),\ch(Y/Z))\otimes\ch(X/Y)$,

\item[(b)] $[D_{Y/Z},I^Z_{XY}]\in
  K(\ch^2(Y/Z),\ch^{-2}(Y/Z))\otimes\ch(X/Y)$,

\item[(c)] $D_{X/Y}I^Z_{XY}\in 
K(\ch^2(Y/Z),\ch(Y/Z))\otimes\ch^{-2}(X/Y)$.

\end{compactenum}

\appendix
\section{Appendix}
\label{s:app}
\renewcommand{\theequation}{A.\arabic{equation}}
\setcounter{equation}{0}

The main part of this appendix is devoted to comments concerning the
generation of $C^*$-algebras of ``energy observables'' by certain
classes of ``elementary'' Hamiltonians. Then we prove a useful
technical result.

\PAR\label{ss:a1}
Let $X$ be a lca group and let $\{U_x\}_{x\in X}$ be a strongly
continuous unitary representation of $X$ on a Hilbert space
$\ch$. Then one can associate to it a Borel regular spectral measure
$E$ on $X^*$ with values projectors on $\ch$ such that
$U_x=\int_{X^*}k(x)E(\rmd k)$ and this allows us to define for each
Borel function $\psi:X^*\to\mbc$ a normal operator on $\ch$ by the
formula $\psi(P)=\int_{X^*} \psi(k)E(\rmd k)$.  The set $\cc^*(X;\ch)$
of all the operators $\psi(P)$ with $\psi\in\Co(X^*)$ is clearly a
non-degenerate $C^*$-algebra of operators on $\ch$.  We say that an
operator $S\in L(\ch)$ is of class $C^0(P)$ if the map $x\mapsto
U_xSU_x^*$ is norm continuous.

\begin{lemma}\label{lm:cop}
Let $S\in L(\ch)$ be of class $C^0(P)$ and let $T\in
\cc^*(X;\ch)$. Then for each $\varepsilon>0$ there is $Y\subset X$
finite and there are operators $T_y\in \cc^*(X;\ch)$ such that
$\|ST-\sum_{y\in Y} T_y U_{y}SU_{y}^*\|<\varepsilon$.
\end{lemma}
\proof It suffices to assume that $T=\psi(P)$ where $\psi$ has a
Fourier transform integrable on $X$, so that $T=\int_X U_x
\what\psi(x) \rmd x$, and then to use a partition of unity on $X$
and the uniform continuity of the map $x\mapsto U_xSU_x^*$ (see the
proof of Lemma 2.1 in \cite{DG1}).  \qed

We say that a subset $\cb$ of $L(\ch)$ is $X$-stable if
$U_xSU_x^*\in\cb$ whenever $S\in\cb$ and $x\in X$. From Lemma
\ref{lm:cop} we see that if $\cb$ is an $X$-stable real linear space
of operators of class $C^0(P)$ then
\[
\cb\cdot \cc^*(X;\ch)= \cc^*(X;\ch)\cdot\cb.
\] 
Since the $C^*$-algebra $\ca$ generated by $\cb$ is also $X$-stable
and consists of operators of class $C^0(P)$
\begin{equation}\label{eq:cop}
\ra\equiv\ca\cdot \cc^*(X;\ch)= \cc^*(X;\ch)\cdot\ca
\end{equation} 
is a $C^*$-algebra. The operators $U_x$ implement a norm continuous
action of $X$ by automorphisms of the algebra $\ca$ so the
$C^*$-algebra crossed product $\ca\rtimes X$ is well defined and the
algebra $\ra$ is a quotient of this crossed product.

A function $h$ on $X^*$ is called \emph{$p$-periodic} for some
non-zero $p\in X^*$ if $h(k+p)=h(k)$ for all $k\in X^*$.

\begin{proposition}\label{pr:cop}
Let $\cv$ be an $X$-stable set of symmetric bounded operators of
class $C^0(P)$ and such that $\lambda\cv\subset\cv$ if
$\lambda\in\mbr$. Denote $\ca$ the $C^*$-algebra generated by $\cv$
and define $\ra$ by \eqref{eq:cop}.  Let $h:X^*\to\mbr$ be
continuous, not $p$-periodic if $p\neq0$, and such that
$|h(k)|\to\infty$ as $k\to\infty$.  Then $\ra$ is the $C^*$-algebra
generated by the self-adjoint operators of the form $h(P+k)+V$ with
$k\in X^*$ and $V\in\cv$.
\end{proposition}
\proof Denote $K=h(P+k)$ and let $R_\lambda=(z-K-\lambda V)^{-1}$
with $z$ not real and $\lambda$ real. Let $\rc$ be the $C^*$-algebra
generated by such operators (with varying $k$ and $V$). By taking
$V=0$ we see that $\rc$ will contain the $C^*$-algebra generated by
the operators $R_0$. By the Stone-Weierstrass theorem this algebra
is $\cc^*(X;\ch)$ because the set of functions $p\to(z-h(p+k))^{-1}$
where $k$ runs over $X^*$ separates the points of $X^*$. The
derivative with respect to $\lambda$ at $\lambda=0$ of $R_\lambda$
exists in norm and is equal to $R_0VR_0$, so $R_0VR_0\in\rc$. Since
$\cc^*(X)\subset\rc$ we get $\phi(P)V\psi(P)\in\rc$ for all
$\phi,\psi\in\Co(X^*)$ and all $V\in\cv$. Since $V$ is of class
$C^0(P)$ we have $(U_x-1)V\psi(P)\sim V(U_x-1)\psi(P)\to0$ in norm
as $x\to0$ from which we get $\phi(P)V\psi(P)\to S\psi(P)$ in norm
as $\phi\to1$ conveniently. Thus $V\psi(P)\in\rc$ for $V,\psi$ as
above. This implies $V_1\cdots V_n\psi(P)\in\rc$ for all
$V_1,\dots,V_n\in\cv$. Indeed, assuming $n=2$ for simplicity, we
write $\psi=\psi_1\psi_2$ with $\psi_i\in\Co(X^*)$ and then Lemma
\ref{lm:cop} allows us to approximate $V_2\psi_1(P)$ in norm with
linear combinations of operators of the form $\phi(P)V^x_2$ where
the $V^x_2$ are translates of $V_2$.  Since $\rc$ is an algebra we
get $V_1\phi(P) V^x_2\psi_2(P)\in\rc$ hence passing to the limit we
get $V_1V_2\psi(P)\in\rc$. Thus we proved $\ra\subset\rc$. The
converse inclusion follows from a series expansion of $R_\lambda$ in
powers of $V$.  \qed

The next two corollaries follow easily from Proposition
\ref{pr:cop}. We take $\ch=L^2(X)$ which is equipped with the usual
representations $U_x,V_k$ of $X$ and $X^*$ respectively. Let
$W_\xi=U_xV_k$ with $\xi=(x,k)$ be the phase space translation
operator, so that $\{W_\xi\}$ is a projective representation of the
phase space $\Xi=X\oplus X^*$.  Fix some classical kinetic energy
function $h$ as in the statement of Proposition \ref{pr:cop} and let
the classical potential $v:X\to\mbr$ be a bounded uniformly
continuous function. Then the quantum Hamiltonian will be
$H=h(P)+v(Q)\equiv K+V$. Since the origins in the configuration and
momentum spaces $X$ and $X^*$ have no special physical meaning one
may argue \cite{Be1,Be2} that $W_\xi H W^*_\xi=h(P-k)+v(Q+x)$ is a
Hamiltonian as good as $H$ for the description of the evolution of
the system. It is not clear to us whether the algebra generated by
such Hamiltonians (with $h$ and $v$ fixed) is in a natural way a
crossed product. On the other hand, it is natural to say that the
coupling constant in front of the potential is also a variable of
the system and so the Hamiltonians $H_\lambda=K+\lambda V$ with any
real $\lambda$ are as relevant as $H$. Then we may apply Proposition
\ref{pr:cop} with $\cv$ equal to the set of operators of the form
$\lambda\tau_xv(Q)$. Thus:

\begin{corollary}\label{co:cop1}
Let $v\in\Cbu(X)$ real and let $\ca$ be the $C^*$-subalgebra of
$\Cbu(X)$ generated by the translates of $v$.  Let $h:X^*\to\mbr$ be
continuous, not $p$-periodic if $p\neq0$, and such that
$|h(k)|\to\infty$ as $k\to\infty$.  Then the $C^*$-algebra generated
by the self-adjoint operators of the form $W_\xi H_\lambda W^*_\xi$
with $\xi\in\Xi$ and real $\lambda$ is the crossed product
$\ca\rtimes X$.
\end{corollary}

Now let $\ct$ be a set of closed subgroups of $X$ such that the
semilattice $\cs$ generated by it (i.e. the set of finite
intersections of elements of $\ct$) consists of pairwise compatible
subgroups. Set $\cc_X(\cs)=\sum^\rmc_{Y\in\cs} \cc_X(Y)$. From
\eqref{eq:reg1} it follows that this is the $C^*$-algebra generated
by  $\sum_{Y\in\ct} \cc_X(Y)$.

\begin{corollary}\label{co:cop2}
Let $h$ be as in Corollary \ref{co:cop1}. Then the $C^*$-algebra
generated by the self-adjoint operators of the form $h(P+k)+v(Q)$
with $k\in X^*$ and $v\in\sum_{Y\in\ct} \cc_X(Y)$ is the
crossed product $\cc_X(\cs)\rtimes X$.
\end{corollary}

\begin{remark}\label{re:cop}
Proposition \ref{pr:cop} and Corollaries \ref{co:cop1} and
\ref{co:cop2} remain true and are easier to prove if we consider the
$C^*$-algebra generated by the operators $h(P)+V$ with all
$h:X^*\to\mbr$ continuous and such that $|h(k)|\to\infty$ as
$k\to\infty$. If in Proposition \ref{pr:cop} we take $\ch=L^2(X;E)$
with $E$ a finite dimensional Hilbert space (describing the spin
degrees of freedom) then the operators $H_0=h(P)$ with $h:X\to L(E)$
a continuous symmetric operator valued function such that
$\|(h(k)+i)^{-1}\|\to 0$ as $k\to\infty$ are affiliated to $\ra$
hence also their perturbations $H_0+V$ where $V$ satisfies the
criteria from \cite{DG3}, for example.
\end{remark}

{\bf Proof of Theorem \ref{th:motiv}:\ } In the remaining part of
the appendix we use the notations of \S\ref{ss:motiv}.

Let $\rc'$ be the $C^*$-algebra generated by the operators of the form
$(z-K-\phi)^{-1}$ where $z$ is a not real number, $K$ is a standard
kinetic energy operator, and $\phi$ is a symmetric field
operator. With the notation \eqref{eq:d} we easily get
$\cc^*(\cs)\subset\rc'$.  If $\lambda\in\mbr$ then $\lambda\phi$ is
also a field operator so $(z-K-\lambda\phi)^{-1}\in\rc'$. By taking
the derivative with respect to $\lambda$ at $\lambda=0$ of this
operator we get $(z-K)^{-1}\phi (z-K)^{-1}\in\rc$. Since
$(z-K)^{-1}=\oplus_X(z-h_X(P))^{-1}$ (recall that $P$ is the momentum
observable independently of the group $X$) and since
$\cc^*(\cs)\subset\rc'$ we get $S\phi(\theta) T\in\rc'$ for all
$S,T\in \cc^*(\cs)$ and $\theta=(\theta_{XY})_{X\supset Y}$,
cf. \S\ref{ss:motiv}.

Let $\rc'_{XY}=\Pi_X\rc'\Pi_Y\subset\rl_{XY}$ be the components of
the algebra $\rc'$ and let us fix $X\supset Y$. Then we get
$\varphi(P) a^*(u) \psi(P)\in\rc'_{XY}$ for all
$\varphi\in\Co(X^*)$, $\psi\in\Co(Y^*)$, and $u\in\ch(X/Y)$.  The
clspan of the operators $a^*(u) \psi(P)$ is $\rt_{XY}$, see
Proposition \ref{pr:def3} and the comments after \eqref{eq:L2a}, and
from \eqref{eq:cyzc} we have $\cc^*(X)\cdot\rt_{XY}=\rt_{XY}$.  Thus
the clspan of the operators $\varphi(P) a^*(u) \psi(P)$ is
$\rt_{XY}$ for each $X\supset Y$ and then we get
$\rt_{XY}\subset\rc'_{XY}$.  By taking adjoints we get
$\rt_{XY}\subset\rc'_{XY}$ if $X\sim Y$.

Now recall that the subspace $\rt^\circ\subset L(\ch)$ defined by
$\rt^\circ_{XY}=\rt_{XY}$ if $X\sim Y$ and $\rt^\circ=\{0\}$ if
$X\not\sim Y$ is a closed self-adjoint linear subspace of $\rt$ and
that $\rt^\circ\cdot\rt^\circ=\rc$, cf. Proposition \ref{pr:tc}. By
what we proved before we have $\rt^\circ\subset\rc'$ hence
$\rc\subset\rc'$. The converse inclusions is easy to prove. This
finishes the proof of Theorem \ref{th:motiv}.

\PAR\label{ss:a2} We prove here a useful technical result. Let
$\ce,\cf,\cg,\ch$ be Hilbert spaces and assume that we have continuous
injective embeddings $\ce\subset\cg$ and $\cf\subset\cg$. Let us equip
$\ce\cap\cf$ with the intersection topology defined by the norm
$(\|g\|_\ce^2+\|g\|_\cf^2)^{1/2}$. It is clear that $\ce\cap\cf$
becomes a Hilbert space continuously embedded in $\cg$.

\begin{lemma}\label{lm:efgh}
The map $K(\ce,\ch)\times K(\cf,\ch) \to K(\ce\cap\cf,\ch)$ which
associates to $S\in K(\ce,\ch)$ and $T\in K(\cf,\ch)$ the operator
$S|_{\ce\cap\cf}+T|_{\ce\cap\cf} \in K(\ce\cap\cf,\ch)$ is
surjective. 
\end{lemma}
\proof It is clear that the map is well defined. Let $R\in
K(\ce\cap\cf,\ch)$, we have to show that there are $S,T$ as in the
statement of the proposition such that $R=
S|_{\ce\cap\cf}+T|_{\ce\cap\cf}$.  Observe that the norm on
$\ce\cap\cf$ has been chosen such that the linear map
$g\mapsto(g,g)\in\ce\oplus\cf$ be an isometry with range a closed
linear subspace $\ci$. Consider $R$ as a linear map $\ci\to\ch$ and
extend it to the orthogonal of $\ci$ by zero. The so defined map
$\wtilde R:\ci\to\ch$ is clearly compact. Let $S,T$ be defined by
$Se=\wtilde R(e,0)$ and $Tf=\wtilde R(0,f)$. Clearly $S\in
K(\ce,\ch)$ and $T\in K(\cf,\ch)$ and if $g\in\ce\cap\cf$ then
\[
Sg+Tg=\wtilde R(g,0)+\wtilde R(0,g)=\wtilde R(g,g)=Rg
\]
which proves the lemma.
\qed

We shall write the assertion of this lemma in the slightly formal
way
\begin{equation}\label{eq:efgh}
K(\ce\cap\cf,\ch)=K(\ce,\ch) + K(\cf,\ch).
\end{equation}
For example, if $E,F$ are Euclidean spaces and $s>0$ is real then
\begin{equation}\label{eq:EFs}
\ch^s(E\oplus F)=\big(\ch^s(E)\otimes\ch(F)\big)\cap
\big(\ch(E)\otimes\ch^s(F)\big)
\end{equation}
hence for an arbitrary Hilbert space $\ch$ we have
\begin{equation}\label{eq:EFH}
K(\ch^s(E\oplus F),\ch)=
K(\ch^s(E)\otimes\ch(F),\ch) + K(\ch(E)\otimes\ch^s(F),\ch).
\end{equation}
If $\ch$ itself is a tensor product $\ch=\ch_E\otimes\ch_F$ then we
can combine this with  \eqref{eq:comtens} and get
\begin{align}\label{eq:EFHEF}
K(\ch^s(E\oplus F),\ch_E\otimes\ch_F) & =
K(\ch^s(E),\ch_E)\otimes K(\ch(F),\ch_F)\\
& + K(\ch(E),\ch_E) \otimes K(\ch^s(F),\ch_F). \nonumber 
\end{align}

\addcontentsline{toc}{section}{
{References}}

\end{document}